\documentclass[aps,onecolumn,prd,showpacs,showkeys,preprintnumbers,superscriptaddress,nobibnotes,floatfix,longbibliography,notitlepage,nofootinbib]{revtex4-1}

\pdfoutput=1
\usepackage{amsmath}
\usepackage{amsfonts}
\usepackage{amssymb}
\usepackage{mathrsfs}
\usepackage{graphicx}
\usepackage{color}

\usepackage{hyperref}
\usepackage{multirow}
\usepackage{upgreek}
\usepackage[capitalise]{cleveref}

\newcommand{\eq}{Eq.}

\newcommand{\Fig}{Fig.}

\newcommand{\mtr}{\mathrm{tr}}
\newcommand{\msh}{\mathrm{sh}}
\newcommand{\mdb}{\mathrm{db}}
\newcommand{\mdbl}{\mathrm{dbl}}
\newcommand{\mesc}{\mathrm{esc}}
\newcommand{\Emin}{E_\mathrm{min}}
\newcommand{\Emax}{E_\mathrm{max}}
\newcommand{\ai}{\{\alpha_{i,\oplus}\}}
\newcommand{\Ntr}{N_\mathrm{tr}}
\newcommand{\Nsh}{N_\mathrm{sh}}
\newcommand{\Ndb}{N_\mathrm{db}}

\begin{document}

\title{Signatures of microscopic black holes and extra dimensions \\ at future neutrino telescopes}

\author{Katherine J. Mack}
\email{kmack@ncsu.edu}
\affiliation{North Carolina State University, Department of Physics, Raleigh, NC 27695-8202, USA}
\affiliation{Perimeter Institute for Theoretical Physics, Waterloo ON N2L 2Y5, Canada}

\author{Ningqiang Song}
\email{ningqiang.song@queensu.ca}
\affiliation{Arthur B. McDonald Canadian Astroparticle Physics Research Institute, Department of Physics, Engineering Physics and Astronomy, Queen's University, Kingston ON K7L 3N6, Canada}
\affiliation{Perimeter Institute for Theoretical Physics, Waterloo ON N2L 2Y5, Canada}

\author{Aaron C. Vincent}
\email{aaron.vincent@queensu.ca}
\affiliation{Arthur B. McDonald Canadian Astroparticle Physics Research Institute, Department of Physics, Engineering Physics and Astronomy, Queen's University, Kingston ON K7L 3N6, Canada}
\affiliation{Perimeter Institute for Theoretical Physics, Waterloo ON N2L 2Y5, Canada}

%\date{\today}

\begin{abstract}
In scenarios with large extra dimensions (LEDs), the fundamental Planck scale can be low enough that collisions between high-energy particles may produce microscopic black holes. High-energy cosmic neutrinos can carry energies much larger than a PeV, opening the door to a higher energy range than Earth-based colliders. Here, for the first time, we identify a number of unique signatures of microscopic black holes as they would appear in the next generation of large-scale neutrino observatories such as IceCube-Gen2 and the Pacific Ocean Neutrino Explorer. These signatures include new event topologies, energy distributions, and unusual ratios of hadronic-to-electronic energy deposition, visible through Cherenkov light echos due to delayed neutron recombination. We find that the next generation of neutrino telescopes can probe LEDs with a Planck scale up to 6 TeV, though the identification of unique topologies could push their reach even further. 
\end{abstract}

\maketitle

\section{Introduction}
\label{sec:intro}
Among the diverse signatures of new physics above the TeV scale, the possibility of producing microscopic black holes in particle collisions remains one of the most alluring. In models with Large Extra Dimensions (LEDs), the large hierarchy between gravity and the electroweak scales can be at least partly explained by confining the Standard Model (SM) gauge interactions to a 3+1-dimensional \textit{brane}, while allowing gravity to ``leak'' into one or more extra 
\textit{bulk} dimensions \cite{ArkaniHamed:1998rs,Antoniadis:1998ig,ArkaniHamed:1998nn,Argyres:1998qn,Randall:1999ee,Randall:1999vf}. This allows a true Planck scale $M_\star$ that can be much lower than the observed $M_{Pl} \sim 10^{18}$ GeV. Such scenarios have been tested extensively in terms of gravitational force tests \cite{Murata:2014nra}, supernova and neutron star cooling \cite{Hanhart:2001fx,Hannestad:2003yd}, and in collider experiments \cite{Sirunyan:2018xwt,Sirunyan:2018wcm}.  If only one LED exists, Solar System-scale modifications of Newtonian gravity \cite{ArkaniHamed:1998rs} prohibit a TeV-scale Planck mass. However, two or more LEDs remain allowed by observation. Current constraints limit the scale $M_\star$ to be above about 3 -- 25 TeV \cite{Murata:2014nra,Hanhart:2001fx,Hannestad:2003yd,Mack:2018fny,Sirunyan:2018xwt,Sirunyan:2018wcm}, depending on the number of extra dimensions. 

A vastly-reduced Planck scale also allows for the creation of microscopic black holes (BHs) in the collision between two high-energy particles with centre of mass (CM) energy $E_{CM} \gtrsim M_\star$, a smoking gun signature of extra dimensions. The hoop conjecture \cite{thorne1995black} tells us that a black hole will form if the impact parameter $b$ between two colliding particles is smaller than twice the horizon radius $r_H$ of a black hole with mass $M_\bullet = E_{CM}$. This has mainly been studied in the context of high-energy colliders \cite{Banks:1999gd,Dimopoulos:2001hw,Giddings:2001bu,Chamblin:2004zg,Harris:2004mf,Cavaglia:2006uk,Alberghi:2006km,Cavaglia:2007ir,Calmet:2008dg,Nayak:2009fv,Gingrich:2009hj,Gingrich:2009da,Landsberg:2014bya,Song:2019lxb} by searching for the high-multiplicity signature from rapid Hawking evaporation \cite{Hawking:1974sw}. The reach of collider experiments is limited by the finite CM collision energy. Cosmic accelerators thus provide an alluring alternative, as high-energy cosmic rays can reach much higher momenta. This has been known for some time, and extensive studies have been performed on the production of BHs in air showers \cite{Emparan:2001kf,Ringwald:2001vk,Anchordoqui:2001cg,Feng:2001ib,Jain:2000pu,Anchordoqui:2001ei,Ahn:2003qn,Ave:2003ew}, particularly if they are observed with energies above the GZK limit. A recent study \cite{Mack:2018fny} explored the observational consequences of micro black hole creation anywhere in our cosmic volume through its potential to trigger vacuum decay, which can place limits on LED models with certain very high fundamental scales.

In this article, we examine in detail the prospects of observing microscopic black holes produced in collisions of high-energy neutrinos with detector nucleons in next-generation neutrino observatories such as IceCube-Gen2. Compared with cosmic ray detectors, neutrino telescopes offer a unique advantage in that events can be completely contained. This allows for a far more detailed accounting of deposited energy and unique topological signatures that can help distinguish ordinary electroweak interactions from BHs. 

Many preliminary studies of this prospect were performed in the decade prior to the detection of the first high-energy astrophysical neutrinos at IceCube, but few have been carried out in the intervening years. Ref.~\cite{Stojkovic:2005fx} studied the connection between black hole production from cosmogenic neutrino-nucleon scattering and proton lifetime, and highlighted two possible solutions to the proton decay problem, which can arise in LED models. Refs.~\cite{Uehara:2001yk,Dutta:2002ca,Kowalski:2002gb,AlvarezMuniz:2002ga,Kisselev:2010zz} mainly focused on the expected number of black hole events in IceCube and other detectors, while Ref.~\cite{Jain:2002kz,Reynoso:2013jya} further studied the zenith dependence of the cosmogenic neutrino flux by including black hole production and decay into neutrinos when neutrinos propagate through the Earth. In addition, Ref.~\cite{Anchordoqui:2006fn} discussed low energy muons with high multiplicity showers as a signature of black holes and Ref.~\cite{Arsene:2013nca} proposed the search for rare muon tracks with large angular separation. Gravity-mediated \textit{elastic} neutrino-nucleus scattering was also considered in Ref.~\cite{Illana:2005pu}:  even without BH formation, such events can lead to new and unique observational signatures. Despite those preliminary studies, the distinction between gravity-mediated and Standard Model interactions remains unclear, and a full phenomenological analysis has never been performed. Our goal in this work is therefore to construct a complete set of observable signatures and quantify the prospects of detection in the next generation of large-scale neutrino observatories. 

At these masses, BH evaporation is effectively instantaneous. Particle emission from BH evaporation follows a thermal distribution with a Hawking temperature $T_H = (n+1)/r_H$, where $n$ is the number of extra dimensions. For a black hole mass of $M_\bullet \sim$ TeV, this corresponds to $T_H \sim$ a few hundred GeV. The BH will only emit a few ($\sim$ 6-20, depending on initial BH mass) particles in its decay. Because of the high Hawking temperature, these evaporation products will be drawn from every Standard Model (and potentially BSM \cite{Song:2019lxb}) degree of freedom. Because interactions are observed in the ``fixed-target'' frame, evaporation products will be highly collimated. Searches for high-multiplicity events become difficult, as electronic and hadronic events lead to immediate cascades, while tracks from longer-lived particles (muons and taus) will lie on top of each other. Missing momentum cannot be detected, because the momentum of the incoming neutrino is unknown. We have identified several key signatures that can indicate the creation and subsequent decay of microscopic black holes in neutrino telescopes:
\begin{enumerate}

    \item \textbf{Flavor composition.} For sufficiently distant sources, the mixing of neutrino flavors via oscillation should result in an approximately even mix of flavors arriving at the detector. While this is true in both Standard Model interactions and the black hole-production scenario, the characteristics of the detected events from which incoming neutrino flavor inferences are made will be different. The topology of a neutrino-nucleus interaction is the principal ingredient in inferring the flavor composition of astrophysical neutrinos. Possible topologies are \emph{cascades} (also called \emph{showers}), muon and tau \emph{tracks}, and at energies $> 10$ PeV, the \emph{double bang} (or double cascade) signature produced by a $\tau^\pm$ decaying electronically or hadronically far enough away from the primary vertex to be separately identified. As BHs can evaporate democratically to all degrees of freedom (as long as their masses are lower than the BH mass and the Hawking temperature), we will show that they yield a very nonstandard ratio of topologies, which can look like unitarity violation in the neutrino oscillation matrix if they are interpreted assuming Standard Model processes -- indicating the possible presence of new physics. %  Standard analysis techniques that infer the flavor composition from expected charged current (CC) and neutral current (NC) neutrino-nucleon interactions only will therefore result in the inference of a different incoming flavor mix, indicating the possible presence of new physics.

    \item \textbf{Topology.} In addition to changing the ratio of topologies, BH events also yield a lower amount of energy deposition into leptonic final states, with respect to the accompanying hadronic shower due to the high multiplicity of BH decay products. This manifests itself as a lower ratio of muon-to-shower energy, and a lower ratio of energies between the second and first showers seen in double-bang events. In addition to the three topologies outlined above, BHs can produce multiple tau leptons, which can decay at different times yielding an $n$-bang signature. If a muon is also produced, it will be highly collimated with the taus, producing a \textit{kebab}-like signature of multiple cascades arrayed on a muon track. If more than one muon is produced from BH decay, multiple tracks are expected in an event. Finally, if one of the decay products remains energetic enough, it may interact with the nucleus and yield another black hole bang in the detector.

    \item \textbf{Cherenkov light timing.} Because a majority of events will take the form of cascades, we will focus on a method of discriminating the proportion of electronic to hadronic energy deposition in shower events. It was shown in Ref. \cite{li:2016kra} that the initial Cherenkov shower produced by neutrino-nucleus deep inelastic scattering (DIS) events is followed by a delayed muon decay signal around $10^{-6}$ s after the initial interaction, and thereafter by a third peak due to neutron capture at $10^{-3}$ s. The size of the neutron peak relative to the total energy can be used as a proxy for the shower composition. We will show that in the \emph{peak ratio/energy deposition plane}, evaporating BHs occupy a specific region, which may allow them to be differentiated from standard charged current and neutral current interactions on a statistical basis.
\end{enumerate}

While it is unlikely to be a clear diagnostic signature in the near term, an additional effect is that as CM energies rise above $M_\star$, the BH formation rates can contribute to a significant rise in the total neutrino-nucleus cross section. This leads to an increase in the rate of interaction with incoming neutrino energy, which could appear as a change in the derived neutrino energy spectrum at the high end. However, this is highly degenerate with other new physics models, as well as with a simple change in the neutrino flux as a function of energy, and relies on a robust reconstruction of the incoming particle energy based on the interactions observed. It may therefore be an interesting target for future searches, but we do not propose it as a diagnostic here.

We begin our analysis by describing production and evaporation of BHs in LED models in neutrino telescopes. In Sec.~\ref{sec:topology}, we examine in detail the specific observational signatures of such events in an IceCube-like detector. Section \ref{sec:flavorcomp} focuses on the reconstructed flavor signature in the presence of LEDs. In Sec.~\ref{sec:trackbangbang} we examine the energy distribution in track and double-bang events, and we end Sec.~\ref{sec:topology} by mentioning new morphological signatures. Sec.~\ref{sec:echo} focuses on Cherenkov light timing signatures, and in Sec.~\ref{sec:detect}, we evaluate the detection prospects at future experiments. Section \ref{sec:vacuumdecay} briefly discusses the implications of micro black hole creation for vacuum stability.  We conclude in Sec. \ref{sec:conclusions}.

\section{Black hole production and decay in neutrino telescopes}
With the exception of neutrinos produced in the core-collapse supernova 1987A, the first positive detection of neutrinos from beyond our Galaxy came in the form of high-energy contained events seen in the IceCube detector \cite{Aartsen:2013bka,Aartsen:2013jdh,Aartsen:2014gkd,Aartsen:2015zva}. Since then, over 100 such events have been catalogued \cite{Schneider:2019ayi}, extending from the tens of TeV (where the atmospheric contribution is expected to sharply fall), up to several PeV in energy. Combined with throughgoing muon signatures of neutrinos interacting outside the detector, these indicate the presence of an isotropic distribution of high energy neutrinos. These appear to follow a power law in energy, and are consistent with an extragalactic origin, likely from high-energy astrophysical processes such as active galactic nuclei. An additional contribution to the extragalactic flux of \textit{cosmogenic} neutrinos is expected around $E_\nu \sim 10^8$ GeV from collisions of ultrahigh energy cosmic rays with intergalactic radiation. 

The IceCube detector consists of a cubic kilometer of ice below the geographic South Pole, instrumented with approximately 5000 digital optical modules (DOMs) designed to detect the Cherenkov light emitted by charged particles produced in high energy collisions between neutrinos and nuclei (and electrons) in the ice. The number of photoelectrons recorded acts as a proxy for the event energy, and the geometrical distribution of triggered DOMs leads to a classification in event topologies, which in turn allows for flavor information to be extracted. 

The center-of-mass energy of a collision between a neutrino and a nucleon at rest is $\sqrt{s}=\sqrt{2m_NE_\nu}$. Therefore, probing a Planck mass $M_\star \gtrsim$ TeV requires neutrino energies of 10 PeV or more. Because of the declining neutrino flux at high energies, and to a lesser degree, the fact that high-energy showers have a large spatial extent, IceCube is likely not large enough to yield a signal with sufficient numbers of fully contained events. We perform our analyses with future experiments in mind, such as IceCube-Gen2 \cite{Aartsen:2014njl}. We will use the IceCube configuration as a template for future detectors, keeping the same detector efficiencies but scaling up detector dimensions.

The black hole production cross section from neutrino collisions with a nucleon $N = n,p$ is \cite{Dai:2007ki}:
\begin{equation}
    \sigma^{\nu N\rightarrow BH} = \int_{M_\star^2/s}^1 du \pi b_{\mathrm{max}}^2 \sum_{i}f_i(u,Q),
    \label{eq:xsec}
\end{equation}
where $b_{\mathrm{max}} = 2r_s^{(d)}(E_{CM})/[1+(d-1)^2/4]^{1/(d-2)}$ is the maximum impact parameter that allows BH production and
\begin{equation}
    r_s^{(d)} = k_d M_\star^{-1}\left(\frac{\sqrt{us}}{M_\star} \right)^{1/(d-2)}
\end{equation}
is the Schwarzschild radius in $d=n+3$ spatial dimensions of a BH with mass $M_\bullet = \sqrt{us}$. The geometric factor $k_d$ is
\begin{equation}
    k_d = \left[2^{d-3}\pi^{(d-6)/2}\frac{\Gamma(d/2)}{d-1}\right]^{1/(d-2)},
\end{equation}
 $f(v,Q)$ is the parton distribution function (PDF) of the proton or neutron, where $v$ is the fraction of energy carried by the individual parton, and $Q$ is the momentum exchange of the collision. The sum is  over all quark flavors and gluons. 
 
 Due to the nonlinearity in the formation of black holes, part of the collisional energy, momentum and angular momentum may be lost in the form of gravitons before the black hole becomes stationary and starts Hawking evaporation. The  energy loss has been studied both analytically and numerically in~\cite{Yoshino:2001ik,Yoshino:2002br,Cardoso:2002pa,Yoshino:2005hi}, and especially in the context of ultra high energy neutrino-nucleon scattering in~\cite{Anchordoqui:2003jr}. It may increase the energy threshold for BH production and reduce the production cross section, see Appendix~\ref{sec:iniloss} for detailed discussion. However, we note that gravitational energy loss is very model dependent---early estimates put this number between 10\% and 30\%, but the situation was summarized by Yoshino \textit{et al.}~\cite{Yoshino:2005hi}, who argued that the nonlinear effects neglected by prior authors called into question the robustness of their results, and that a full, rigorous calculation remained out of reach. In the absence of a self-consistent quantum gravity theory, we nevertheless set the initial energy loss to be 0 and keep in mind that it may affect the BH production rate accordingly. The particles emitted by Hawking radiation follow the a blackbody spectrum, modified by a ``greybody factor'', which depends on the particle spin and mass, and the angular momentum of the BH. Because evaporated particles are drawn from all kinematically and thermally accessible degrees of freedom, a mix of leptonic and hadronic emission is expected. Hadronic showers will dominate, because of the six quark flavors, three colors, and eight gluon degrees of freedom per helicity. 
 
 In the last stage of the evaporation when the black hole mass drops below $M_\mathrm{min}$ where the semiclassical treatment is no longer applicable, we assume the black hole bursts into a minimum number of particles which conserve the energy, momentum and all gauge charges of the black hole~\cite{Dai:2007ki,Song:2019lxb}. Some authors have proposed that an evaporating black hole might not fully decay but rather leave behind a Planck-mass relic~\cite{Markov:1982,Hossenfelder:2003dy}, which could carry a small electric charge. This would alter the spectrum at the final stages of evaporation, and may produce distinct signatures, such as a delayed flash due to subsequent accretion and evaporation. We leave the exploration of this interesting possibility for future work, and assume total evaporation for the scenarios that follow.

% The most conservative type of constraint we can place is based on the total event rate predicted by LED models.

We employ a set of numerical tools that we modify to model the production, evaporation, and energy deposition by microscopic black holes in the detector material.  We first modify the  BlackMax \cite{Dai:2007ki,dai:2009by,Frolov:2002gf} code to accept collisions between high-energy neutrinos and nuclei at rest in the detector volume. BlackMax is a Monte Carlo simulator designed to model the production and evolution of black holes at colliders. It includes relevant greybody factors of evaporation products, and can model the effects of BH rotation, brane splitting between fermions, brane tension and bulk recoil from gravition emission. For the purpose of this work, we consider only the case without rotation, splitting, brane tension, or recoil, as this case includes the simplest black hole production models (and is in some cases a conservative assumption). Rotation would likely result in a higher evaporation flux \cite{Frost:2010zz,Harris:2003eg}, though we note that superradiance may enhance the emission probability of gravitons for highly rotating blackholes~\cite{Stojkovic:2004hp}. Rotation would also make the emission less isotropic, but this is unlikely to have a significant impact due to the expectation of a high level of beaming. In the absence of greybody factors for gravitons for rotating black holes, we assume non-rotating black holes for conservative study. Whether or not brane tension needs to be considered depends on the extra dimension model -- the ADD model \cite{ArkaniHamed:1998rs} does not include brane tension, whereas the Randall-Sundrum model \cite{Randall:1999ee,Randall:1999vf} does; the most stringent existing limits either restrict their analysis to the former \cite{Sirunyan:2017jix,Sirunyan:2018wcm,Sirunyan:2018xwt} or explicitly consider both possibilities \cite{Aaboud:2016ewt}. In terms of the possibility of split branes, we assume for simplicity that the entire Standard Model is confined to the brane and only gravity can pass into the bulk.

\begin{figure}[!htb]
   { 
    \includegraphics[width=0.6\textwidth]{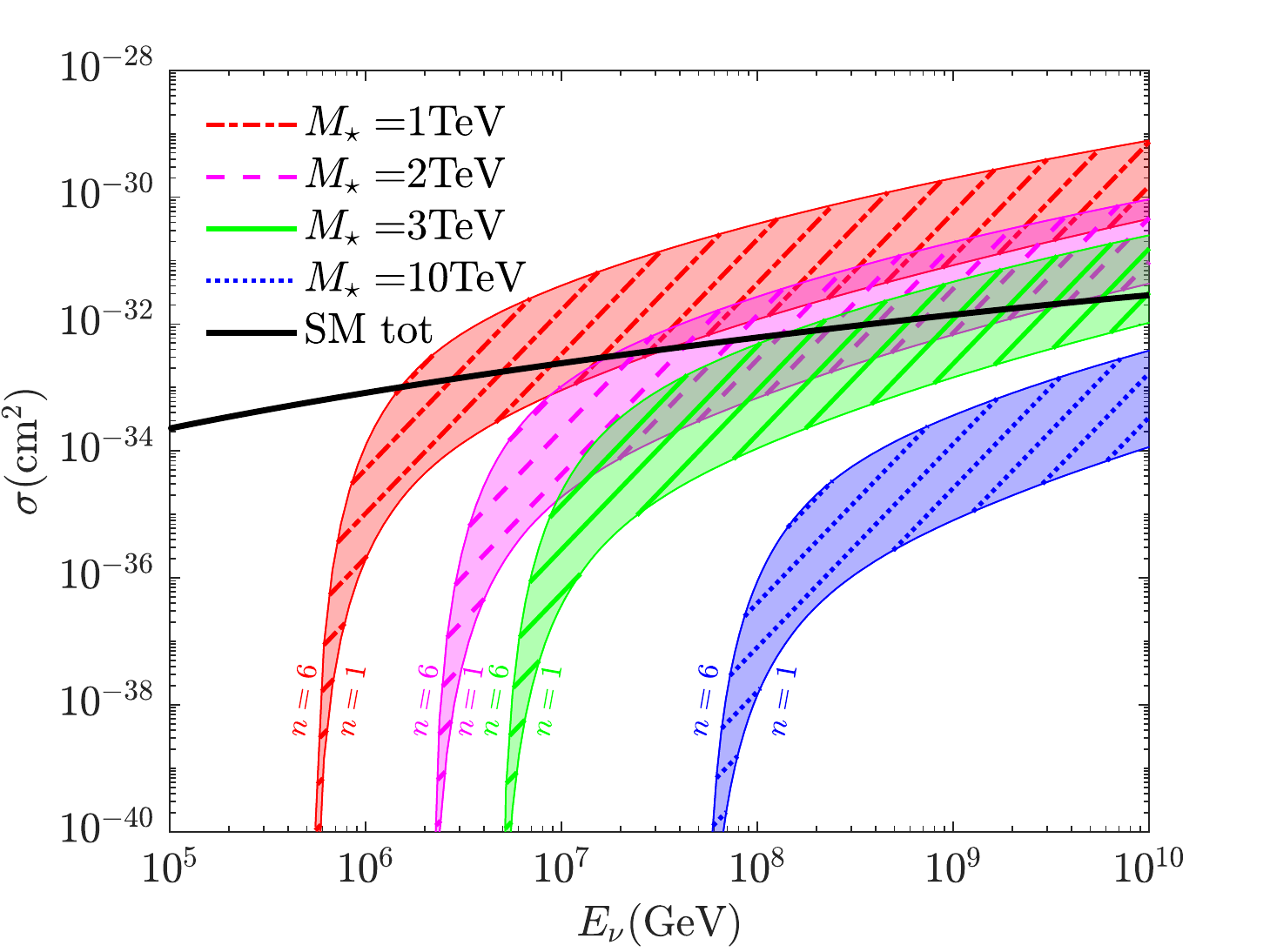}
    }
    \caption{The black hole production cross section from neutrino-nucleon interactions as a function of the incoming neutrino energy. The hatched colored bands correspond to $M_\star=$ 1~TeV (dash-dotted red), 2~TeV (dashed magenta), 3~TeV (solid green) and 10~TeV (dotted blue) respectively. For each band, the lower limit represents $n=1$ extra dimensions and the upper bound has $n=6$. The total Standard Model (SM) neutral current and charged current cross section is shown by the black line.}
    \label{fig:xs}
\end{figure}

We show in Fig.~\ref{fig:xs} the neutrino-nucleon cross section as a function of neutrino energy for a range of Planck masses $M_\star$, assuming a minimum black hole mass $M_\mathrm{min}=M_\star$. 
The cross section increases with the number of extra dimensions $n$ but decreases quickly as the Planck scale $M_\star$ is raised. At $M_\star=10\ $TeV the black hole production is subdominant up to $E_\nu =$ 10~EeV, even assuming 6 extra dimensions. In the same figure, the black line shows the SM-only expectation including neutral current (NC, $Z$ boson exchange) and charged current (CC, $W^\pm$ boson exchange) interactions. We refer the reader to Ref. \cite{palomares-Ruiz:2015mka} for a complete discussion of how these processes are modeled.
Once a spectrum of evaporation products has been produced, we pass the results to PYTHIA 8 in order to model decay of heavy states and hadronization of coloured particles \cite{Sjostrand:2014zea}. We employ the CT14NNLO \cite{Dulat:2015mca} PDFs, implemented with LHAPDF6 \cite{Buckley:2014ana}. The latter step gives the list of four-momenta of propagating particles originating at the interaction vertex.

We will focus on the case of $n = 6$ extra dimensions for the results presented below. A smaller number of LEDs has the straightforward effect of lowering the BH production cross section, and thus the total event rate, but will not change the observational signatures, which are due to the high-multiplicity Hawking evaporation products. 

Finally, to account for energy deposition of these interaction products in the detector volume as Cherenkov light, the PYTHIA results are fed into FLUKA \cite{Ferrari:2005zk,Bohlen:2014buj}, a multiparticle transport code that accounts for interaction of high-energy particles in matter. We return to the details of our FLUKA simulations in Sec. \ref{sec:echo}. We now turn to specific signatures of BHs that can be identified with our simulations. 

\section{Event Topology}
\label{sec:topology}
The spatial distribution of deposited electromagnetic energy in the detector volume is known as the event topology. Different topologies are due to the difference in final-state products from neutrino-nucleus interactions in the ice. Because microscopic black holes will evaporate to a subset of every SM degree of freedom, the set of final-state particles -- and thus the event topology -- can be expected to be markedly different from the SM case.

\emph{Showers} (or cascades) occur when energy is immediately deposited in a roughly spherical distribution leading to a large, but contained, signal. \emph{Tracks} are produced by muons (or, at very high energies, taus, though their energy deposition rate is much lower) as they propagate away from the interaction vertex, eventually escaping the detector. A third topology, the \emph{double-bang} (or double cascade), is expected if charged tau production and decay both occur within the detector, with a large enough spatial separation to identify two separate but correlated cascades.
Timing information can additionally be used to narrow down the origin of each topology. This has been used for example to search for \textit{double-pulse} events, in which the double-bang from tau decay cannot be resolved in space. We return to the use of timing in Sec. \ref{sec:echo}.

In the following, we will employ a set of criteria to determine the observed topology of each event that we simulate. 
\begin{enumerate}
\item If one or more high-energy muon is produced with a total energy above a certain threshold (1 TeV in the following sections), the event is classified as a \textbf{track}. These muons can be produced directly from the BH evaporation, or from the decay of intermediate tau leptons. Besides muons, we also consider high energy taus (5~PeV or higher) produced in neutrino-nucleon interaction which decay outside the detector as track events. 
\item The \textbf{double bang} signature is produced when the tau created from a $\nu_\tau$ charged current interaction decays electronically or hadronically inside the detector, leading to a visible second shower. We require the tau to travel farther than 100~m before it decays to clearly separate the second shower from the primary one~\cite{Cowen:2007ny}. Following~\cite{usner:2017aio} we construct the energy asymmetry
\begin{equation}
    E_A \equiv \frac{E_1-E_2}{E_1+E_2}
\end{equation}{}
 where $E_1$ and $E_2$ are, respectively, the energies of the first and second shower. (So when $E_1>E_2$, $E_A>0$, and vice versa.) We require $-0.98\le E_A \le 0.3$. We will relax our condition on $E_A$ in Sec.~\ref{sec:bangbang} and Sec. \ref{sec:detect}.
\item Events that do not meet criteria 1 or 2 are classified as \textbf{showers}.
\end{enumerate}
We begin this section by first asking how a modified distribution of topologies from BH creation changes the inferred flavor composition at high energies. In Sec. \ref{sec:trackbangbang}, we refine our analysis to ask how the energy distribution in track and double-bang events is modified by black hole creation. We discuss new, non-standard topologies in Sec. \ref{sec:newtopologies}.

\subsection{Flavor Composition}
\label{sec:flavorcomp}
In SM interactions, event topology is directly related to the flavor of the progenitor neutrino. $\nu_\tau$'s can produce double bangs and $\nu_\mu$'s yield muon tracks, as can  $\nu_\tau$'s if the $\tau^\pm$ decays leptonically. All flavors can produce cascades via NC interactions, immediate decay of $\tau^\pm$ leptons, or direct production of $e^\pm$'s which deposit their energy almost immediately. Every topology is associated with an initial shower due to the momentum transferred to the nucleus and subsequent hadronic shower, making misidentification a constant concern. 

\begin{figure}[!htb]
    % \centering
    \includegraphics[width=0.6\textwidth]{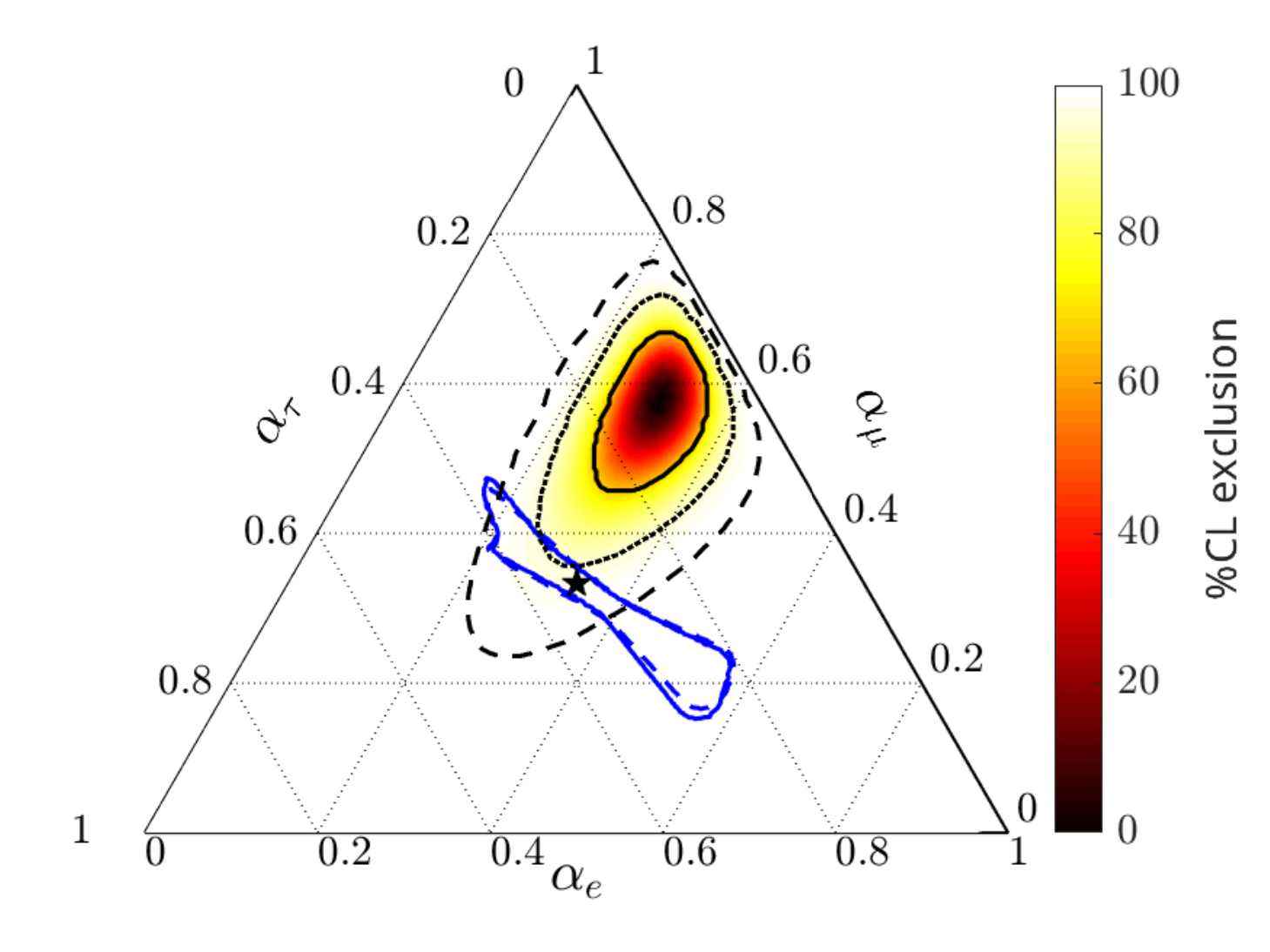}
    
    \caption{Allowed regions for the neutrino flavor composition at the Earth assuming black hole production in neutrino-nucleon interactions in IceCube. The solid, dotted and dashed black contours show the 68\%, 95\% and 99.7\% CL respectively. The 3$\sigma$ regions for Standard Model interactions (without BH formation) after neutrino oscillation with any flavor combination at the source are shown in blue contours. The solid blue line is obtained from normal ordering and dashed line is from inverted ordering. Neutrino oscillation parameters are taken from NuFIT 4.1~\cite{nufit4.1,Esteban:2018azc}. The black star in the centre indicates the $(1:1:1)$ flavor ratio.}
    \label{fig:ternary}
\end{figure}

The flavor composition inferred from a sample of neutrino events is a well-known diagnostic for new physics \cite{Baerwald:2012kc,Mena:2014sja,palomares-Ruiz:2015mka,Arguelles:2015dca,Bustamante:2015waa,Gonzalez-Garcia:2016gpq,Farzan:2018pnk,Ahlers:2018yom,Arguelles:2019rbn}. As neutrinos propagate over large, uncorrelated distances to Earth, their flavors mix incoherently, leading to a signal that has been averaged over all flavors. Out of the possible combinations of $(\nu_e:\nu_\mu:\nu_\tau)$, only a small region is allowed after oscillation. Once experimental uncertainties in the mixing parameters are included, this leads to an allowed region delimited by the blue contour in the central area of the ternary plot in Fig. \ref{fig:ternary}. A reconstructed flavor composition outside of this region therefore indicates some non-unitarity in the oscillation matrix, the presence of new physics, or a misunderstanding of experimental uncertainties. 

We therefore begin by asking: in the presence of BH creation from neutrino-nucleus interactions, what is the reconstructed flavor composition when the sample is analyzed assuming Standard Model processes only? We first simulate a number of BH events, and classify the topology of each event into showers (sh), tracks (tr), and double bangs (db), according to the final state particles from BH evaporation. We then proceed in a similar way to Ref. \cite{Mena:2014sja},  computing $P(\{\Xi\} | (\alpha_e:\alpha_\mu : \alpha_\tau))$, the probability of a given set of observed topologies $\{\Xi\}$ conditional on each possible flavor composition, in order to reconstruct the SM interpretation of the observed events.

% For illustration, we consider a detector with an efficiency similar to IceCube's, but with a much larger effective volumearea to allow for sufficient high-energy events to be recorded. 

\begin{figure}[!htb]
\begin{tabular}{c c}
    \includegraphics[width =0.49 \textwidth]{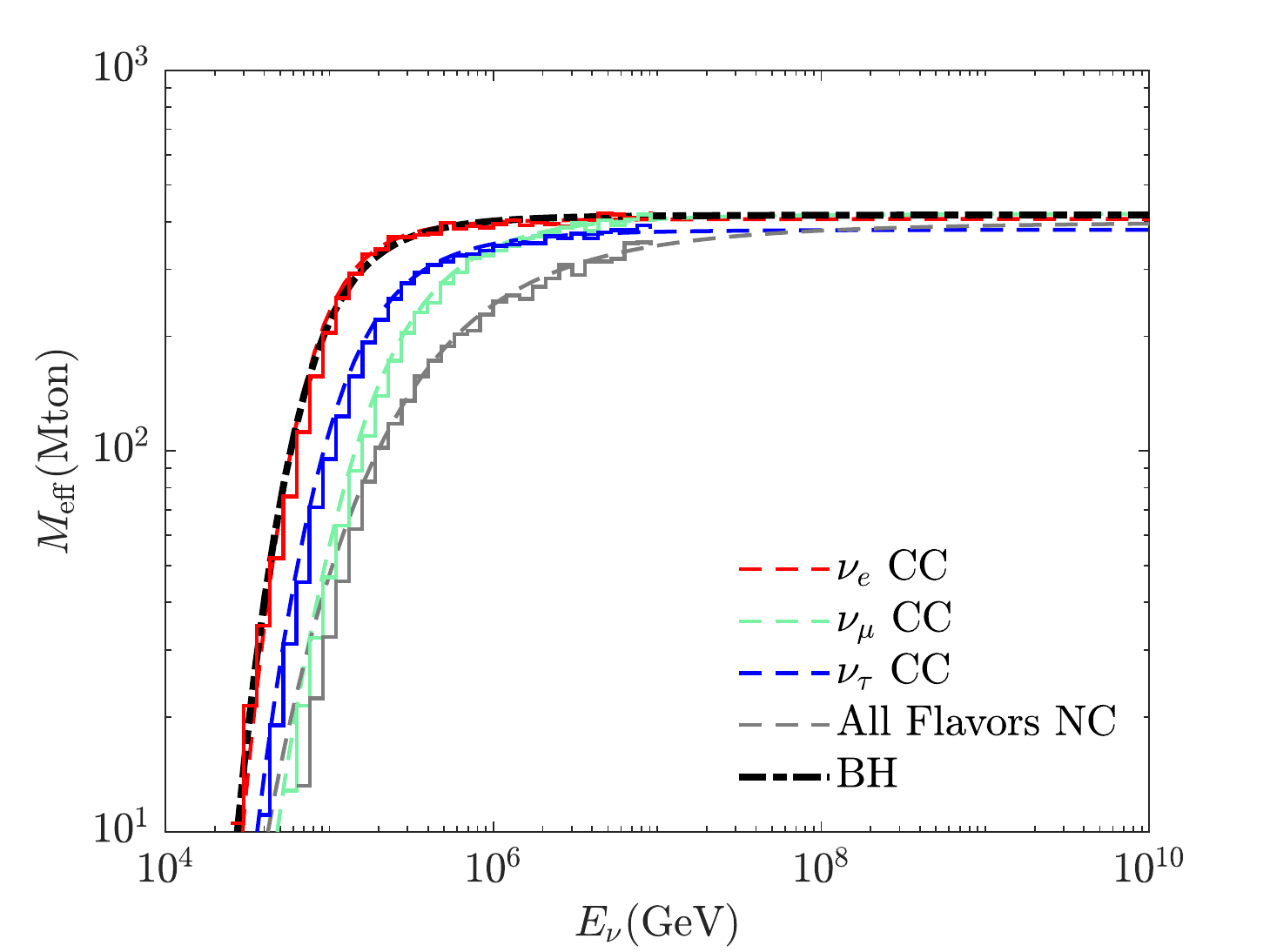} &
  \includegraphics[width = 0.49\textwidth]{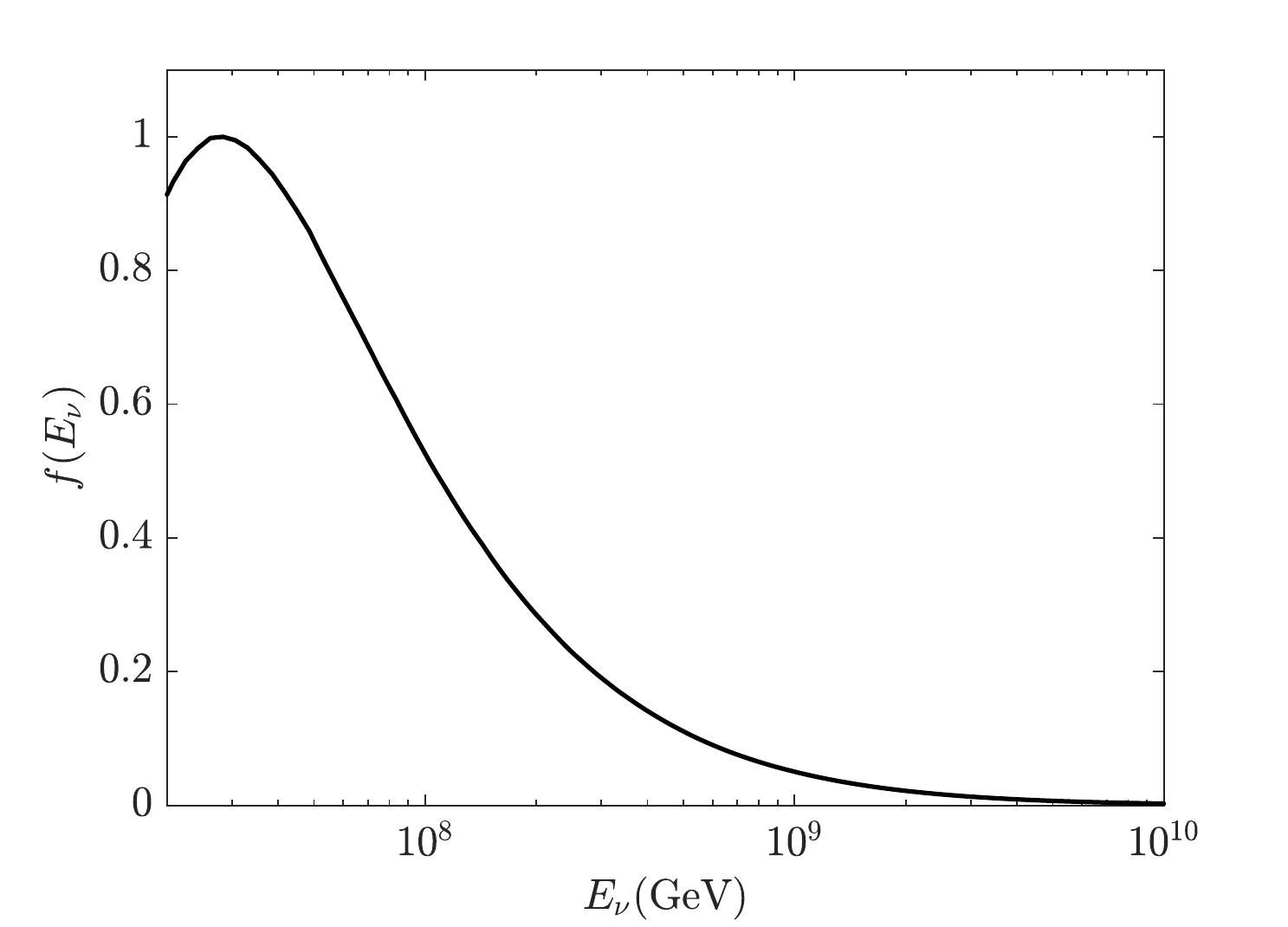}
  \end{tabular}
    \caption{\textbf{Left:} Effective mass as a function of the incoming neutrino energy for $\nu_e$ CC (red), $\nu_\mu$ CC (green), $\nu_\tau$ CC (blue), NC (gray) and black hole productions (black), for IceCube. In this section we use the same efficiency function, but rescale the overall normalization by a factor of 7.9/0.644 to simulate the larger volume of IceCube-Gen2. The solid lines show the  effective mass from the IceCube collaboration and the dashed lines are obtained from the fit. The effective mass for black hole productions is taken from the $M_\mathrm{eff}(E_\mathrm{true})$ found in~\cite{palomares-Ruiz:2015mka}. \textbf{Right:} Weighting function $f(E_\nu)$ which governs the black hole production rate’s dependence on incoming neutrino energy $E_\nu$ (see Eq.~\eqref{eq:weights}).  The maximum of the distribution is normalized to 1. In producing results in this section, we draw 1000 black hole events randomly according to this distribution. We assume the production of non-rotating black holes on a tensionless brane. For this specific case, the minimum black hole mass is set to $M_\mathrm{min}=M_\star=3\ $TeV.}
    \label{fig:MefffEnu}
\end{figure}

In the Standard Model, the total number of expected events for a topology $\Xi =$ (sh, tr, db) is generally given by
% \begin{equation}
%     N^\Theta=TN_A\int_{E_\mathrm{min}}^{E_\mathrm{max}} M_\mathrm{eff}(E_\nu)Att(E_\nu)\dfrac{d\phi(E_\nu)}{dE_\nu}\int_{y_\mathrm{min}}^{y_\mathrm{max}}dy\dfrac{d\sigma^{\nu N \rightarrow \Theta}(E_\nu,y)}{dy}\,,
% \end{equation}
\begin{equation}
    N_{\nu_i}^{\Xi,\mathrm{CC}}=TN_A \int_{E_\mathrm{min}}^{E_\mathrm{max}} dE_\nu M^{\nu_i,\mathrm{CC}}_\mathrm{eff}(E_\nu)Att(E_\nu)\dfrac{d\phi_{\nu_i}(E_\nu)}{dE_\nu}\sigma^{\nu_i N \rightarrow \Xi}(E_\nu)\,,\label{eq:NXi} 
    \end{equation}
    for CC events. For NC events, only showers are produced:
    \begin{equation}
        N^{\mathrm{sh,NC}}=TN_A \int_{E_\mathrm{min}}^{E_\mathrm{max}} dE_\nu M^{\mathrm{NC}}_\mathrm{eff}(E_\nu)Att(E_\nu)\dfrac{d\phi_{6\nu}(E_\nu)}{dE_\nu}\sigma^{\mathrm{NC}}(E_\nu)\,,\label{eq:NNC} 
\end{equation}
where $i=e,\mu,\tau$ indicates different neutrino flavors, $T$ is the exposure time, $N_A$ is Avogadro's number, and $M^{\nu_i,\Theta}_\mathrm{eff}(E_\nu)$ is the effective mass of the detector, for a given neutrino flavor and process. The latter can be interpreted as the quantity of target material times a detector efficiency. $Att$ is the attenuation factor due to the absorption and regeneration of neutrinos in the Earth. We focus on ultra high energy neutrinos with $20\ \mathrm{PeV}\le E_\nu \le 10\ \mathrm{EeV}$ where the black hole production cross section can be comparable to or larger than the Standard Model cross sections. The minimum deposited energy is assumed to be 20~PeV. At such high energies the Earth is almost opaque to neutrinos. Thus only downgoing neutrinos need be considered, and $Att = 1/2$. $M_\mathrm{eff}$ can be obtained by fitting the IceCube effective mass shown in Ref.~\cite{Aartsen:2013jdh}, where we use a fitting function as in~\cite{palomares-Ruiz:2015mka}
\begin{equation}
    M_\mathrm{eff}(x)=\rho_\mathrm{ice}\dfrac{cx^q}{1+dx^q}\,.
    \label{eq:Meff}
\end{equation}
$\rho_\mathrm{ice}=916.7\ \mathrm{kg} \, \mathrm{m}^{-3}$ and $x=\log_{10}(E_\nu/10\, \mathrm{TeV})$. The results are shown in the left panel of \Fig~\ref{fig:MefffEnu} and the best-fit parameters are listed in Tab.~\ref{tab:Mefffit}.
\begin{table}[!htb]
    \centering
    \setlength{\tabcolsep}{2em} % for the horizontal padding
    {\renewcommand{\arraystretch}{1.2}% for the vertical padding
    \begin{tabular}{c  c c c} \toprule
           best-fit& $c$ (km$^3$)&$d$&$q$\\ \colrule
         $\nu_e$ CC&0.59&1.3&5.2  \\ 
         $\nu_\mu$ CC&0.071&0.16&4.8 \\ 
         $\nu_\tau$ CC&0.18&0.42&4.7 \\ 
         All Flavors NC &0.059&0.13&3.5 \\ 
         BH&0.50&1.1&4.6 \\ \botrule 
    \end{tabular}}
    \caption{Best-fit parameters of $M_\mathrm{eff}$ used in Eq. \eqref{eq:Meff} for different types of interactions in IceCube.}
    \label{tab:Mefffit}
\end{table}
The fit agrees well with the IceCube effective masses, especially for neutrinos above 1 PeV. We also extrapolate the effective mass up to 10~EeV where the detector sensitivity is saturated. The effective mass for black hole production is taken to be the effective mass as a function of the true deposited energy obtained from~\cite{palomares-Ruiz:2015mka}, since neutrinos deposit almost all of their energy in the detector. 

We wish to evaluate the detection prospects of a detector the size of the planned IceCube-Gen2~\cite{vanSanten:2017chb}. We model it as a 7.9~km$^3$ cylindrical detector with a height of 1.25~km and a radius of 1.42~km, about 10 times larger than the current IceCube detector. We assume 10 years of exposure, which is equivalent to $\sim 100$ IceCube-years. We then classify the events into three categories according to their topologies: tracks, showers and double bangs.

\begin{figure}[!htb]
    \centering
    \includegraphics[width=0.6\textwidth]{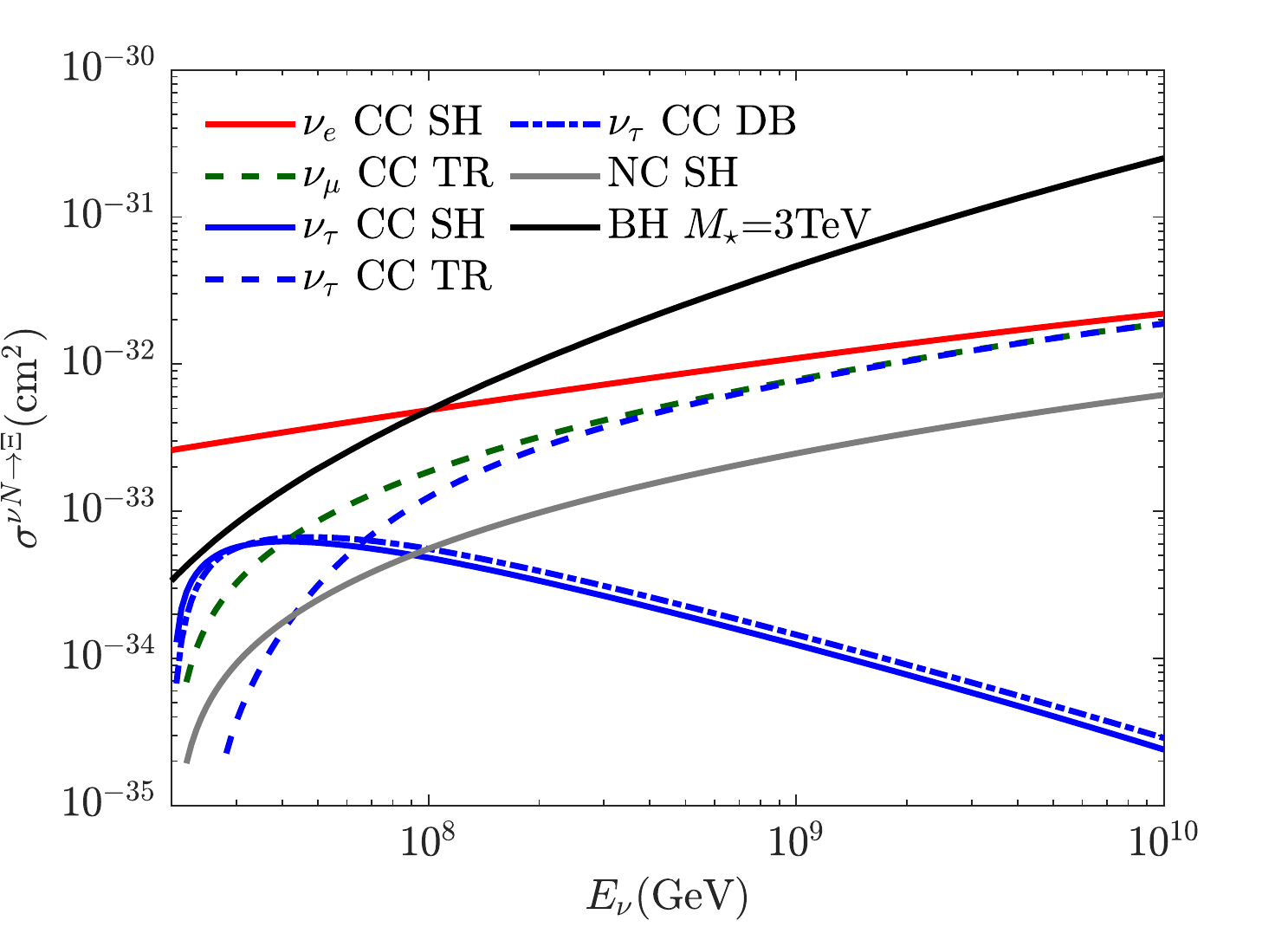}
    \caption{Total neutrino-nucleon interaction cross section with different topologies as a function of incoming neutrino energies. Standard Model showers, tracks and double bangs are shown in solid, dashed and dashed-dotted lines respectively, noting that black hole events will produce all three topologies. The contributions from neutrinos and antineutrinos have been averaged over. In producing the cross sections we require a minimum energy deposition of 20~PeV. See Appendix~\ref{sec:xss} for details.}
    \label{fig:xsreweight}
\end{figure}

The cosmogenic neutrino spectrum can be approximated as a powerlaw with a spectral index $\gamma$, estimated to be between 2 and 3. For better event reconstruction we assume the optimistic neutrino energy spectrum with the spectral index $\gamma=2$. The total all-flavor astrophysical neutrino flux is therefore given by
\begin{equation}
    \dfrac{d\phi_{6\nu}}{dE_\nu}=4\pi\phi_\mathrm{astro}\left(\dfrac{E_\nu}{100\ \mathrm{TeV}}\right)^{-2}\times 10^{-18}\ (\mathrm{GeV}^{-1}\mathrm{cm}^{-2}\mathrm{s}^{-1})
    \label{eq:nuflux}
\end{equation}
with the same normalization as in~\cite{Schneider:2019ayi} where $\phi_\mathrm{astro}=6.45$. We also assume the neutrino flux has no angular dependence. The expected number of events with different topologies in Standard Model processes are obtained from \cref{eq:NXi,eq:NNC}. The total cross section for different neutrino-nucleon scattering processes is shown in Fig.~\ref{fig:xsreweight}. We refer the reader to Appendix~\ref{sec:xss} for more technical details.
%%NOrway

Black hole production is independent of the incoming neutrino flavor. To obtain the event topology from black hole decay, we simulate 1000 black hole events using modified BlackMax. We then pass the evaporation products to PYTHIA 8 for hadronization and heavy particle decay. Black hole production as a function of neutrino energy is weighted according to the all-flavor neutrino flux ${d\phi_{6\nu}}/{dE_\nu}$, the production cross section and the flavor-independent detector efficiency based on \cite{palomares-Ruiz:2015mka} for a true deposited energy of $E_\nu$ is:
\begin{equation}
    f(E_\nu)=\dfrac{d\phi_{6\nu}}{dE_\nu}\sigma_\nu^{BH}(E_\nu)M_{\mathrm{eff}}(E_\nu)\,,
    \label{eq:weights}
\end{equation}
which is also shown in the right panel of Fig.~\ref{fig:MefffEnu}.

The black hole production cross section $\sigma_\nu^{BH}$ is obtained assuming $M_\star=3\ $TeV and 6 extra dimensions (see \Fig~\ref{fig:xs}). Taus are allowed to randomly travel a certain distance before decay, according to the probability distribution of their lifetime. We implement the criteria outlined above to select the track, shower and double bang events (see Appendix~\ref{sec:xss} for more details). We also require the muon energy to be higher than 1~TeV~\cite{Berghaus:2009jb} to avoid the low energy muons produced in the hadronization secondaries. If both muon track and tau double bang are seen, we consider it as a double bang event.

The expected number of events in Standard Model interactions and black hole production is summarized in Tab.~\ref{tab:Nsummary}, as given by 
\begin{equation}
    N_{BH}^{\Xi}=\dfrac{1}{2}TN_A\int_{\Emin}^{\Emax}dE_\nu  M_{\nu}^{BH}(E_\nu)\dfrac{d\phi_{6\nu}}{dE_\nu} \sigma_{\nu}^{BH}(E_\nu)f^\Xi\,,
    \label{eq:NBH}
\end{equation}  
where $f^\Xi$ with $\Xi=\mtr,\msh,\mdb$ is the fraction of track, shower and double bang events obtained from the black hole simulations described above.
\begin{table}[!htb]
    \centering
    \setlength{\tabcolsep}{2em} % for the horizontal padding
    {\renewcommand{\arraystretch}{1.2}% for the vertical padding
    \begin{tabular}{c c c c} \toprule
           & \emph shower&track&double bang\\ \colrule
         $\nu_e$ SM&28.58&0&0  \\ 
         $\nu_\mu$ SM&2.31&8.31&0 \\ 
         $\nu_\tau$ SM&5.07&5.39&2.83 \\ \colrule
         All Flavor Total SM&35.96&13.70&2.83\\ 
         All Flavor Total BH &62.96&36.36&0.20 \\ \botrule
    \end{tabular}}
    \caption{Expected number of events with different topologies from Standard Model and black hole production. We assume $M_\star=3$~TeV and 6 extra dimensions in the latter case. Double bangs are only produced in $\nu_\tau$ CC and black hole events. The neutrino flux is taken from \eq~\eqref{eq:nuflux}.}
    \label{tab:Nsummary}
\end{table}
Black holes produce more showers than tracks since only a limited number of particles are emitted in black hole evaporation, and muons and taus only make up a small fraction of the emission. We note that as the black hole mass increases, more particles can be emitted, which significantly increases the probability of muon production. As we can see from the right panel of Fig.~\ref{fig:MefffEnu}, the energy distribution of incoming neutrinos which produce black holes features a non-negligible tail at 100~PeV or higher energies, which explains the larger track-to-shower ratio for black holes. It should also be noted that $f^\mdb$ decreases from 6\% to about 0.2\% once the energy asymmetry condition is implemented, meaning that the tau emitted is usually less energetic compared with the particles which initiate the primary shower. 

We would like to see how the black hole production in such a future detector would affect the reconstruction of the flavor composition at the Earth. To do so we generate mock data as the sum of Standard Model and black hole events from Tab.~\ref{tab:Nsummary} assuming $(1:1:1)$ favor ratio, i.e. $N_\mtr=50$, $N_\msh=99$ and $N_\mdb=3$. We then reconstruct the flavor composition assuming Standard Model-only interactions. We follow a similar procedure as in 
Ref.~\cite{Mena:2014sja} by constructing the likelihood
\begin{equation}
    \mathcal{L}(\ai,N_a|\Ntr,\Nsh,\Ndb)=\prod\limits_\Xi e^{-(\sum\limits_i N_{\nu_i}^\Xi N_a)}\dfrac{(\sum\limits_i N_{\nu_i}^\Xi N_a)^{N_\Xi}}{N_\Xi !}\,,
\end{equation}
where $i=e,\mu,\tau$ and $\Xi=\mtr,\msh,\mdb$. $N_{\nu_i}^\Xi$ is the expected number of events for a topology from the Standard Model interaction of neutrino flavor $i$, given in Tab.~\ref{tab:Nsummary}. The atmospheric neutrino background is negligible in this energy range (20~PeV to 10~EeV)~\cite{Okumura:2017wtz}. $N_a$ is a normalization factor for the neutrino flux and can be treated as a nuisance parameter which we vary to maximize the likelihood given a certain flavor composition. We construct the test statistic
\begin{equation}
    \lambda(\Ntr,\Nsh,\Ndb|\ai)=-2\ln\left(\dfrac{\mathcal{L}(\ai,N_{a,\mathrm{max}}|\Ntr,\Nsh,\Ndb)}{\mathcal{L}(\ai_\mathrm{max},N_{a,\mathrm{max}}|\Ntr,\Nsh,\Ndb)}\right)\,,
\end{equation}
where in the denominator $\ai$ is also varied to obtain the maximum likelihood for all possible flavor compositions. Since $\lambda$ asymptotically approaches a $\chi^2$ distribution~\cite{Mena:2014sja}, we translate it to $p-$values using Wilks' theorem. The results are shown in Fig.~\ref{fig:ternary}.
Due to the paucity of double bang events in black hole decay, the preferred region is away from large $\nu_\tau$ flavor content. In addition, the larger track-to-shower ratio imposes a slight preference for muon neutrinos over electron neutrinos. The flavor combination obtained from SM interactions and oscillations only is marginally contained in the $3\sigma$
region and the $(1:1:1)$ flavor ratio is excluded at 95\% CL.

We end this section by noting that the choice of energy threshold in constructing the ternary plot in Fig \ref{fig:ternary} can have a significant effect on the best fit location. This is in contrast to an SM-only data sample which by construction should be robust to such choices. It is clear that a significant measurement of a flavor composition that is outside the allowed oscillation region can provide a hint for the new physics processes we are considering here. To obtain a more robust prediction of the signatures of new physics, we now turn to a detailed analysis of the track and double-bang distributions. 

\subsection{Track and double-bang energy distributions}
\label{sec:trackbangbang}
\subsubsection{Track events}
\label{sec:track}
If one of the evaporation products is a muon, a black hole event looks like a track. In contrast with SM events, in which most of the momentum is carried away by the outgoing lepton $\ell$ (i.e. distributions are peaked at large $(1-y) \equiv E_\ell/E_\nu$), a lepton produced by BH decay will only carry $\sim 1/N$ of the energy, where $N$ is the total number of evaporation products. The rest of the neutrino energy is invariably released in the (mostly) hadronic shower of products that do not escape very far from the interaction vertex.

\begin{figure}
\begin{tabular}{c c}
    \includegraphics[width =0.49 \textwidth]{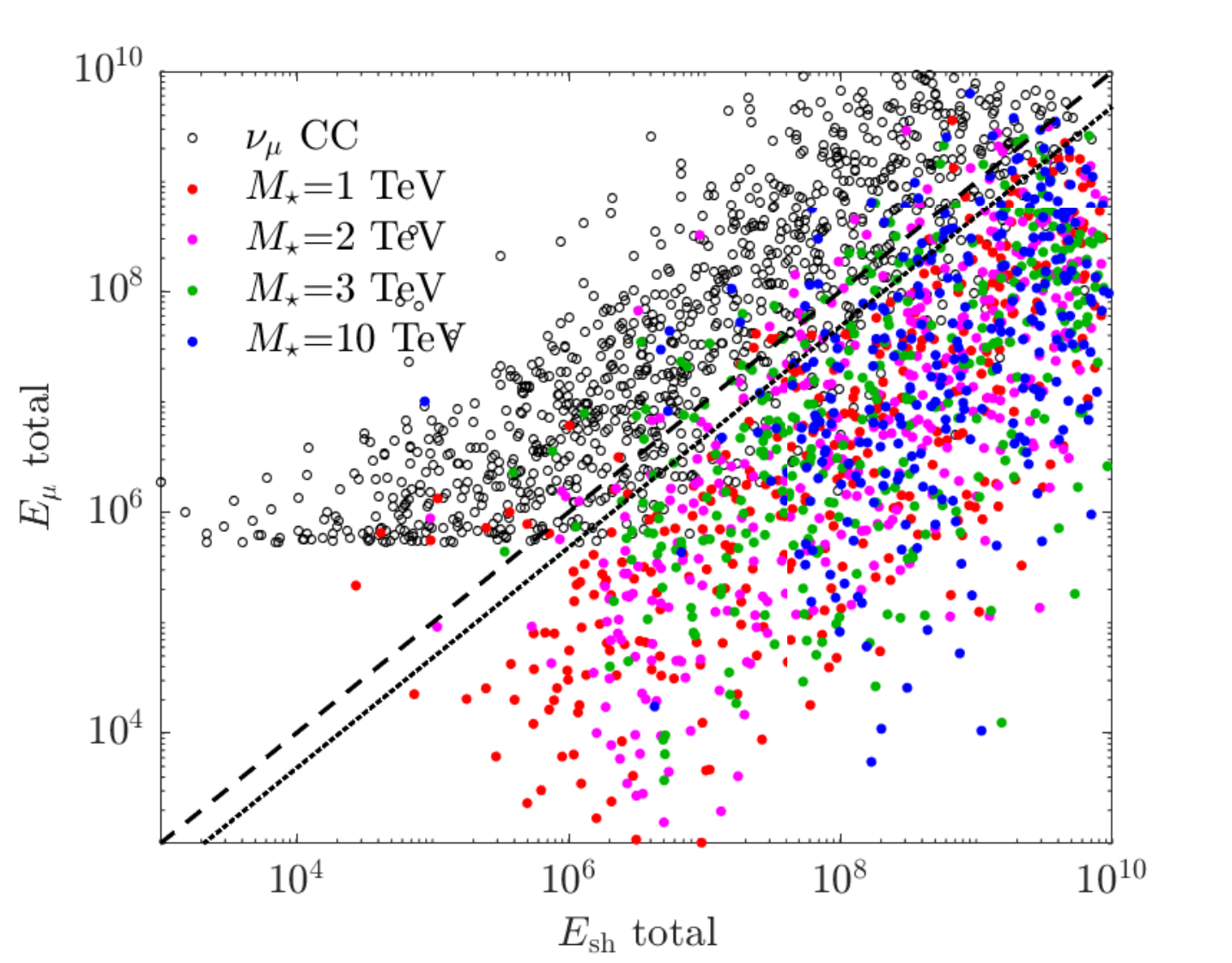} &
  \includegraphics[width = 0.49\textwidth]{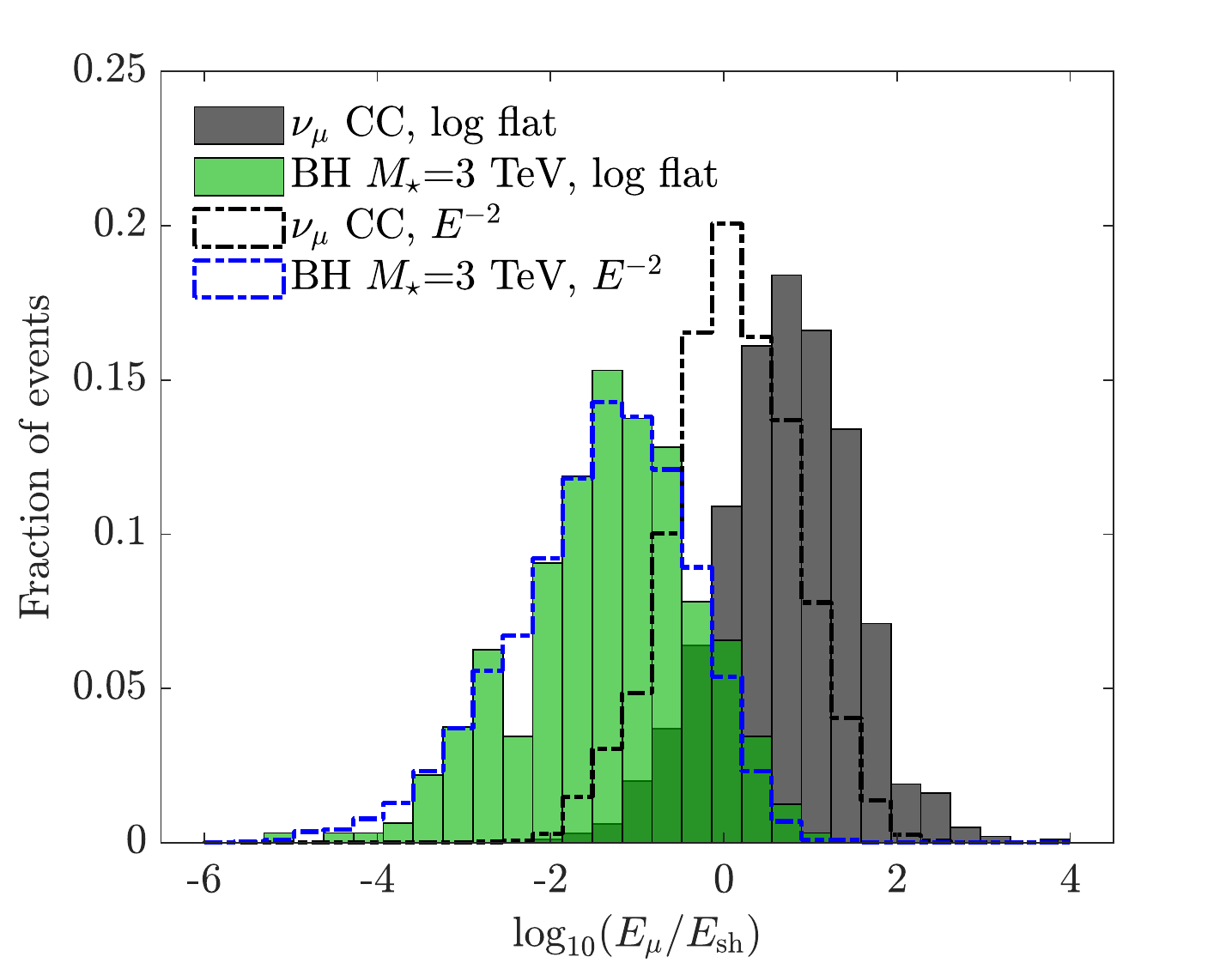}
  \end{tabular}

    \caption{\textbf{Left:} Total energy of muons versus total energy of shower particles in black hole simulations (colored dots) and $\nu_\mu$ CC events (black circles). Red, magenta, green and blue dots correspond to simulations with $M_\star=$1~TeV, 2~TeV, 3~TeV and 10~TeV, respectively. Simulations assume $n = 6$ extra dimensions. Lowering $n$ reduces the production cross section but does not affect the distributions. The dashed line depicts where $E_\mu=E_\mathrm{sh}$, and the dotted line $E_\mu=0.5E_\mathrm{sh}$  delineates the region in which 90\% of $\nu_\mu$ CC events are contained. Each case is based on 1000 simulations, sampled flatly in $\log_{10}(E_{\nu})$. We require $E_\mu>1\ $TeV for an event to be considered a track event, and black holes not producing muons are not shown. We also neglect the possible additional contributions from taus as evaporation products (See Fig. \ref{fig:E1vsE2}). \textbf{Right:} Distribution of the ratio $\log(E_\mu/E_\msh)$ in the total muon energy to total shower energy ratio across all energy ranges for for CC events (black histogram) and BH events with $M_\star = 3$~TeV (green histogram) which produce muon tracks during evaporation for log flat neutrino flux. Dashed black and blue lines show the distribution of $\log(E_\mu/E_\msh)$ with $E^{-2}$ spectrum instead and a minimum energy deposition of 20~PeV as described in Sec.~\ref{sec:flavorcomp}. See Sec.~\ref{sec:detect} for details.}
    \label{fig:EtrEsh}
\end{figure}

We show the shower energy and muon energy in Fig.~\ref{fig:EtrEsh} by simulating 1000 black hole production events at different Planck scales using BlackMax and 1000 $\nu_\mu$ CC events. As before we pass the BlackMax simulations to Pythia 8 for hadronization. $E_\mathrm{sh}$ is the total energy of the evaporation products that initiate showers, and $E_\mu$ is the total energy of the muons produced in a black hole decay. We address tau production and decay in the next subsection. The $\nu_\mu$ charged current events are drawn randomly according to the differential cross section $d\sigma(E_{\nu_\mu})/dE_\mu$. Here $E_\mu$ is the energy of muon produced in the $\nu_\mu$-nucleon scattering and $E_\mathrm{sh} = E_{\nu_\mu}-E_\mu$.  The energy of muons can in principle be reconstructed from the energy of the track as a function of track length~\cite{Toscano:2019qwd}, though stochastic energy loss can lead to uncertainty in the reconstruction.  Samples are drawn uniformly in log space, in order to illustrate our results across the whole energy range, up to 10 EeV.  The minimum energy for black hole simulation (to achieve the necessary CM energy) is $E_\mathrm{min} = M_\star^2/2m_p$ where $m_p$ is the proton mass. The charged current events that we simulate start at a neutrino-proton CM collision energy of 1~TeV ($E_\nu=0.53$~PeV).

The left panel of Fig.~\ref{fig:EtrEsh} vividly shows a butterfly shape with $\nu_\mu$ CC events on one wing and black hole track events on the other. Since almost all of the incoming neutrino's momentum is transferred to the muon in $\nu_\mu$ CC, the black circles are mostly above the $E_\mu=E_\mathrm{sh}$ line. In the black hole case, only one muon (or, rarely, two) is emitted during evaporation, or is produced in the subsequent hadronization process. It carries only a fraction of the energy of the incoming neutrino, which is why most of the colored dots are below that line, as expected. This tendency is clearly seen also in the right panel of Fig.~\ref{fig:EtrEsh} where we show the distribution of track energy to shower energy ratio: the ratio in the black hole case extends to small values whereas it peaks at $E_\mu \simeq 10E_\mathrm{sh}$ in $\nu_\mu$ CC. The trend is independent of incoming neutrino energy and $M_\star$. This enables us to distinguish black hole tracks and $\nu_\mu$ CC events with $E_\mu/E_\mathrm{sh}$ on a statistical basis. We note that the fraction of track events produced by BHs also decreases with increasing $M_\star$ due to a lower total number of particles emitted. For 
$M_\star=$1~TeV, 2~TeV, 3~TeV and 10~TeV the fraction is $34\%$, $34\%$, $32\%$, $28\%$, respectively.

\subsubsection{Double bang-like event}
\label{sec:bangbang}
\begin{figure}
\begin{tabular}{c c}
    \includegraphics[width =0.49 \textwidth]{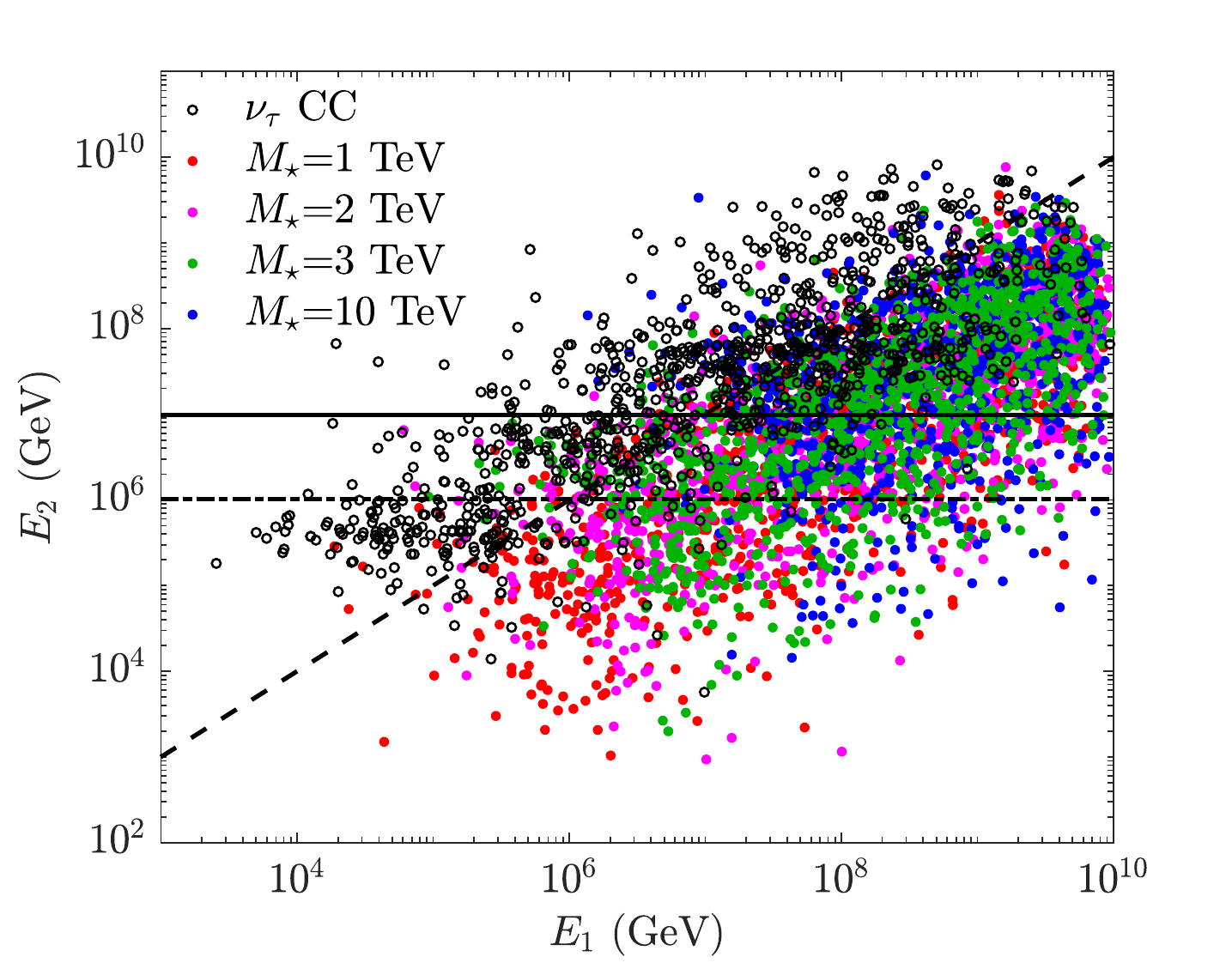} &
  \includegraphics[width = 0.49\textwidth]{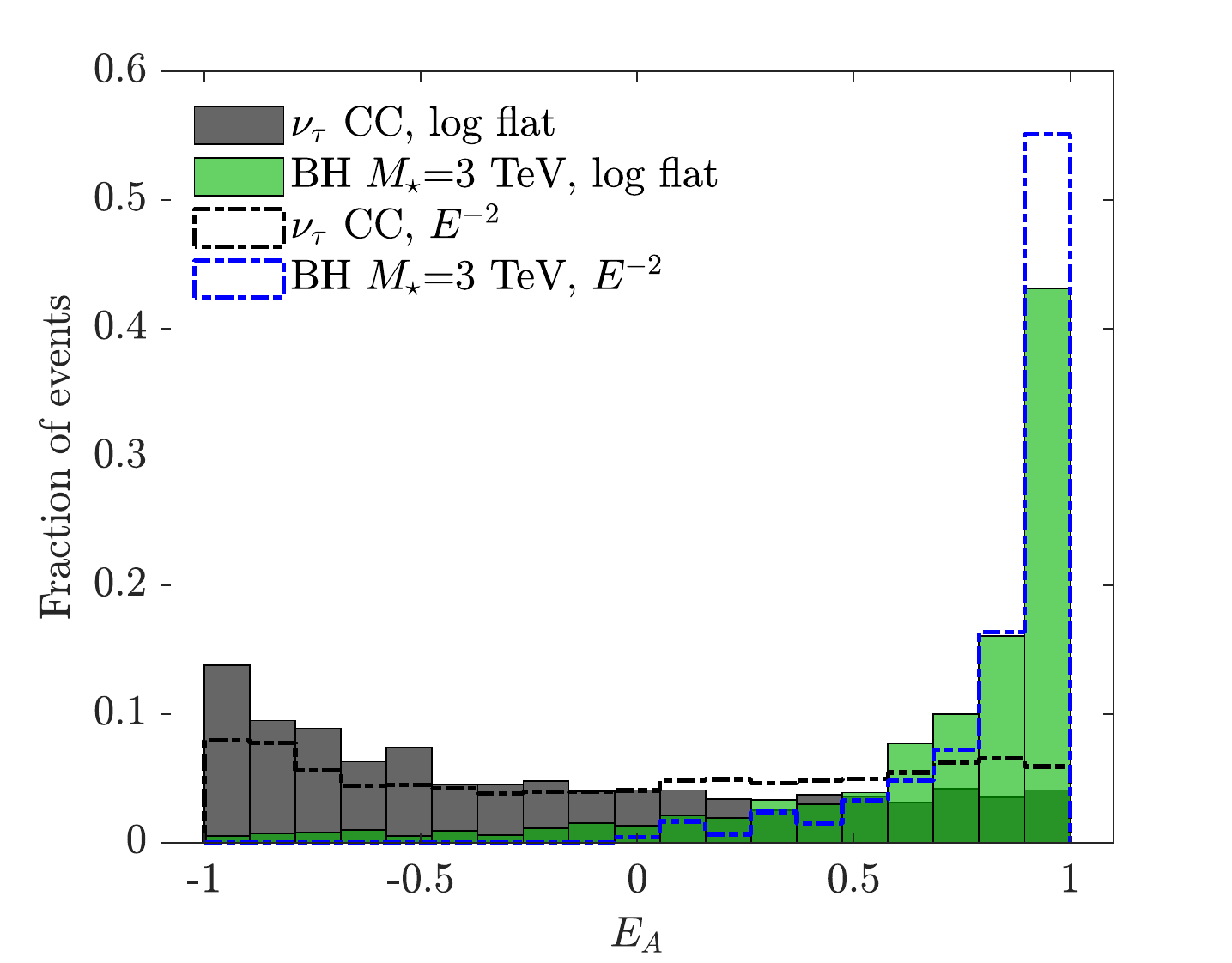}
  \end{tabular}

    \caption{\textbf{Left:} The energy of the first cascade versus the energy of the second cascade in black hole simulations (colored dots) and $\nu_\tau$ charged current events (black circles). Red, magenta, green and blue dots correspond to simulations in 6 extra dimensions with $M_\star=$1~TeV, 2~TeV, 3~TeV, and 10~TeV, respectively. The dashed line depicts the $E_1=E_2$ threshold, the solid line shows the region above which most of the taus leave the detector, and the dash-dotted line shows the region below which most of the taus are supposed to decay within 100~m. Events in between would appear as ``double bang-like.'' In each case, 1000 double bang-like events are simulated in the same way as Fig.~\ref{fig:EtrEsh}. We require $E_\tau>1\ $TeV to produce a clear second cascade. Tau energy loss is not included. \textbf{Right:} Histogram of the energy asymmetry factor $E_A$ in $\nu_\tau$ CC (shaded black) and $M_\star=3$~TeV black hole evaporation (shaded green) for a log flat neutrino spectrum with infinite detector volume. Dashed lines show the corresponding cases using $E^{-2}$ neutrino flux and a minimum energy deposition of 20~PeV. For these, in contrast with the shaded bars, only tau decays inside the detector are selected and the detector geometry is described in Sec.~\ref{sec:flavorcomp}. See Sec.~\ref{sec:detect} for details.}
    \label{fig:E1vsE2}
\end{figure}

If a tau lepton is one of the evaporation products and it decays hadronically or electronically inside the detector, a black hole event produces two cascades separated by a distance $c$ times the tau's decay time. To be classified as a double bang event, the energy of the cascades has to satisfy the energy asymmetry condition $-0.98\leq E_A \leq 0.3$, as described in the beginning of Sec.~\ref{sec:topology}. Here, we instead refer to all two-cascade events -- whether or not they satisfy the asymmetry condition -- as ``double bang-like'' events (abbreviated later as dbl). We simulate $\nu_\tau$ CC and black hole events in the same way as in our track analysis, and show the results in Fig.~\ref{fig:E1vsE2}. Simulated taus are forced to decay regardless of the distance they travel and their energy loss is neglected. Only hadronic and electronic decay events are selected and the total number of events is 1000 in each case. We also draw a horizontal solid line above which taus are more likely to leave the detector and a dash-dotted line below which most of the taus will decay within 100~m. Double-bang like events will be detectable in between. The solid line is obtained by assuming $E_2=E_\tau/2$ where $E_\tau c\tau_0/m_\tau =\langle L\rangle$ (see Appendix~\ref{sec:xss} for $\langle L\rangle$) and the dash-dotted line is obtained with $L=100$~m correspondingly.

Since we focus on high energy incoming neutrinos (PeV or higher), most of the taus produced in $\nu_\tau$ CC will leave the detector before decay. This is not necessarily the case however for black hole decay, where taus are only one or a few of the evaporation products which can be less energetic. The difference results in different energy distributions in the two cascades: $\nu_\tau$ CC is characterized by a small cascade, followed by a large shower, whereas black hole yields a large shower, followed by a smaller cascade. The distribution of the energy asymmetry is shown in the right panel of Fig.~\ref{fig:E1vsE2}. It is manifested that $E_A$ is peaked at large positive values for black hole events and $\nu_\tau$ CC is peaked at the negative end. The distinction is consistent with the discussion of double bang events in Sec.~\ref{sec:flavorcomp}. Because the distributions do overlap, the double bang distribution can be used to exclude a BH component, but will require more statistics to infer the presence of BHs on its own.

\subsection{New unique topologies}
\label{sec:newtopologies}
\begin{figure}
    \centering
    \includegraphics[width=\textwidth]{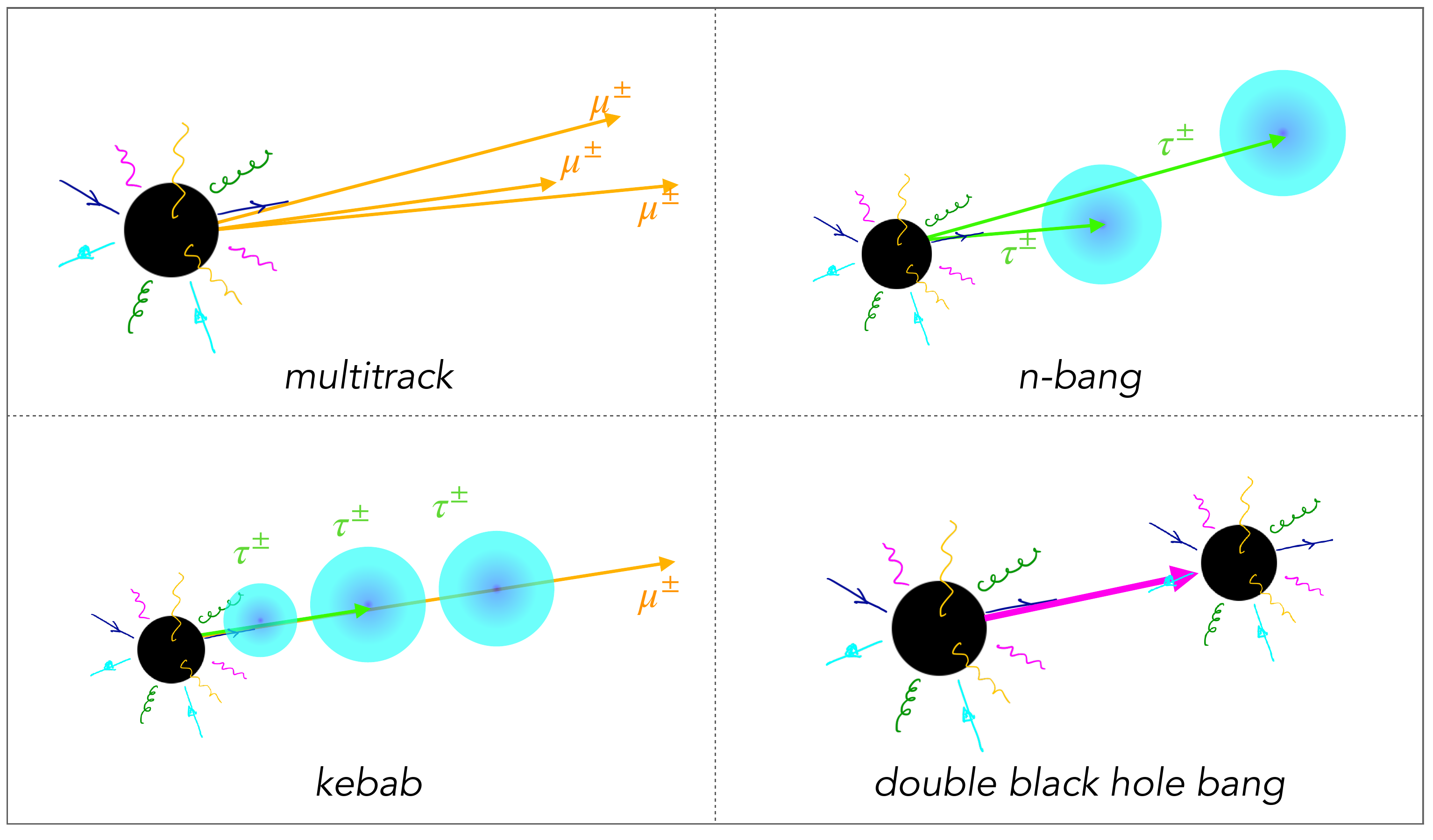}
    \caption{New event topologies that can arise from neutrino-nucleus interactions that result in the creation of a microscopic black hole. See text for technical description. Here we represent the black hole as a black circle. Because most evaporation products will be hadronic, electronic, or low-energy products of weak decays, there will always be a cascade at the interaction vertex. High-energy muons which propagate large distances in the ice are shown as orange arrows. The blue circles represent hadronic or electronic decays of energetic tau leptons resulting in time-delayed cascade (``bang'') signatures. Note that opening angles are greatly exaggerated: because the collisions are highly boosted, these will be well below $10^{-2}$ degrees.}
    \label{fig:weirdtopologies}
\end{figure}

In addition to a modification of the production rates of events of the topologies usually seen in neutrino events, the high multiplicity of BH evaporation products can lead to new and unique event topologies:
\begin{enumerate}
    \item High-multiplicity \textbf{multitrack} events. For about 12\% of black hole events, more than one muon or tau track can be produced from BH evaporation. However, we find that the angular separation between tracks is below $0.02^\circ$ (typically $\sim 0.001^\circ$) due to beaming -- much lower than the angular resolution that can optimistically be achieved with IceCube~\cite{Bradascio:2019eub}, around $0.1^\circ$. 
    \item The \textbf{$n$-bang} is the result of multiple tau leptons decaying electronically or hadronically at different distances from the primary interaction vertex, leading to several cascades in a row. We require the mutual separation of the bangs to be larger than 100~m. $n$-bangs account for about 0.2\% of total BH events for $E^{-2}$ neutrino spectrum between 20~PeV and 10~EeV assuming $M_\star=3$~TeV in IceCube-Gen2.
    \item The \textbf{kebab} can occur when high-energy muons are produced simultaneously with one or more tau leptons. Tracks can either be produced by muons from BH evaporation or from tau decay within the detector, or high energy taus (above 5~PeV) escaping the detector. Kebabs account for about 3\% of total BH events with the same flux assumption. Note that there is about 1\% overlap between multitrack and kebab.
    \item In the event that $E_\nu \gg M_\star$, one may even see a \textbf{double black hole bang}, wherein a high-energy evaporation product interacts again in the detector, producing a second BH-induced shower. 
\end{enumerate}
These are illustrated in Fig. \ref{fig:weirdtopologies}.

Unless one of these smoking-gun signatures is seen, we have shown in this section that event topology allows for a limited identification of BH events: track and double-bang production rates can be different from the SM case, and more importantly, the ratio of track-to-shower energy (or second-to-first cascade, in the case of double-bangs) can further help distinguish BHs from SM events. In the next section, we turn to shower-type events, which represent the majority of predicted observations, and ask whether BHs lead to a unique signature there as well.

% (It seems that black holes with TeV mass or higher evaporate really quickly so that only shower events are seen in IceCube, but we can possibly look at the angular distribution of the Cherenkov photons $\frac{dN_\gamma}{d\Phi}$ or the energy distribution of the photons $\frac{dN_\gamma}{dE}$ or both. In the case where electromagnetic and hadronic showers can be distinguished, we can probably look at the ratio between $\frac{dN_h}{dE}$ (hadronic) and $\frac{dN_em}{dE}$ (electromagnetic).)

\section{Cherenkov timing and light echos}
\label{sec:echo}
In both SM and black hole events, showers are the most common topology. The main feature of a BH event is the rapid production of a large number of high-energy particles, drawn randomly from the SM. Including all polarization states and colors, a large number of those SM degrees of freedom lie in the strong (quark + gluon) sector. This means that shower-type events caused by evaporating BHs will be characterized by a high-energy hadronic shower.

In the SM, charged  current $\nu_e$ interactions produce an electromagnetic cascade, largely resulting in electrons and positrons, and a very few hadrons, which cascade to low energies. Electromagnetic cascades typically occur at large values of $(1-y) = E_e/E_\nu$ (where $y$ here is the Bjorken $y$ parameter commonly used to indicate the fraction of the neutrino energy transferred to the quark), as most of the incoming neutrino's momentum is transferred to the electron. In contrast, neutral current events are seen via their \textit{hadronic cascade}, which can lead to large numbers of neutrons and pions in final state. These are typically much lower in energy than the incoming neutrino, as most of the energy remains with the outgoing neutrino.

Electromagnetic and hadronic showers are not spatially distinguishable. However, given that their compositions are quite different, one can search for an ``afterglow'' from hadronic showers, as copious neutrons and muons decay and recombine with nuclei in the ice. Ref.~\cite{li:2016kra} found that this leads to a distinctive three-peak signature in the time-evolution of the Cherenkov light produced by a high-energy neutrino. The first peak, which crests around $10^{-7}$~s is the largest, and comes from the immediate energy deposition by primary particles. The second peak is visible between 1 and 10 $\mu$s after the first, and is due to the decay of low-energy muons produced in the shower. The third, reaching its maximum between 0.1 and 1 ms, is due to neutron capture. This three-peak signature is predicted for all types of interaction, but because CC events produce electrons which lead to very few muons (which come from meson decay) and neutrons, the amplitude of the second and third peaks should be much lower than in hadronic showers.

As in the previous section, we perform our black hole simulations using our modified version of BlackMax, and pass these results to Pythia 8 for hadronization and heavy particle decay. This output is then fed into FLUKA~\cite{ferrari2005fluka,Battistoni:2007zzb}, which models subsequent propagation, decay, and interaction with the detector material, including the production of Cherenkov photons. We note one improvement with respect to the analysis in Ref. \cite{li:2016kra}: rather than treating hadronic events as a single hadron, we will include the full effects of hadronization with the aid of Pythia, which leads to slightly more overlap between NC and CC events. 

\begin{figure}[!htb]
    \begin{tabular}{c c}
    \includegraphics[width=0.49\textwidth]{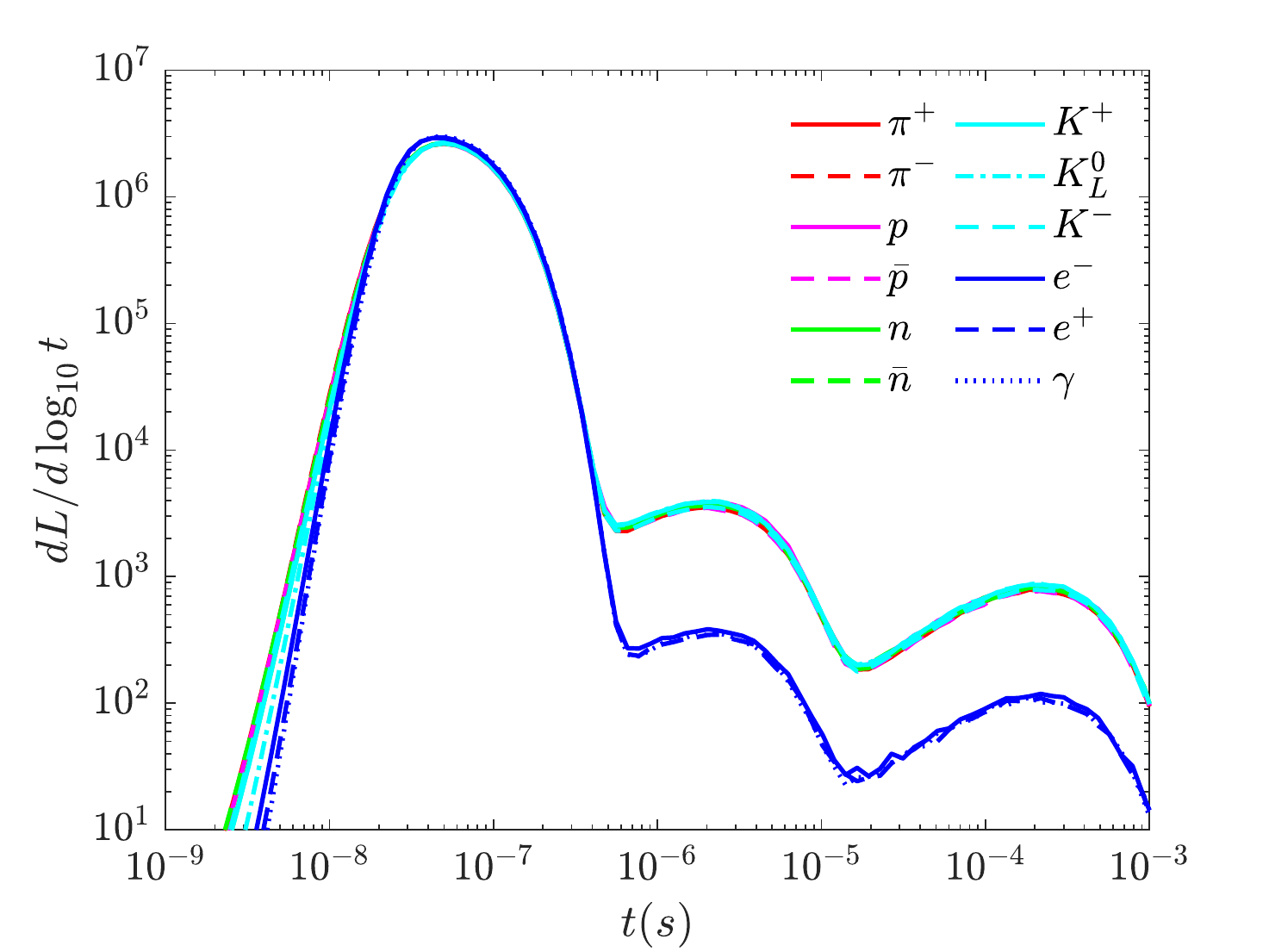} &
    \includegraphics[width=0.49\textwidth]{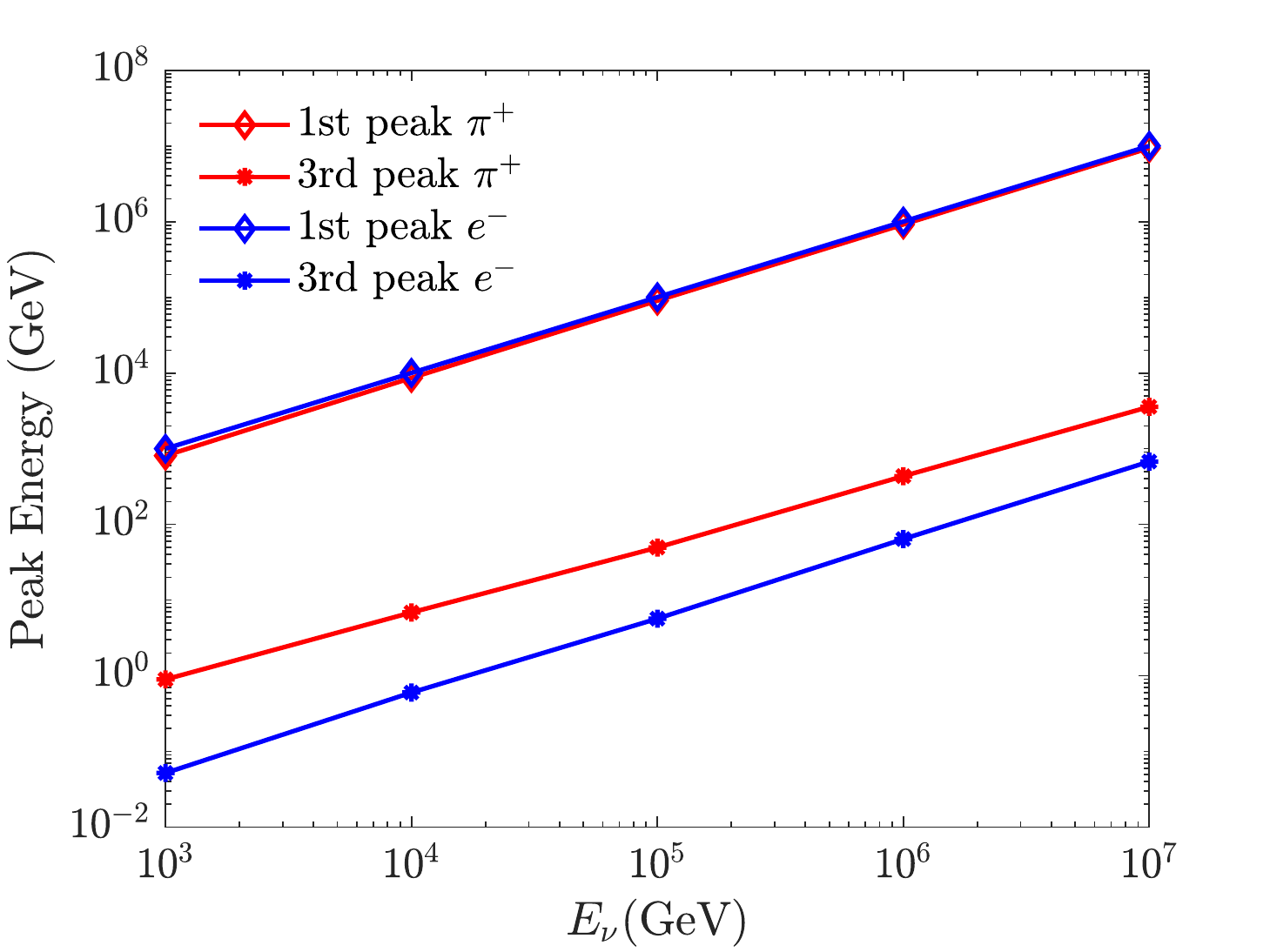}
    \end{tabular}
    \caption{\textbf{Left:} Spectrum of Cherenkov light produced as a function of time by propagating a 1~PeV particle in ice using FLUKA. The y-axis is in arbitrary units. Relatively long-lived particles are simulated including $\pi^+$ (solid red), $\pi^-$ (dashed red), proton (solid magenta), antiproton(dashed magenta), neutron (solid green), antineutron (dashed green), $K^+$ (solid cyan), $K^-$ (dashed cyan), $K_L^0$ (long-lived eigenstate, dash-dotted cyan), electron (solid blue), positron (dashed blue) and photon (dotted blue) -- these last three overlap on the lower curve, whereas the others overlap on the upper one. \textbf{Right:} Energy of the first peak (line with diamonds) and third peak (line with asterisks) in the Cherenkov light spectrum as a function of particle energy. $\pi^+$ is shown in red and electron is shown in blue. The peak energies are normalized so that the energy of the first peak at 10~PeV equals the energy of the injected particle.}
    \label{fig:spectrumall}
\end{figure}

The time evolution curves of the Cherenkov light produced when a single 1~PeV color-neutral particle is injected into the ice are shown in the left panel of Fig.~\ref{fig:spectrumall}. The shower signatures clearly fall into two categories: hadrons, which produce larger second and third peaks, and electrons and photons which lead to electromagnetic showers with suppressed muon and neutron production. Different particles within each category are nearly indistinguishable from each other, but the echo energy from the hadronic shower is typically about 10 times larger than that from the electromagnetic shower. The behavior with energy scales monotonically, as shown in the right-hand panel of Fig.~\ref{fig:spectrumall}, noting that the second and third peak are perfectly correlated (as also shown in \cite{li:2016kra}). We therefore use $\pi^+$ as a proxy for hadrons, and electrons for electromagnetic showers. Because of the larger time separation, we will focus on the third and first peaks. To estimate the total peak energy, we count the photons before 1~$\mu$s for the first peak, and after 16~$\mu$s as the third peak. The primary peak from the decay of any tau event that is not long-lived enough to be seen as a double-cascade will thus not overlap with the third echo. The peak energies scale almost linearly with the particle energy. In Tab.~\ref{tab:peakfit}, we provide coefficients for the peak energy fit to a power law: $E_p=b(E/\mathrm{GeV})^a$ GeV, where $E_p$ and $E$ are respectively  the energy of the peak and the injected particle. 

\begin{table}[!htb]
    \centering
    \setlength{\tabcolsep}{2em} % for the horizontal padding
    {\renewcommand{\arraystretch}{1.2}% for the vertical padding
    \begin{tabular}{c c c} \toprule
           & a&b\\ \colrule
         1st peak $\pi^+$ &1.02&0.738  \\ 
         3rd peak $\pi^+$ &0.901&$1.71\times 10^{-3}$\\ 
         1st peak $e^-$ &1.00&1.00 \\ 
         3rd peak $e^-$ &1.03&$4.45\times 10^{-5}$ \\\botrule
    \end{tabular}}
    \caption{Best-fit parameters of the relation $E_p=b(E/\mathrm{GeV})^a$ GeV, where $E_p$ is the energy of the peak shown in Fig.~\ref{fig:spectrumall} and $E$ is the energy of the injected particle.}
    \label{tab:peakfit}
\end{table}

The energy of the peak is nearly proportional to the energy of the particle, except for the third peak of $\pi^+$, which grows less efficiently as the particle energy increases. Although we have not simulated the propagation of particles with energy higher than 10~PeV due to computational requirements\footnote{FLUKA is not parallelized, and such a computation requires more than one CPU month.}, we expect that the relation in Tab.~\ref{tab:peakfit} will continue to hold up to at least 10~EeV in the absence of new interactions. We can also construct a new variable $r_{31}\equiv E_{p_3}/E_{p_1}$, the energy ratio between the third and first peaks. Using the parameters in Tab. \ref{tab:peakfit}, we find:
\begin{equation}
    \log_{10}r_{31}=-0.114\log_{10}E_{p_1}-2.65
    \label{eq:NCline}
\end{equation}
for $\pi^+$, i.e. the peak ratio decreases with increasing particle energy. We will refer to Eq.~\eqref{eq:NCline} as the ``Hadron Line'', as it forms a distinctive separation on the $r_{31}$ -- $E_{p_1}$ plane.

\begin{figure}[!htb]
    \begin{tabular}{c c}
    \includegraphics[width=0.49\textwidth]{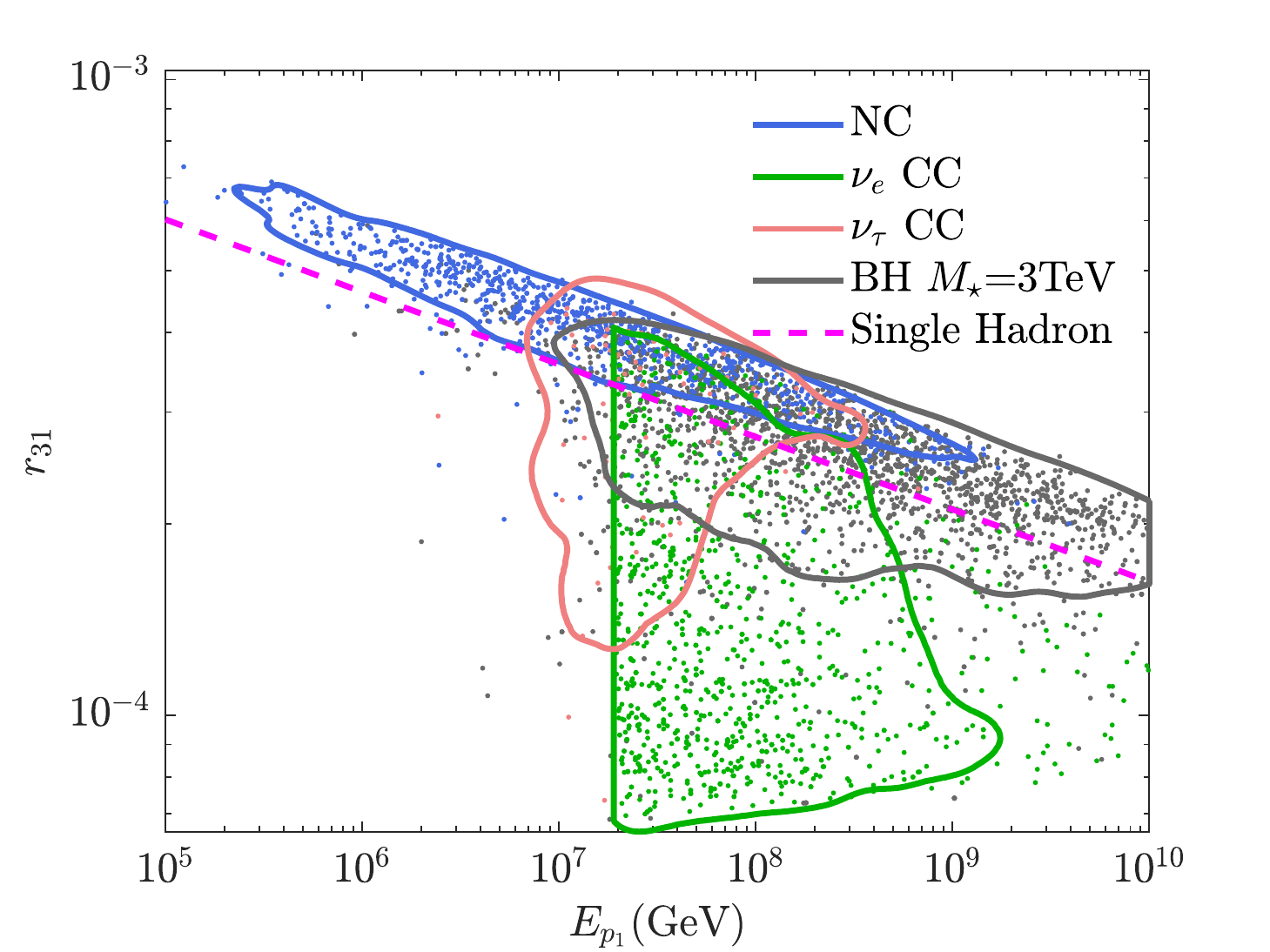} &
    \includegraphics[width=0.49\textwidth]{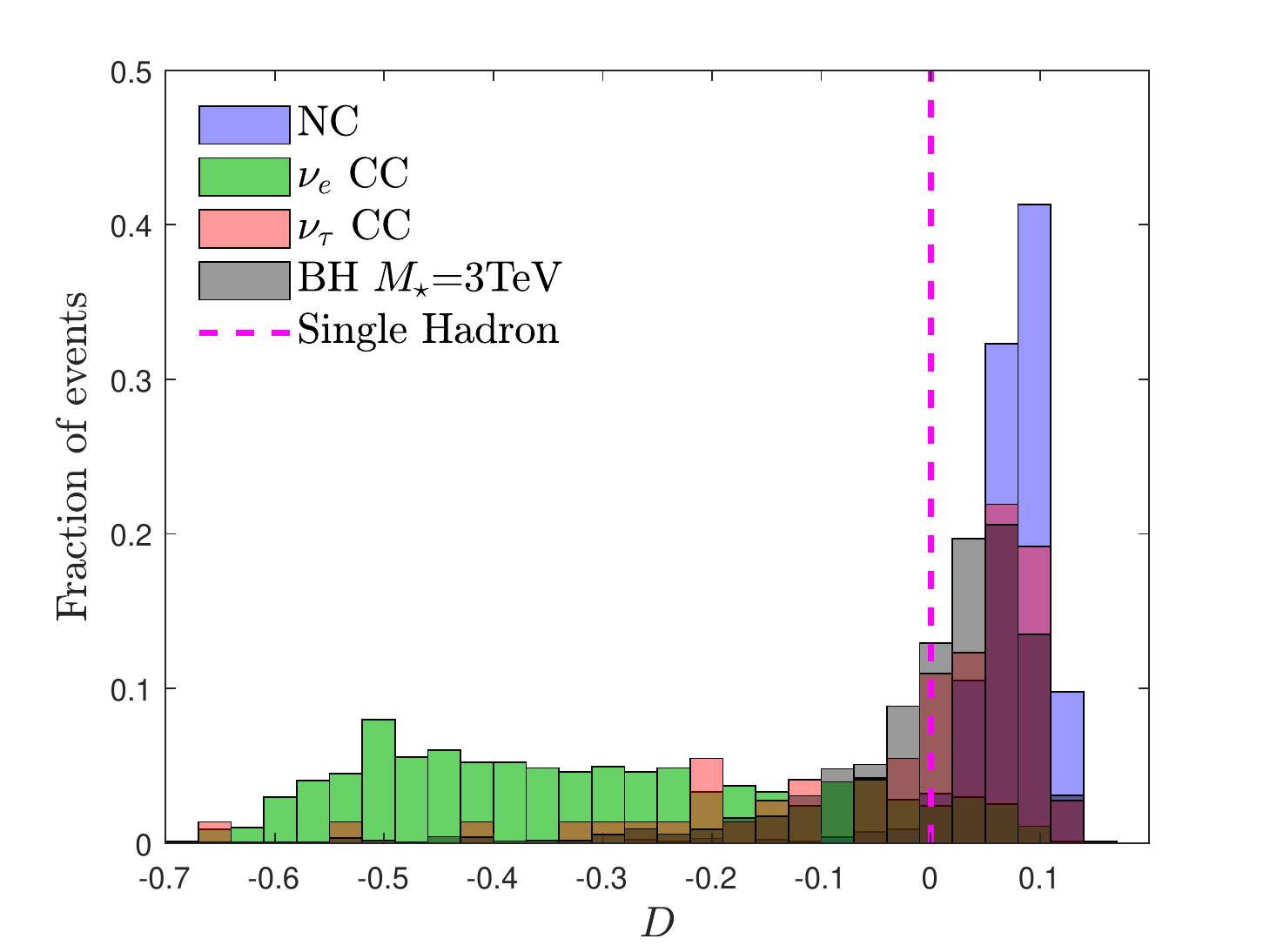}
    \end{tabular}
    \caption{\textbf{Left:} Third peak to first peak energy ratio as a function of the energy of the first peak. Neutrinos from 20~PeV to 10~EeV following the $E^{-2}$ spectrum is simulated assuming (1:1:1) flavor composition. Blue, green, red and black dots depict NC, $\nu_e$ CC, $\nu_\tau$ CC and black hole production events with $M_\star=3$ TeV and 6 extra dimensions. The solid lines enclose the 90\% regions correspondingly. The dashed magenta line is the Hadron Line of Eq.~\eqref{eq:NCline}. The number of simulations is proportional to the number of events predicted in the detector, i.e. $\int dE_\nu\sigma(E_\nu)M_\mathrm{eff}(E_\nu)\frac{d\phi_\nu}{dE_\nu}$. For SM interactions an up quark is hadronized as part of the final state particles. In both SM and BH cases, only shower events and tau decay within 100~m are selected. \textbf{Right:} Distance of events from the Hadron Line defined in Eq.~\eqref{eq:D} for NC (blue), $\nu_e$ CC (green), $\nu_\tau$ CC (red) and black hole with $M_\star=3$ TeV (black). The single Hadron Line is again shown in dashed magenta.}
    \label{fig:pratios}
\end{figure}

However, actual NC events do not produce single mesons. Rather, hadronization will lead to a jet of multiple mesons and baryons as well as electrons and photons, causing the final signature to look like a linear combination of the hadronic and electromagnetic peak structures. To model this correctly, we simulate NC events by hadronizing an up quark with the momentum transfer expected in an NC event using Pythia 8. An $E^{-2}$ neutrino spectrum is assumed from 20~PeV to 10~EeV as in Sec.~\ref{sec:flavorcomp}. The result is shown in the left panel of Fig.~\ref{fig:pratios}. We see that most of the NC events are above the Hadron Line described in Eq.~\ref{eq:NCline}. This is because the amplitude of each peak scales with the number of hadrons in a jet, but the ratio remains constant. This pushes the events to the right of the figure, though the occasional production of $\pi^0$ particles, which decay to photons, can lower the peak ratio in individual events. 

In Fig.~\ref{fig:pratios}, we also show $\nu_e$ CC events after hadronization. Since most of the momentum goes into the lepton in such events, they mainly fall below the Hadron Line, the value of $r_{13}$ being determined by $y$.

On the other hand, black holes, which burst into a mixture of hadrons and electromagnetic particles, lie right in between. However, when compared to NC events, they cluster at higher energies. This is because the NC rate goes as $y d\phi/dE_\nu$, whereas the black hole production cross section has no dependence on transferred momentum. NC events, which peak at low $y$, thus require a higher neutrino energy to produce a shower with the same energy. This is strongly suppressed by the power law flux.

For $\nu_\tau$ CC, we consider tau decays within 100~m to be showers (rather than double bang events) since in such case it will be challenging to distinguish the second cascade from the primary one. Note that for such high energy neutrinos (20~PeV or higher) tau leptons from $\nu_\tau$ CC are much more likely to leave a track or produce two cascades than a single shower. We nonetheless include them in the analysis. As above, we simulate $\nu_\tau$ CC and tau decay with Pythia 8. We see from the figure that the few $\nu_\tau$ CC shower events cluster at energies between 20~PeV and 400~PeV, due to the tau decay requirement. Depending on the decay channel, $\nu_\tau$ CC showers can reside above or below the Hadron Line. 

Next, we look at the distribution of events according to their distance from the Hadron Line:
\begin{equation}
    D\equiv\dfrac{0.114\log_{10}E_{p_1}+\log_{10}r_{31}+2.65}{\sqrt{0.114^2+1}}\,.
    \label{eq:D}
\end{equation}
$D$ is positive above the Hadron Line and negative below. We show the histogram of $D$ in the right panel of Fig.~\ref{fig:pratios}. As expected, NC events are mostly positive while $\nu_e$ CC are negative, and $\nu_\tau$ CC and black hole events mainly fall into the region $[-0.1, 0.1]$.

\section{Detection prospects}
\label{sec:detect}

We now determine the number of observed events required to ascertain black hole production based on  showers (Cherenkov echo information) from Sec.~\ref{sec:echo}, muon tracks (total track energy to shower energy ratio) from Sec.~\ref{sec:track} and double-bang events (energy asymmetry) from Sec.~\ref{sec:bangbang}.

\subsection{Showers}
\begin{figure}[!htb]
    \includegraphics[width=0.8\textwidth]{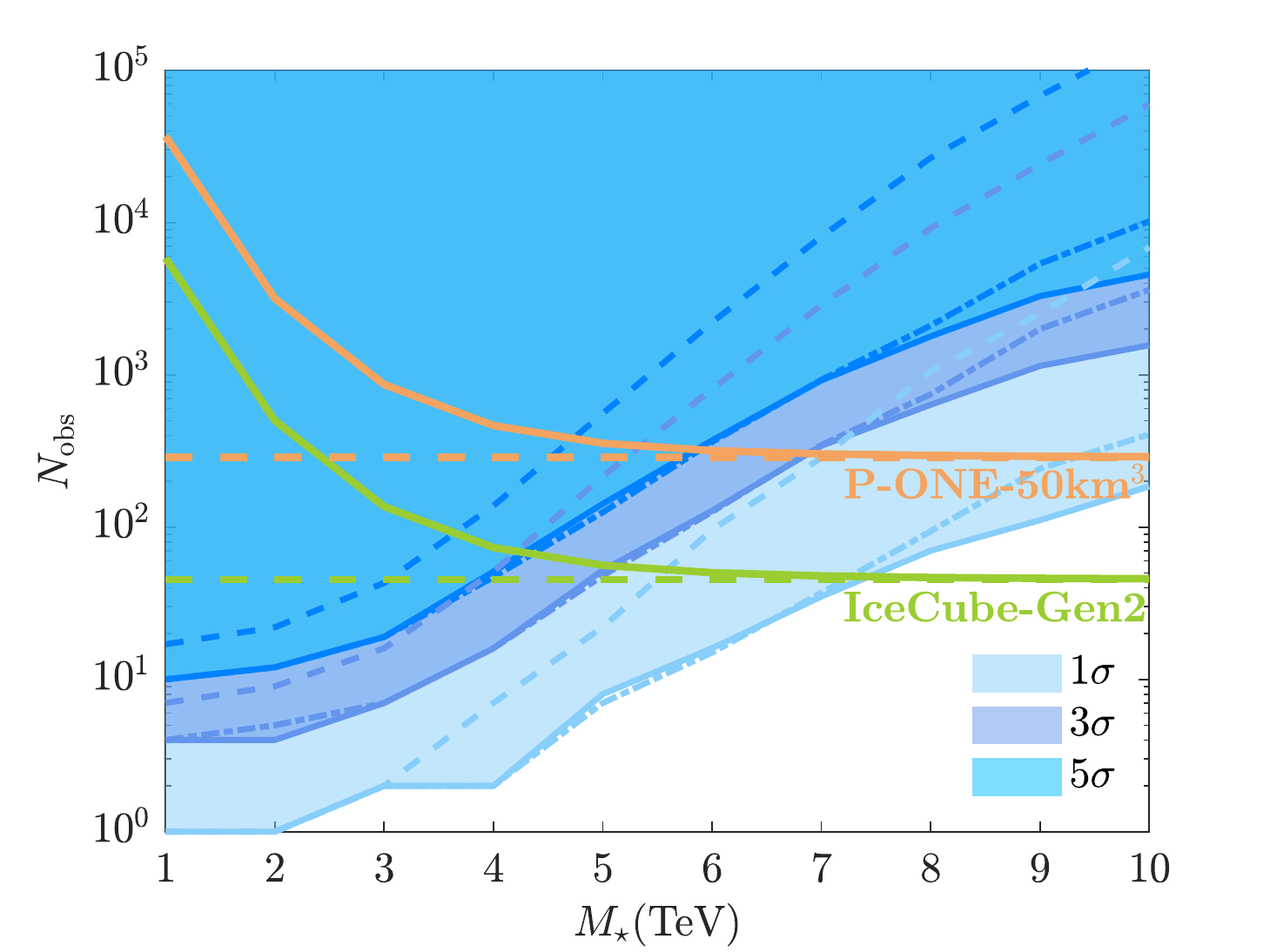}
    \caption{Total number of observation required to discriminate black hole production. The shaded regions from light to dark correspond to $1\sigma$, $3\sigma$ and $5\sigma$ CL with all shower, muon track and double bang-like events included. The dashed blue lines from bottom to top show the $1\sigma$, $3\sigma$ and $5\sigma$ CL using only information from showers while the dash-dotted blue lines use both showers and muon tracks. The dashed green line depicts the expected total number of observations at IceCube-Gen2 assuming SM only interactions and solid green line shows the expected observations with both SM and BH interactions. The dashed and solid orange lines show the expect observations at the P-ONE in-ocean neutrino telescope~\cite{Bedard:2018zml,nelles2019} with 50~km$^3$ effective volume and 10 years of exposure in SM and SM+BH cases.}
    \label{fig:Nobs}
\end{figure}

We begin with shower events. We again assume an isotropic, $E_{\nu}^{-2}$ neutrino spectrum with $(1:1:1)$ flavor composition and require a minimum energy deposition of 20~PeV (corresponding to a CM energy of 4.5 TeV). Note that unlike in Sec.~\ref{sec:flavorcomp}, events in which taus decay in the detector after traveling farther than 100~m are classified as double-bang like events (dubbed ``DBL''), regardless of the energy asymmetry $E_A$. Events where the tau decays within 100~m are still considered as showers, unless the decay results in a muon track. As mentioned before, because of the high neutrino energy, showers from $\nu_\tau$ CC only make up about 2\% of total shower events in SM. Combining NC, $\nu_e$ CC and $\nu_\tau$ CC shower events in the Standard Model case, we obtain the following probability distribution for the distance $D$ from the Hadron Line Eq.~\eqref{eq:D}:
\begin{equation}
    f^\mathrm{SM}(D)=f^\mathrm{NC}(D)N^{\msh,\mathrm{NC}}+f_{\nu_e}^\mathrm{CC}(D)N_{\nu_e}^{\msh,\mathrm{CC}}+f_{\nu_\tau}^\mathrm{CC}(D)N_{\nu_\tau}^{\msh,\mathrm{CC}}\,,
\end{equation}
where $N_{\nu_e}^{\msh,\mathrm{CC}}$ and $N^{\msh,\mathrm{NC}}$ are given by \cref{eq:NXi,eq:NNC}, $N_{\nu_\tau}^{\msh,\mathrm{CC}}=0.79$. $f_{\nu_e}^\mathrm{CC}$ is represented by the green histogram in the right panel of Fig.~\ref{fig:pratios} and $f^\mathrm{NC}$ and $f_{\nu_\tau}^\mathrm{CC}$ are similar to the blue and red histograms regardless of the minimum energy deposition. In the presence of LEDs, the distribution includes both SM and BH interactions:
\begin{equation}
    f^\mathrm{SM+BH}(D,M_\star)=f^\mathrm{NC}(D)N^{\msh,\mathrm{NC}}+f_{\nu_e}^\mathrm{CC}(D)N_{\nu_e}^{\msh,\mathrm{CC}}+f_{\nu_\tau}^\mathrm{CC}(D)N_{\nu_\tau}^{\msh,\mathrm{CC}}+f^\mathrm{BH}(D,M_\star)N_\mathrm{BH}^\msh\,,
\end{equation}
where $N_\mathrm{BH}^\msh$ is given by Eq.~\eqref{eq:NBH} and $f^\mathrm{BH}$ selects the events satisfying the minimum energy deposition condition, as in the NC case. We provide a plot of these combined distributions in Fig.~\ref{fig:pdfd} (Appendix \ref{sec:pdfd}). For low $M_\star$, black holes dominate the events, while for $M_\star>8$~TeV, the distribution is barely distinguishable from the SM case. Next we generate $N_\mathrm{obs}^\msh$ shower events from SM+BH distribution $f^\mathrm{SM+BH}(D,M_\star)$, and compare the unbinned Poisson likelihood when reconstructing the events using SM and SM+BH distributions with the same $M_\star$. We allow the flux normalization to vary and calculate the test statistic
\begin{equation}
    \lambda=-2\ln\left(\dfrac{\mathcal{L}^\mathrm{SM}(N_{a,\max})}{\mathcal{L}^\mathrm{SM+BH}(N_{a,\max})}\right)\,,
    \label{eq:teststats}
\end{equation}
where the likelihood is defined as:
\begin{equation}
    \mathcal{L}(N_a)=e^{-N_a}\prod\limits_{i=1}^{N_\mathrm{obs}^\msh} N_a f(D_i),
\end{equation}
and $N_{a,\max}$ is chosen to maximize the likelihood in each case. From $\lambda$ we then compute $p$-values assuming a $\chi^2$ distribution (Wilks' theorem). The results are shown as dashed lines in Fig.~\ref{fig:Nobs}. We find the number of observations required to discover BHs increases dramatically with $M_\star$. For $M_\star=3$~TeV at least 2, 16, and 45 events are necessary to claim a discovery at the $1\sigma$, $3\sigma$ and $5\sigma$ CL respectively, while the numbers grow to $7 \times 10^3$, $6 \times 10^4$ and $1.6 \times 10^5$ events for $M_\star=10$~TeV. 

\subsection{Muon tracks}

We then proceed to add information from muon track events.  Defining $\xi\equiv \log_{10}(E_\mu/E_\msh)$, the distributions of $\xi$ assuming an $E_\nu^{-2}$ neutrino spectrum are shown by the dashed lines in the right panel of Fig.~\ref{fig:EtrEsh}. The distribution of the BH events is not very different from the case where they were evenly sampled across every decade in energy. However, the SM $\nu_\mu$ CC events shift to the left, due to the minimum energy cut we impose on the shower energy, making the distributions harder to distinguish from one another. The distribution including both SM and BH events reads
\begin{equation}
    f^\mathrm{SM+BH}(\xi,M_\star)=f_{\nu_\mu}^\mathrm{CC}(\xi)N_{\nu_\mu}^{\mtr,\mathrm{CC}}+f^\mathrm{BH}(\xi,M_\star)N_\mathrm{BH}^\mtr\,,
    \label{eq:pdfalltr}
\end{equation}
where $N_{\nu_\mu}^{\mtr,\mathrm{CC}}$ and $N_\mathrm{BH}^\mtr$ are given by \cref{eq:NXi,eq:NBH}. In the SM case we simply have:
\begin{equation}
   f^\mathrm{SM}(\xi)=f_{\nu_\mu}^\mathrm{CC}(\xi).  
\end{equation}
The likelihood using both shower and muon track events is now
\begin{equation}
    \mathcal{L}(N_a)=e^{-N_a}\left( \prod\limits_{i=1}^{N_\mathrm{obs}^\msh} N_a p^\msh f(D_i) \right) \left(\prod\limits_{i=1}^{N_\mathrm{obs}^\mtr} N_a p^\mtr f(\xi_i)\right)\,,
\end{equation}
where $p^\msh$ and $p^\mtr$ are the proportion of shower and muon track events. In the SM,
\begin{equation}
    p^\msh=\dfrac{N_{\nu_e}^{\msh,\mathrm{CC}}+N^{\msh,\mathrm{NC}}}{N_{\nu_e}^{\msh,\mathrm{CC}}+N^{\msh,\mathrm{NC}}+N_{\nu_\mu}^{\mtr,\mathrm{CC}}}
    \label{eq:pshsm}
\end{equation} 
and $p^\mtr=1-p^\msh$. With BH events,
\begin{equation}
    p^\msh=\dfrac{N_{\nu_e}^{\msh,\mathrm{CC}}+N^{\msh,\mathrm{NC}}+N_\mathrm{BH}^\msh}{N_{\nu_e}^{\msh,\mathrm{CC}}+N^{\msh,\mathrm{NC}}+N_\mathrm{BH}^\msh+N_{\nu_\mu}^{\mtr,\mathrm{CC}}+N_\mathrm{BH}^\mtr}
    \label{eq:pshall}
\end{equation}
and $p^\mtr$ is given by the same relation. We again generate mock data sets with $N_\mathrm{obs}^\msh$ and $N_\mathrm{obs}^\mtr$ from the SM+BH distribution and build the test statistic using Eq.~\eqref{eq:teststats}. The results are shown as dash-dotted lines in the right panel of Fig.~\ref{fig:Nobs}. Compared with the shower-only analysis, this leads to a slight increase in sensitivity at low $M_\star$, up to nearly an order of magnitude improvement as $M_\star$ rises towards 10 TeV. This is mainly because the shower signatures become indistinguishable from one another, whereas the BH events continue to generate an excess of lower-energy muons with respect to the total shower energy in each interaction. 

\subsection{Double bang-like events}

Finally we add double-bang information so that all topologies are included. The distribution of the two-bang energy asymmetry in the case of an $E_\nu^{-2}$  incoming neutrino spectrum is shown with dashed lines in the right panel of Fig.~\ref{fig:E1vsE2}. Since more energetic taus are prone to escaping the detector instead of creating a second bang, the probability of large energy asymmetry $E_A$ increases in both the SM and BH cases compared to the log flat distribution. We then arrive at our full likelihood:
\begin{equation}
    \mathcal{L}(N_a)=e^{-N_a}\left(\prod\limits_{i=1}^{N_\mathrm{obs}^\msh} N_a p^\msh f(D_i)\right)\left(\prod\limits_{i=1}^{N_\mathrm{obs}^\mtr} N_a p^\mtr f(\xi_i)\right)\left(\prod\limits_{i=1}^{N_\mathrm{obs}^\mdbl} N_a p^\mdbl f(E_{A_i})\right)\,.
\end{equation}
In the SM case $f^\mathrm{SM}(E_A)=f_{\nu_\tau}^\mathrm{CC}(E_A)$. For SM+BH the distribution is similar to Eq.~\eqref{eq:pdfalltr}. The frequency of each topology is recalculated in analogy to \cref{eq:pshsm,eq:pshall} by including the expected double-bang-like events ($N_{\nu_\tau}^{\mdbl,\mathrm{CC}}=4.77$). The number of observations required in this case are shown with the shaded regions (solid lines) in Fig.~\ref{fig:Nobs}. We find that adding double bang information improves the sensitivity mildly only at $M_\star>7$~TeV, due to the limited fraction of double bang-like events in the total observations.

We also draw green and orange lines in Fig.~\ref{fig:Nobs} corresponding to the expected number of observed neutrinos at future experiments with 10 years of operation in SM only (dashed) and SM+BH (solid) cases. To compute these sensitivities for IceCube-Gen2 (green), we use the effective mass discussed in Sec \ref{sec:flavorcomp}. We also include sensitivities for the proposed Pacific Ocean Neutrino Explorer (P-ONE), the full-scale follow-up to STRAW (Strings for Absorption length In Water\cite{Bedard:2018zml}), located off Vancouver Island. P-ONE has a modular approach, and by deploying several clusters of strings, up to 50 cubic kilometers could be instrumented \cite{nelles2019}. For this projection we take an efficiency similar to that of IceCube-Gen2, but increase the effective volume to 50 km$^3$, by extending the horizontal area of the detector\footnote{This is not quite equivalent to simply scaling the volume, as the likelihood of observing a double-bang event before the tau escapes the detector depends on the geometry.}. This scenario is depicted as orange lines in Fig.~\ref{fig:Nobs}.

IceCube-Gen2 allows for the discovery of BHs with $M_\star<5.1$~TeV (4.3~TeV) at 3$\sigma$ ($5\sigma$) CL and P-ONE improves the sensitivity limit to 6.9~TeV (5.9~TeV).

\begin{figure}[!htb]
    \centering
    \includegraphics[width=0.8\textwidth]{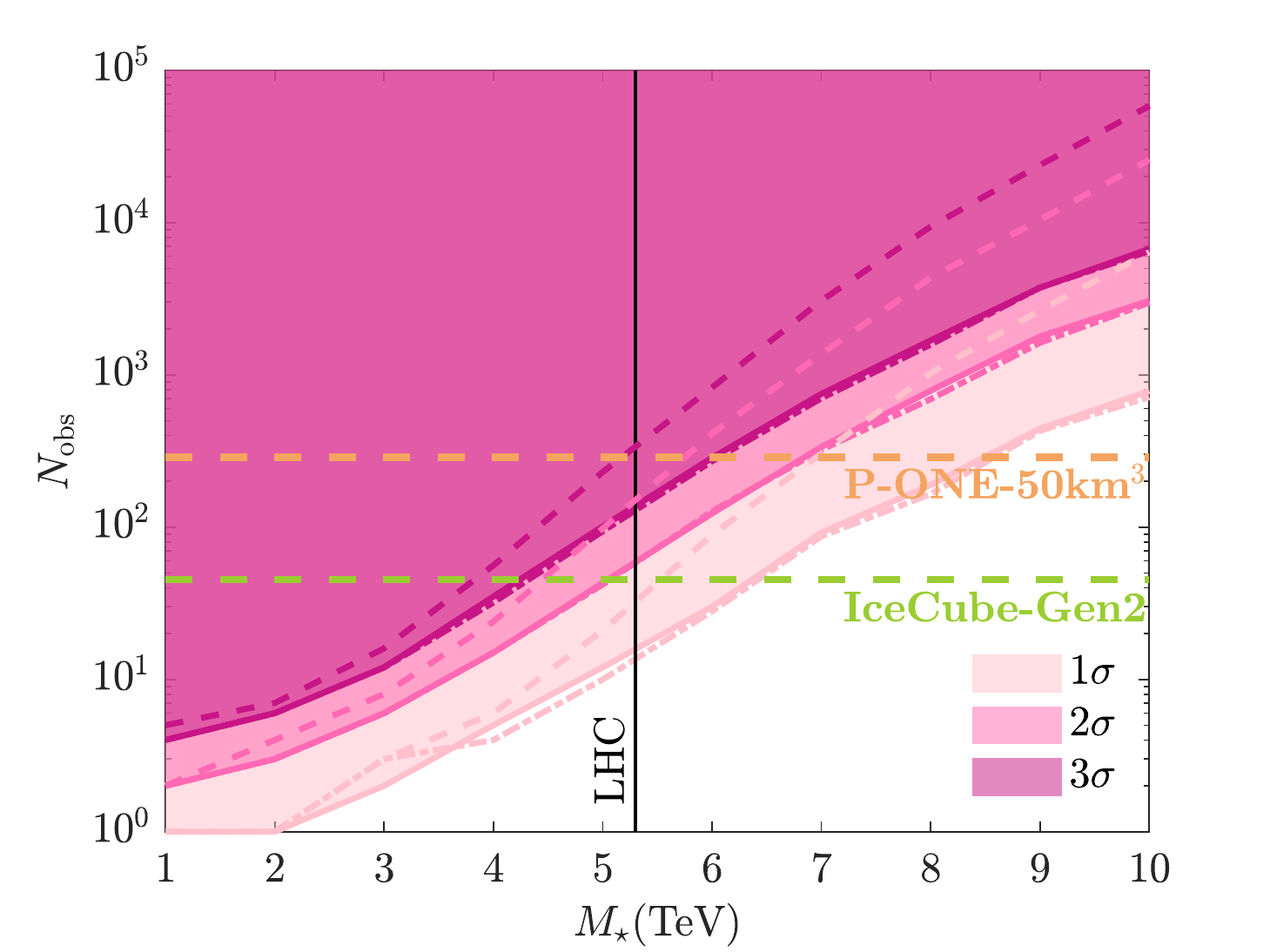}
    \caption{Total number of observations required to exclude black hole production as a function of $M_\star$. The shaded regions from light to dark correspond to $1\sigma$, $2\sigma$ and $3\sigma$ CL with all showers , muon tracks and dobule bang events included. Dashed pink lines from bottom to top show the $1\sigma$, $2\sigma$ and $3\sigma$ CL using only information from showers while the dash-dotted pink lines use both showers and muon tracks. Dashed green and orange lines depict the expected total number of observations at IceCube-Gen2 and P-ONE within 10 years assuming SM only interactions. LHC bound on $M_\star$ at 95\% CL for $n=6$ extra dimensions is shown in black for comparison.}
    \label{fig:Nobsexclude}
\end{figure}

In Fig.~\ref{fig:Nobsexclude} we show the number of observed events required to \textit{exclude} (rather than discover) black hole production for certain $M_\star$. To obtain this we follow the above steps, but generate mock observations from SM-only processes. The test statistic is defined as the inverse of Eq.~\eqref{eq:teststats}, i.e.
\begin{equation}
    \lambda=-2\ln\left(\dfrac{\mathcal{L}^\mathrm{SM+BH}(N_{a,\max})}{\mathcal{L}^\mathrm{SM}(N_{a,\max})}\right)\,,
    \label{eq:teststatssm}
\end{equation}
with the same likelihood functions as above. Comparing with Fig.~\ref{fig:Nobs} we find, unsurprisingly, that more observations are required to exclude BHs than discover them at the same confidence level, since a BH may lead to more extreme values of observables where the probability of SM is small, while the reverse is not true. We also show the total expected number of SM observations in IceCube-Gen2 and P-ONE with horizontal lines. IceCube-Gen2 allows for the exclusion of BH with $M_\star<4.3$~TeV (5.1~TeV) at $3\sigma$ ($2\sigma$) and P-ONE pushes the limit to 6~TeV (6.9~TeV). This can be compared with the current LHC limit $M_\star>5.3$~TeV at $2\sigma$~\cite{Sirunyan:2017jix} for $n=6$ extra dimensions. While IceCube-Gen2 is expected to exclude the Planck scale comparable to LHC, P-ONE has the potential to exclude higher $M_\star$ with larger detector volume and exposure.

% To include muon and tau tracks in the analysis:
% \begin{itemize}
%     \item One can infer the energy of track particle from $\frac{dE}{dx}$ as a function of $x$
%     \item $\frac{dE}{dx}$ for muon track is roughly 10 times larger than tau track
%     \item The energy deposition from tau track in SM $\nu_\tau$ CC is roughly the same as the muon track from black hole decay, which makes it more difficult to distinguish SM tracks and BH tracks
%     \item The likelihood of a track with $E_\mathrm{dep}$ is $f(E_\mathrm{dep}|\mathcal{M})=f(E_\mathrm{dep},\mu|\mathcal{M})+f(E_\mathrm{dep},\tau|\mathcal{M})$ for $\mathcal{M}=$SM and $\mathcal{M}=$BH
%     \item One has to do spectral analysis to construct the likelihood properly, for that one needs $\frac{dN}{dE_\mathrm{dep}}$, see reference B of 1502.02649 for details. If the production rate of black hole is not negligible, it will increase the number of low $E_\mathrm{dep}$ tracks versus high $E_\mathrm{dep}$ tracks. This degenerates with a harder $\nu_\tau$ spectrum or a softer $\nu_\mu$ spectrum.
% \end{itemize}

%\begin{enumerate}
%    \item Neutral current (NC) events all lie in a single line
%    \item Charged-current events lie below the NC line, due to the partial contribution to the echo from the %hadronic shower
%    \item black hole events lie above the line, because of the large number of lower-energy hadronic business
%\end{enumerate}

\section{Vacuum Instability}
\label{sec:vacuumdecay}
Several authors have explored the potentially catastrophic consequences of black hole evaporation in the context of Higgs vacuum metastability \cite{Hiscock:1987hn,Berezin:1987ea,Moss:1984zf,Cheung:2013sxa,Gregory:2013hja,Burda:2015yfa,Burda:2016mou,Cuspinera:2019jwt,Cuspinera:2018woe}. If our Higgs vacuum is indeed metastable, then there exists an instability scale $\Lambda_I$ (typically estimated to be $\sim 10^{19}-10^{20}$ eV) above which the Higgs quartic coupling term becomes negative. This opens up the possibility for stochastic vacuum decay via a tunneling event, for which the decay timescale is, in general, expected to be much longer than the age of the Universe \cite{Andreassen:2017rzq,Chigusa:2018uuj}. However, this decay time can be dramatically reduced in the presence of an evaporating black hole, if the initial mass of the black hole is above the instability scale. In \cite{Mack:2018fny}, the authors showed that this can be used to place constraints on extra dimensional theories in which cosmic ray collisions can create black holes. Similarly, black holes created through the collision of astrophysical neutrinos with nuclei could, in principle, also create black holes of masses above $\Lambda_I$. If these are observed via their unique signatures as described in the previous sections, their presence will constitute indirect evidence for new physics beyond the Standard Model (either associated with the extra dimensional theory or in addition to it) that provides Higgs potential corrections that stabilize the vacuum. If black holes are created in collisions with detector nucleons, this will be a sign that black holes are forming from collisions elsewhere in the cosmos with even higher CM collision energies, and that these, too, are unable to trigger vacuum decay. In this way, an observation of a black hole evaporation signature in a next-generation neutrino telescope would have implications for the ultimate stability of our cosmos in addition to providing clues about the dimensionality of spacetime.

\section{Conclusions}
\label{sec:conclusions}

\subsection{Signatures of new physics}

Astrophysical neutrinos can reach energies up to tens of EeV. In the presence of large extra dimensions, microscopic black holes can be produced in neutrino-nucleus scattering if the center of mass energy of the collision is above the Planck scale in the bulk. This opens up a myriad of possibilities in the next generation of neutrino telescopes. For the first time, we have investigated the new signatures of black hole production and decay in future telescopes such as IceCube-Gen2.

\begin{itemize}
\item \textbf{Flavor composition.}
We first examined how the flavor composition of high energy neutrinos detected at the Earth could change due to the production of black holes. We focused on neutrino energies between 20~PeV and 10~EeV with a minimum energy deposition of 20~PeV and built the likelihood based on the number of shower, track and double bang events in 10 years of IceCube-Gen2 run. We stress that we include tau tracks and double bangs in the analysis, aiming at a complete flavor composition reconstruction framework at ultra high energies. With the assumption of BH production with $M_\star=3$~TeV and 6 extra dimensions, we find the purely tau neutrino case would be strongly disfavored due to the lack of double bangs in BH events and the flavor contour would prefer mostly muon neutrino scenario. The (1:1:1) flavor ratio in that scenario is excluded at 95\% CL. 

\item \textbf{Track muon-to-shower ratios.}
We also show the signatures of tracks and double bangs from BH production. We have shown that the tracks feature a lower muon energy-to-shower energy ratio in the BH case compared with SM $\nu_\mu$ CC, regardless of $M_\star$, since muons from Hawking radiation carry only a fraction of BH energy. 

\item \textbf{Double bang energy cascade ratios.}
When this applies to double bang events, we find BH decay suppresses the energy ratio between the second shower and the first shower, resulting in a large energy asymmetry factor $E_A$ contrasted with SM $\nu_\tau$ CC. 

\item \textbf{Unique topologies.}
We also find that BH production may lead to unique topologies that do not exist in the SM: the multitrack, the $n$-bang, the kebab and the double black hole bang. These are described in Sec.~\ref{sec:newtopologies} and, if seen, may be a clear sign of new physics.

%For example, BH may decay to multiple muons and taus, leaving multiple tracks in the detector. The angular separation between tracks is quite small and experimentally challenging to resolve. BH may also produce several taus which subsequently decay to hadrons or electrons inside the detector, which we call ``n-bang''. ``Kebab'' appears when tau decays are associated with additional muon or tau track. If the black hole decay product is energetic enough to interact with nucleon and produce another black hole, we may have ``double black hole bang''. These new topologies are quite rare, however, the observation of at least one event will be a smoking-gun of black hole production.

\item \textbf{Cherenkov light timing.}
In addition, the time evolution of Cherenkov light enables us to break the degeneracy between SM and BH in shower events. We have simulated the propagation of different particles in ice to obtain the timing information. We find that the neutron echo-to-primary peak ratio of a hadron falls onto a single Hadron Line, which is roughly 10 times larger than the ratio of electrons and photons. We then built a new variable to quantify the distance of peak ratio to the Hadron Line. We found that NC events cluster above the hadron line and $\nu_e$ CC are found below, while BH events lie in between.
\end{itemize}

\subsection{Exclusion and detection prospects}

To forecast the power of future detectors to test extra dimension models, we computed the number of observed high energy neutrino events required to discover or exclude black hole production as a function of the Planck scale $M_\star$ based on the BH shower, track, and double bang signatures discussed above. We found for $M_\star<5.1$~TeV, IceCube-Gen2 can resolve BH production at $3\sigma$ CL within 10 years. A fully upgraded Pacific ocean detector with a 50~km$^3$ effective volume will be sensitive to $M_\star<6.9$~TeV. In terms of exclusion, the limits are 4.3~TeV and 6~TeV instead. See Table~\ref{tab:limits} for a summary of our projected limits. We note that it could be challenging to distinguish muon tracks from tau tracks merely from topology. Nonetheless, muons and taus suffer from different energy loss when propagating, where the average energy loss rate of a tau is typically one order of magnitude lower than that of a muon at energies above 10~TeV~\cite{koehne2013proposal}. This is to say, tau track events from the SM are characterized by a lower $E_{\mtr}/E_{\msh}$ energy ratio than muon tracks. Therefore, full energy dependent analyses of the track-to-shower energy ratio combining $\nu_\mu$ CC and $\nu_\tau$ CC, in cases in which black hole production is present or not, are desired. This is left for future work. 

We have been using $M_\star=3$~TeV as a benchmark throughout the paper to illustrate the ample signatures brought about by large extra dimensions in neutrino telescopes. However, we note that all the signatures remain valid for higher Planck scale. We also note that  stringent limits on $M_\star$~\cite{Sirunyan:2017jix} have already been obtained from proton-proton collisions at the LHC, which is comparable to the exclusion limit of IceCube-Gen2 but inferior to P-ONE. Compared with hadron colliders, IceCube and other neutrino telescopes can serve as a study of quantum gravity in neutrino-nucleon scattering, with different microphysics, particle detection capabilities, and systematics. For example, in split-brane models~\cite{ArkaniHamed:1999dc} where quarks and leptons are confined to different slices of the brane, cross sections to produce black holes in proton-proton collisions and neutrino-proton collisions are vastly different. Such models can only be directly constrained by comparing black hole production in colliders and neutrino telescopes. Furthermore, considering the expected highest astrophysical neutrino energy of tens of EeV, we might eventually explore the Planck scales higher than 100 TeV in future upgrades of neutrino telescopes, merely by increasing detector exposure, which will likely occur before the Future Circular Collider. Any positive signal can thus provide additional motivation and guidance for that project.

%and may thus mimic what is expected from BH muon tracks, as we can see from Fig.~\ref{fig:EtrEsh}. One possible solution to this problem is to make energy dependent analysis, i.e. compare the the event rate per energy loss in SM and BH production,
%\begin{equation}
%    \dfrac{dN}{dE_\mathrm{loss}}(E_\msh)=\dfrac{dN^\mu}{dE%_\mathrm{loss}}(E_\msh,\mathcal{M})+\dfrac{dN^\tau}{dE_\ma%thrm{loss}}(E_\msh,\mathcal{M})\,,
%\end{equation}
%where $\mathcal{M}=\ $SM, SM+BH. We leave this for further work, for now assuming that a muon track can be identified and its energy can be well reconstructed.

\begin{table}[!htb]
    \centering
    \setlength{\tabcolsep}{2em} % for the horizontal padding
    {\renewcommand{\arraystretch}{1.2}% for the vertical padding
    \begin{tabular}{c c c} \toprule
         Experiment & $M_\star$ (TeV), detection & $M_\star$ (TeV), exclusion \\ 
         & $3\sigma$ ($5\sigma$) & $2\sigma$ ($3\sigma$) \\ \colrule
        IceCube-Gen2 & $<5.1$ ($<4.3$) & $<5.1$ $(<4.3)$  \\ 
         P-ONE & $<6.9$ ($<5.9$) & $<6.9$ $(<6.0)$ \\\botrule
    \end{tabular}}
    \caption{Projections for the 6-extra-dimension fundamental scale that can be detected or excluded with upcoming experiments. We consider detectors with the proposed properties of IceCube-Gen2 and P-ONE with 10 years of running time. See Figure \ref{fig:Nobs} for discovery limits and \ref{fig:Nobsexclude} for exclusion limits for a range of fundamental scales.}
    \label{tab:limits}
\end{table}

We have focused on future ``plum-pudding'' configuration IceCube-like Cherenkov telescopes. Other types of experiments are also planned. The IceCube-Gen2 radio array~\cite{Aartsen:2019swn} should have an effective volume 20 times larger than IceCube-Gen2. Although the muon track energy and time evolution of Chenrenkov light are not available in the radio detector, radio telescopes are still a powerful tool to determine the shower energy ratio in double bang events, which can be complementary to IceCube-Gen2 BH searches. Neutrinos may interact right below the horizon or inside mountain ranges, producing tau leptons which leave the rock and subsequently decay in the air. These neutrinos are called Earth-skimming neutrinos. Gamma ray and Cherenkov telescopes like MAGIC~\cite{Mallamaci:2019mmo}, HAWC\cite{Vargas:2019kbb}, Pierre Auger
Observatory~\cite{Aab:2019auo}, the planned EUSO-SPB2 and POEMMA~\cite{Otte:2019lbq} are sensitive to the Cherenkov emission induced by air showers from tau decay, and future radio telescopes, e.g. GRAND~\cite{Tueros:2019phm}, are able to detect the radio emission from shower originating from Earth-skimming neutrinos with unprecedented sensitivity. We stress that these neutrino telescopes will considerably enlarge the effective volume for BH $n$-bang search and thus improve the sensitivity for BH discovery. A detailed analysis is also left for further work.

If LEDs exist in nature, the next generation of neutrino telescopes could allow us a glimpse beyond our four-dimensional brane. With a plethora of new and unique phenomena, the observation of microscopic black holes in neutrino-nucleon interaction is not only interesting in itself, but also opens up the window to new physics above the Planck scale, including insights into the stability of the Standard Model. Near future neutrino and cosmic ray telescopes may shine light on its mystery, and enable us to explore the dimensions beyond the scope of our world.

\section*{Acknowledgements}
We thank Shirley Li for help with FLUKA, and Shivesh Mandalia for helping coax PYTHIA into accepting neutrinos. We further acknowledge Dejan Stojkovic for insightful comments. We also thank the anonymous referees for insightful comments. NS thanks Francis Halzen for inspiring discussions about the project. This work was supported by the Arthur B. McDonald Canadian Astroparticle Physics Research Institute. Computations were performed at the Centre for Advanced Computing, and supported in part by the Canada Foundation for Innovation. Research at Perimeter Institute is supported by the Government of Canada through the Department of Innovation, Science, and Economic Development, and by the Province of Ontario through the Ministry of Research and Innovation.

\bibliographystyle{JHEPmod}
\bibliography{bigbib}

\providecommand{\href}[2]{#2}\begingroup\raggedright\begin{thebibliography}{100}

\bibitem{ArkaniHamed:1998rs}
N.~Arkani-Hamed, S.~Dimopoulos, and G.~R. Dvali, {\it {The Hierarchy problem
  and new dimensions at a millimeter}},  {\em Phys. Lett.} {\bf B429} (1998)
  263--272, [\href{http://arxiv.org/abs/hep-ph/9803315}{{\tt hep-ph/9803315}}].

\bibitem{Antoniadis:1998ig}
I.~Antoniadis, N.~Arkani-Hamed, S.~Dimopoulos, and G.~R. Dvali, {\it {New
  dimensions at a millimeter to a Fermi and superstrings at a TeV}},  {\em
  Phys. Lett.} {\bf B436} (1998) 257--263,
  [\href{http://arxiv.org/abs/hep-ph/9804398}{{\tt hep-ph/9804398}}].

\bibitem{ArkaniHamed:1998nn}
N.~Arkani-Hamed, S.~Dimopoulos, and G.~R. Dvali, {\it {Phenomenology,
  astrophysics and cosmology of theories with submillimeter dimensions and TeV
  scale quantum gravity}},  {\em Phys. Rev.} {\bf D59} (1999) 086004,
  [\href{http://arxiv.org/abs/hep-ph/9807344}{{\tt hep-ph/9807344}}].

\bibitem{Argyres:1998qn}
P.~C. Argyres, S.~Dimopoulos, and J.~March-Russell, {\it {Black holes and
  submillimeter dimensions}},  {\em Phys. Lett.} {\bf B441} (1998) 96--104,
  [\href{http://arxiv.org/abs/hep-th/9808138}{{\tt hep-th/9808138}}].

\bibitem{Randall:1999ee}
L.~Randall and R.~Sundrum, {\it {A Large mass hierarchy from a small extra
  dimension}},  {\em Phys. Rev. Lett.} {\bf 83} (1999) 3370--3373,
  [\href{http://arxiv.org/abs/hep-ph/9905221}{{\tt hep-ph/9905221}}].

\bibitem{Randall:1999vf}
L.~Randall and R.~Sundrum, {\it {An Alternative to compactification}},  {\em
  Phys. Rev. Lett.} {\bf 83} (1999) 4690--4693,
  [\href{http://arxiv.org/abs/hep-th/9906064}{{\tt hep-th/9906064}}].

\bibitem{Murata:2014nra}
J.~Murata and S.~Tanaka, {\it {A review of short-range gravity experiments in
  the LHC era}},  {\em Class. Quant. Grav.} {\bf 32} (2015) 033001,
  [\href{http://arxiv.org/abs/1408.3588}{{\tt arXiv:1408.3588}}].

\bibitem{Hanhart:2001fx}
C.~Hanhart, J.~A. Pons, D.~R. Phillips, and S.~Reddy, {\it {The Likelihood of
  GODs' existence: Improving the SN1987a constraint on the size of large
  compact dimensions}},  {\em Phys. Lett.} {\bf B509} (2001) 1--9,
  [\href{http://arxiv.org/abs/astro-ph/0102063}{{\tt astro-ph/0102063}}].

\bibitem{Hannestad:2003yd}
S.~Hannestad and G.~G. Raffelt, {\it {Supernova and neutron star limits on
  large extra dimensions reexamined}},  {\em Phys. Rev.} {\bf D67} (2003)
  125008, [\href{http://arxiv.org/abs/hep-ph/0304029}{{\tt hep-ph/0304029}}].
  [Erratum: Phys. Rev.D69,029901(2004)].

\bibitem{Sirunyan:2018xwt}
CMS: A.~M. Sirunyan {\em et.~al.}, {\it {Search for black holes and sphalerons
  in high-multiplicity final states in proton-proton collisions at $
  \sqrt{s}=13 $ TeV}},  {\em JHEP} {\bf 11} (2018) 042,
  [\href{http://arxiv.org/abs/1805.06013}{{\tt arXiv:1805.06013}}].

\bibitem{Sirunyan:2018wcm}
CMS: A.~M. Sirunyan {\em et.~al.}, {\it {Search for new physics in dijet
  angular distributions using proton–proton collisions at $\sqrt{s}=$ 13 TeV
  and constraints on dark matter and other models}},  {\em Eur. Phys. J.} {\bf
  C78} (2018) 789, [\href{http://arxiv.org/abs/1803.08030}{{\tt
  arXiv:1803.08030}}].

\bibitem{Mack:2018fny}
K.~J. Mack and R.~McNees, {\it {Bounds on extra dimensions from micro black
  holes in the context of the metastable Higgs vacuum}},  {\em Phys. Rev.} {\bf
  D99} (2019) 063001, [\href{http://arxiv.org/abs/1809.05089}{{\tt
  arXiv:1809.05089}}].

\bibitem{thorne1995black}
K.~Thorne, {\em Black Holes \& Time Warps: Einstein's Outrageous Legacy
  (Commonwealth Fund Book Program)}.
\newblock WW Norton \& Company, 1995.

\bibitem{Banks:1999gd}
T.~Banks and W.~Fischler, {\it {A Model for high-energy scattering in quantum
  gravity}},  \href{http://arxiv.org/abs/hep-th/9906038}{{\tt hep-th/9906038}}.

\bibitem{Dimopoulos:2001hw}
S.~Dimopoulos and G.~L. Landsberg, {\it {Black holes at the LHC}},  {\em Phys.
  Rev. Lett.} {\bf 87} (2001) 161602,
  [\href{http://arxiv.org/abs/hep-ph/0106295}{{\tt hep-ph/0106295}}].

\bibitem{Giddings:2001bu}
S.~B. Giddings and S.~D. Thomas, {\it {High-energy colliders as black hole
  factories: The End of short distance physics}},  {\em Phys. Rev.} {\bf D65}
  (2002) 056010, [\href{http://arxiv.org/abs/hep-ph/0106219}{{\tt
  hep-ph/0106219}}].

\bibitem{Chamblin:2004zg}
A.~Chamblin, F.~Cooper, and G.~C. Nayak, {\it {SUSY production from TeV scale
  blackhole at CERN LHC}},  {\em Phys. Rev.} {\bf D70} (2004) 075018,
  [\href{http://arxiv.org/abs/hep-ph/0405054}{{\tt hep-ph/0405054}}].

\bibitem{Harris:2004mf}
C.~M. Harris, {\em {Physics beyond the standard model: Exotic leptons and black
  holes at future colliders}}.
\newblock PhD thesis, Cambridge U., 2004.
\newblock \href{http://arxiv.org/abs/hep-ph/0502005}{{\tt hep-ph/0502005}}.

\bibitem{Cavaglia:2006uk}
M.~Cavaglia, R.~Godang, L.~Cremaldi, and D.~Summers, {\it {Catfish: A Monte
  Carlo simulator for black holes at the LHC}},  {\em Comput. Phys. Commun.}
  {\bf 177} (2007) 506--517, [\href{http://arxiv.org/abs/hep-ph/0609001}{{\tt
  hep-ph/0609001}}].

\bibitem{Alberghi:2006km}
G.~L. Alberghi, R.~Casadio, and A.~Tronconi, {\it {Quantum Gravity Effects in
  Black Holes at the LHC}},  {\em J. Phys.} {\bf G34} (2007) 767--778,
  [\href{http://arxiv.org/abs/hep-ph/0611009}{{\tt hep-ph/0611009}}].

\bibitem{Cavaglia:2007ir}
M.~Cavaglia, R.~Godang, L.~M. Cremaldi, and D.~J. Summers, {\it {Signatures of
  black holes at the LHC}},  {\em JHEP} {\bf 06} (2007) 055,
  [\href{http://arxiv.org/abs/0707.0317}{{\tt arXiv:0707.0317}}].

\bibitem{Calmet:2008dg}
X.~Calmet, W.~Gong, and S.~D.~H. Hsu, {\it {Colorful quantum black holes at the
  LHC}},  {\em Phys. Lett.} {\bf B668} (2008) 20--23,
  [\href{http://arxiv.org/abs/0806.4605}{{\tt arXiv:0806.4605}}].

\bibitem{Nayak:2009fv}
G.~C. Nayak, {\it {Dark Matter Production at LHC from Black Hole Remnants}},
  {\em Phys. Part. Nucl. Lett.} {\bf 8} (2011) 337--341,
  [\href{http://arxiv.org/abs/0901.3358}{{\tt arXiv:0901.3358}}].

\bibitem{Gingrich:2009hj}
D.~M. Gingrich, {\it {Quantum black holes with charge, colour, and spin at the
  LHC}},  {\em J. Phys.} {\bf G37} (2010) 105008,
  [\href{http://arxiv.org/abs/0912.0826}{{\tt arXiv:0912.0826}}].

\bibitem{Gingrich:2009da}
D.~M. Gingrich, {\it {Monte Carlo event generator for black hole production and
  decay in proton-proton collisions}},  {\em Comput. Phys. Commun.} {\bf 181}
  (2010) 1917--1924, [\href{http://arxiv.org/abs/0911.5370}{{\tt
  arXiv:0911.5370}}].

\bibitem{Landsberg:2014bya}
G.~Landsberg, {\it {Black Holes at the Large Hadron Collider}},  {\em Fundam.
  Theor. Phys.} {\bf 178} (2015) 267--292.

\bibitem{Song:2019lxb}
N.~Song and A.~C. Vincent, {\it {Discovery and spectroscopy of dark matter and
  dark sectors with microscopic black holes}},
  \href{http://arxiv.org/abs/1907.08628}{{\tt arXiv:1907.08628}}.

\bibitem{Hawking:1974sw}
S.~W. Hawking, {\it {Particle Creation by Black Holes}},  {\em Commun. Math.
  Phys.} {\bf 43} (1975) 199--220. [,167(1975)].

\bibitem{Emparan:2001kf}
R.~Emparan, M.~Masip, and R.~Rattazzi, {\it {Cosmic rays as probes of large
  extra dimensions and TeV gravity}},  {\em Phys. Rev.} {\bf D65} (2002)
  064023, [\href{http://arxiv.org/abs/hep-ph/0109287}{{\tt hep-ph/0109287}}].

\bibitem{Ringwald:2001vk}
A.~Ringwald and H.~Tu, {\it {Collider versus cosmic ray sensitivity to black
  hole production}},  {\em Phys. Lett.} {\bf B525} (2002) 135--142,
  [\href{http://arxiv.org/abs/hep-ph/0111042}{{\tt hep-ph/0111042}}].

\bibitem{Anchordoqui:2001cg}
L.~A. Anchordoqui, J.~L. Feng, H.~Goldberg, and A.~D. Shapere, {\it {Black
  holes from cosmic rays: Probes of extra dimensions and new limits on TeV
  scale gravity}},  {\em Phys. Rev.} {\bf D65} (2002) 124027,
  [\href{http://arxiv.org/abs/hep-ph/0112247}{{\tt hep-ph/0112247}}].

\bibitem{Feng:2001ib}
J.~L. Feng and A.~D. Shapere, {\it {Black hole production by cosmic rays}},
  {\em Phys. Rev. Lett.} {\bf 88} (2002) 021303,
  [\href{http://arxiv.org/abs/hep-ph/0109106}{{\tt hep-ph/0109106}}].

\bibitem{Jain:2000pu}
P.~Jain, D.~W. McKay, S.~Panda, and J.~P. Ralston, {\it {Extra dimensions and
  strong neutrino nucleon interactions above 10**19-eV: Breaking the GZK
  barrier}},  {\em Phys. Lett.} {\bf B484} (2000) 267--274,
  [\href{http://arxiv.org/abs/hep-ph/0001031}{{\tt hep-ph/0001031}}].

\bibitem{Anchordoqui:2001ei}
L.~Anchordoqui and H.~Goldberg, {\it {Experimental signature for black hole
  production in neutrino air showers}},  {\em Phys. Rev.} {\bf D65} (2002)
  047502, [\href{http://arxiv.org/abs/hep-ph/0109242}{{\tt hep-ph/0109242}}].

\bibitem{Ahn:2003qn}
E.-J. Ahn, M.~Ave, M.~Cavaglia, and A.~V. Olinto, {\it {TeV black hole
  fragmentation and detectability in extensive air showers}},  {\em Phys. Rev.}
  {\bf D68} (2003) 043004, [\href{http://arxiv.org/abs/hep-ph/0306008}{{\tt
  hep-ph/0306008}}].

\bibitem{Ave:2003ew}
M.~Ave, E.-J. Ahn, M.~Cavaglia, and A.~V. Olinto, {\it {Probing TeV gravity
  with extensive air - showers}},  in {\em {Proceedings, 28th International
  Cosmic Ray Conference (ICRC 2003): Tsukuba, Japan, July 31-August 7, 2003}}
  (2003) 1641--1644, [\href{http://arxiv.org/abs/astro-ph/0306344}{{\tt
  astro-ph/0306344}}].

\bibitem{Stojkovic:2005fx}
D.~Stojkovic, G.~D. Starkman, and D.-C. Dai, {\it {Why black hole production in
  scattering of cosmic ray neutrinos is generically suppressed}},  {\em Phys.
  Rev. Lett.} {\bf 96} (2006) 041303,
  [\href{http://arxiv.org/abs/hep-ph/0505112}{{\tt hep-ph/0505112}}].

\bibitem{Uehara:2001yk}
Y.~Uehara, {\it {Production and detection of black holes at neutrino array}},
  {\em Prog. Theor. Phys.} {\bf 107} (2002) 621--624,
  [\href{http://arxiv.org/abs/hep-ph/0110382}{{\tt hep-ph/0110382}}].

\bibitem{Dutta:2002ca}
S.~I. Dutta, M.~H. Reno, and I.~Sarcevic, {\it {On black hole detection with
  the OWL / Airwatch telescope}},  {\em Phys. Rev.} {\bf D66} (2002) 033002,
  [\href{http://arxiv.org/abs/hep-ph/0204218}{{\tt hep-ph/0204218}}].

\bibitem{Kowalski:2002gb}
M.~Kowalski, A.~Ringwald, and H.~Tu, {\it {Black holes at neutrino
  telescopes}},  {\em Phys. Lett.} {\bf B529} (2002) 1--9,
  [\href{http://arxiv.org/abs/hep-ph/0201139}{{\tt hep-ph/0201139}}].

\bibitem{AlvarezMuniz:2002ga}
J.~Alvarez-Muniz, J.~L. Feng, F.~Halzen, T.~Han, and D.~Hooper, {\it {Detecting
  microscopic black holes with neutrino telescopes}},  {\em Phys. Rev.} {\bf
  D65} (2002) 124015, [\href{http://arxiv.org/abs/hep-ph/0202081}{{\tt
  hep-ph/0202081}}].

\bibitem{Kisselev:2010zz}
A.~V. Kisselev, {\it {High-energy cosmic neutrinos and extra spatial
  dimensions}},  {\em Phys. Atom. Nucl.} {\bf 73} (2010) 996--1014. [Yad.
  Fiz.73,1033(2010)].

\bibitem{Jain:2002kz}
P.~Jain, S.~Kar, D.~W. McKay, S.~Panda, and J.~P. Ralston, {\it {Angular
  dependence of neutrino flux in KM**3 detectors in low scale gravity models}},
   {\em Phys. Rev.} {\bf D66} (2002) 065018,
  [\href{http://arxiv.org/abs/hep-ph/0205052}{{\tt hep-ph/0205052}}].

\bibitem{Reynoso:2013jya}
M.~M. Reynoso and O.~A. Sampayo, {\it {Effects of large extra dimensions on
  cosmogenic neutrino fluxes}},  {\em J. Phys.} {\bf G40} (2013) 055202.
  [Erratum: J. Phys.G40,079501(2013)].

\bibitem{Anchordoqui:2006fn}
L.~A. Anchordoqui, M.~M. Glenz, and L.~Parker, {\it {Black Holes at IceCube
  Neutrino Telescope}},  {\em Phys. Rev.} {\bf D75} (2007) 024011,
  [\href{http://arxiv.org/abs/hep-ph/0610359}{{\tt hep-ph/0610359}}].

\bibitem{Arsene:2013nca}
N.~Arsene, X.~Calmet, L.~I. Caramete, and O.~Micu, {\it {Back-to-Back Black
  Holes decay Signature at Neutrino Observatories}},  {\em Astropart. Phys.}
  {\bf 54} (2014) 132--138, [\href{http://arxiv.org/abs/1303.4603}{{\tt
  arXiv:1303.4603}}].

\bibitem{Illana:2005pu}
J.~I. Illana, M.~Masip, and D.~Meloni, {\it {TeV gravity at neutrino
  telescopes}},  {\em Phys. Rev.} {\bf D72} (2005) 024003,
  [\href{http://arxiv.org/abs/hep-ph/0504234}{{\tt hep-ph/0504234}}].

\bibitem{li:2016kra}
S.~W. Li, M.~Bustamante, and J.~F. Beacom, {\it {Echo Technique to Distinguish
  Flavors of Astrophysical Neutrinos}},
  \href{http://arxiv.org/abs/1606.06290}{{\tt arXiv:1606.06290}}.

\bibitem{Aartsen:2013bka}
IceCube: M.~G. Aartsen {\em et.~al.}, {\it {First observation of PeV-energy
  neutrinos with IceCube}},  {\em Phys. Rev. Lett.} {\bf 111} (2013) 021103,
  [\href{http://arxiv.org/abs/1304.5356}{{\tt arXiv:1304.5356}}].

\bibitem{Aartsen:2013jdh}
IceCube: M.~G. Aartsen {\em et.~al.}, {\it {Evidence for High-Energy
  Extraterrestrial Neutrinos at the IceCube Detector}},  {\em Science} {\bf
  342} (2013) 1242856, [\href{http://arxiv.org/abs/1311.5238}{{\tt
  arXiv:1311.5238}}].

\bibitem{Aartsen:2014gkd}
IceCube: M.~G. Aartsen {\em et.~al.}, {\it {Observation of High-Energy
  Astrophysical Neutrinos in Three Years of IceCube Data}},  {\em Phys. Rev.
  Lett.} {\bf 113} (2014) 101101, [\href{http://arxiv.org/abs/1405.5303}{{\tt
  arXiv:1405.5303}}].

\bibitem{Aartsen:2015zva}
IceCube: M.~G. Aartsen {\em et.~al.}, {\it {The IceCube Neutrino Observatory -
  Contributions to ICRC 2015 Part II: Atmospheric and Astrophysical Diffuse
  Neutrino Searches of All Flavors}},  in {\em {Proceedings, 34th International
  Cosmic Ray Conference (ICRC 2015): The Hague, The Netherlands, July 30-August
  6, 2015}} (2015) [\href{http://arxiv.org/abs/1510.05223}{{\tt
  arXiv:1510.05223}}].

\bibitem{Schneider:2019ayi}
A.~Schneider, {\it {Characterization of the Astrophysical Diffuse Neutrino Flux
  with IceCube High-Energy Starting Events}},  in {\em {36th International
  Cosmic Ray Conference (ICRC 2019) Madison, Wisconsin, USA, July 24-August 1,
  2019}} (2019) [\href{http://arxiv.org/abs/1907.11266}{{\tt
  arXiv:1907.11266}}].

\bibitem{Aartsen:2014njl}
IceCube: M.~G. Aartsen {\em et.~al.}, {\it {IceCube-Gen2: A Vision for the
  Future of Neutrino Astronomy in Antarctica}},
  \href{http://arxiv.org/abs/1412.5106}{{\tt arXiv:1412.5106}}.

\bibitem{Dai:2007ki}
D.-C. Dai, G.~Starkman, {\em et.~al.}, {\it {BlackMax: A black-hole event
  generator with rotation, recoil, split branes, and brane tension}},  {\em
  Phys. Rev.} {\bf D77} (2008) 076007,
  [\href{http://arxiv.org/abs/0711.3012}{{\tt arXiv:0711.3012}}].

\bibitem{Yoshino:2001ik}
H.~Yoshino, Y.~Nambu, and A.~Tomimatsu, {\it {Hoop conjecture for colliding
  black holes: Nontime symmetric initial data}},  {\em Phys. Rev.} {\bf D65}
  (2002) 064034, [\href{http://arxiv.org/abs/gr-qc/0109016}{{\tt
  gr-qc/0109016}}].

\bibitem{Yoshino:2002br}
H.~Yoshino and Y.~Nambu, {\it {High-energy headon collisions of particles and
  hoop conjecture}},  {\em Phys. Rev.} {\bf D66} (2002) 065004,
  [\href{http://arxiv.org/abs/gr-qc/0204060}{{\tt gr-qc/0204060}}].

\bibitem{Cardoso:2002pa}
V.~Cardoso, O.~J.~C. Dias, and J.~P.~S. Lemos, {\it {Gravitational radiation in
  D-dimensional space-times}},  {\em Phys. Rev.} {\bf D67} (2003) 064026,
  [\href{http://arxiv.org/abs/hep-th/0212168}{{\tt hep-th/0212168}}].

\bibitem{Yoshino:2005hi}
H.~Yoshino and V.~S. Rychkov, {\it {Improved analysis of black hole formation
  in high-energy particle collisions}},  {\em Phys. Rev.} {\bf D71} (2005)
  104028, [\href{http://arxiv.org/abs/hep-th/0503171}{{\tt hep-th/0503171}}].
  [Erratum: Phys. Rev.D77,089905(2008)].

\bibitem{Anchordoqui:2003jr}
L.~A. Anchordoqui, J.~L. Feng, H.~Goldberg, and A.~D. Shapere, {\it {Updated
  limits on TeV scale gravity from absence of neutrino cosmic ray showers
  mediated by black holes}},  {\em Phys. Rev.} {\bf D68} (2003) 104025,
  [\href{http://arxiv.org/abs/hep-ph/0307228}{{\tt hep-ph/0307228}}].

\bibitem{Markov:1982}
{\em {Quantum gravitation. Publications of the 2nd seminar: quantum theory of
  gravitation. Moscow, 13 - 15 October 1981.}}, 1982.

\bibitem{Hossenfelder:2003dy}
S.~Hossenfelder, M.~Bleicher, S.~Hofmann, H.~Stoecker, and A.~V. Kotwal, {\it
  {Black hole relics in large extra dimensions}},  {\em Phys. Lett.} {\bf B566}
  (2003) 233--239, [\href{http://arxiv.org/abs/hep-ph/0302247}{{\tt
  hep-ph/0302247}}].

\bibitem{dai:2009by}
D.-C. Dai, C.~Issever, {\em et.~al.}, {\it {Manual of BlackMax. A black-hole
  event generator with rotation, recoil, split branes, and brane tension.
  Version 2.02}},  {\em Comput. Phys. Commun.} {\bf 236} (2019) 285--301,
  [\href{http://arxiv.org/abs/0902.3577}{{\tt arXiv:0902.3577}}].

\bibitem{Frolov:2002gf}
V.~P. Frolov and D.~Stojkovic, {\it {Black hole as a point radiator and recoil
  effect in the brane world}},  {\em Phys. Rev. Lett.} {\bf 89} (2002) 151302,
  [\href{http://arxiv.org/abs/hep-th/0208102}{{\tt hep-th/0208102}}].

\bibitem{Frost:2010zz}
J.~A. Frost, {\it {Phenomenology of rotating extra-dimensional black holes at
  hadron colliders}},  {\em AIP Conf. Proc.} {\bf 1200} (2010) 595--598.

\bibitem{Harris:2003eg}
C.~M. Harris and P.~Kanti, {\it {Hawking radiation from a (4+n)-dimensional
  black hole: Exact results for the Schwarzschild phase}},  {\em JHEP} {\bf 10}
  (2003) 014, [\href{http://arxiv.org/abs/hep-ph/0309054}{{\tt
  hep-ph/0309054}}].

\bibitem{Stojkovic:2004hp}
D.~Stojkovic, {\it {Distinguishing between the small ADD and RS black holes in
  accelerators}},  {\em Phys. Rev. Lett.} {\bf 94} (2005) 011603,
  [\href{http://arxiv.org/abs/hep-ph/0409124}{{\tt hep-ph/0409124}}].

\bibitem{Sirunyan:2017jix}
CMS: A.~M. Sirunyan {\em et.~al.}, {\it {Search for new physics in final states
  with an energetic jet or a hadronically decaying $W$ or $Z$ boson and
  transverse momentum imbalance at $\sqrt{s}=13\text{ }\text{ }\mathrm{TeV}$}},
   {\em Phys. Rev.} {\bf D97} (2018) 092005,
  [\href{http://arxiv.org/abs/1712.02345}{{\tt arXiv:1712.02345}}].

\bibitem{Aaboud:2016ewt}
ATLAS: M.~Aaboud {\em et.~al.}, {\it {Search for TeV-scale gravity signatures
  in high-mass final states with leptons and jets with the ATLAS detector at
  $\sqrt{s}=13$ TeV}},  {\em Phys. Lett.} {\bf B760} (2016) 520--537,
  [\href{http://arxiv.org/abs/1606.02265}{{\tt arXiv:1606.02265}}].

\bibitem{palomares-Ruiz:2015mka}
S.~Palomares-Ruiz, A.~C. Vincent, and O.~Mena, {\it {Spectral analysis of the
  high-energy IceCube neutrinos}},  {\em Phys. Rev.} {\bf D91} (2015) 103008,
  [\href{http://arxiv.org/abs/1502.02649}{{\tt arXiv:1502.02649}}].

\bibitem{Sjostrand:2014zea}
T.~Sjöstrand, S.~Ask, {\em et.~al.}, {\it {An Introduction to PYTHIA 8.2}},
  {\em Comput. Phys. Commun.} {\bf 191} (2015) 159--177,
  [\href{http://arxiv.org/abs/1410.3012}{{\tt arXiv:1410.3012}}].

\bibitem{Dulat:2015mca}
S.~Dulat, T.-J. Hou, {\em et.~al.}, {\it {New parton distribution functions
  from a global analysis of quantum chromodynamics}},  {\em Phys. Rev.} {\bf
  D93} (2016) 033006, [\href{http://arxiv.org/abs/1506.07443}{{\tt
  arXiv:1506.07443}}].

\bibitem{Buckley:2014ana}
A.~Buckley, J.~Ferrando, {\em et.~al.}, {\it {LHAPDF6: parton density access in
  the LHC precision era}},  {\em Eur. Phys. J.} {\bf C75} (2015) 132,
  [\href{http://arxiv.org/abs/1412.7420}{{\tt arXiv:1412.7420}}].

\bibitem{Ferrari:2005zk}
A.~Ferrari, P.~R. Sala, A.~Fasso, and J.~Ranft, {\it {FLUKA: A multi-particle
  transport code (Program version 2005)}}, .

\bibitem{Bohlen:2014buj}
T.~T. Böhlen, F.~Cerutti, {\em et.~al.}, {\it {The FLUKA Code: Developments
  and Challenges for High Energy and Medical Applications}},  {\em Nucl. Data
  Sheets} {\bf 120} (2014) 211--214.

\bibitem{Cowen:2007ny}
IceCube: D.~F. Cowen, {\it {Tau neutrinos in IceCube}},  {\em J. Phys. Conf.
  Ser.} {\bf 60} (2007) 227--230.

\bibitem{usner:2017aio}
IceCube: M.~Usner, {\it {Search for Astrophysical Tau Neutrinos in Six Years of
  High-Energy Starting Events in IceCube}},  {\em PoS} {\bf ICRC2017} (2018)
  974.

\bibitem{Baerwald:2012kc}
P.~Baerwald, M.~Bustamante, and W.~Winter, {\it {Neutrino Decays over
  Cosmological Distances and the Implications for Neutrino Telescopes}},  {\em
  JCAP} {\bf 1210} (2012) 020, [\href{http://arxiv.org/abs/1208.4600}{{\tt
  arXiv:1208.4600}}].

\bibitem{Mena:2014sja}
O.~Mena, S.~Palomares-Ruiz, and A.~C. Vincent, {\it {Flavor Composition of the
  High-Energy Neutrino Events in IceCube}},  {\em Phys. Rev. Lett.} {\bf 113}
  (2014) 091103, [\href{http://arxiv.org/abs/1404.0017}{{\tt
  arXiv:1404.0017}}].

\bibitem{Arguelles:2015dca}
C.~A. Argüelles, T.~Katori, and J.~Salvado, {\it {New Physics in Astrophysical
  Neutrino Flavor}},  {\em Phys. Rev. Lett.} {\bf 115} (2015) 161303,
  [\href{http://arxiv.org/abs/1506.02043}{{\tt arXiv:1506.02043}}].

\bibitem{Bustamante:2015waa}
M.~Bustamante, J.~F. Beacom, and W.~Winter, {\it {Theoretically palatable
  flavor combinations of astrophysical neutrinos}},  {\em Phys. Rev. Lett.}
  {\bf 115} (2015) 161302, [\href{http://arxiv.org/abs/1506.02645}{{\tt
  arXiv:1506.02645}}].

\bibitem{Gonzalez-Garcia:2016gpq}
M.~C. Gonzalez-Garcia, M.~Maltoni, I.~Martinez-Soler, and N.~Song, {\it
  {Non-standard neutrino interactions in the Earth and the flavor of
  astrophysical neutrinos}},  {\em Astropart. Phys.} {\bf 84} (2016) 15--22,
  [\href{http://arxiv.org/abs/1605.08055}{{\tt arXiv:1605.08055}}].

\bibitem{Farzan:2018pnk}
Y.~Farzan and S.~Palomares-Ruiz, {\it {Flavor of cosmic neutrinos preserved by
  ultralight dark matter}},  {\em Phys. Rev.} {\bf D99} (2019) 051702,
  [\href{http://arxiv.org/abs/1810.00892}{{\tt arXiv:1810.00892}}].

\bibitem{Ahlers:2018yom}
M.~Ahlers, M.~Bustamante, and S.~Mu, {\it {Unitarity Bounds of Astrophysical
  Neutrinos}},  {\em Phys. Rev.} {\bf D98} (2018) 123023,
  [\href{http://arxiv.org/abs/1810.00893}{{\tt arXiv:1810.00893}}].

\bibitem{Arguelles:2019rbn}
C.~A. Argüelles, M.~Bustamante, {\em et.~al.}, {\it {Fundamental physics with
  high-energy cosmic neutrinos today and in the future}},  in {\em {36th
  International Cosmic Ray Conference (ICRC 2019) Madison, Wisconsin, USA, July
  24-August 1, 2019}} (2019) [\href{http://arxiv.org/abs/1907.08690}{{\tt
  arXiv:1907.08690}}].

\bibitem{vanSanten:2017chb}
IceCube Gen2: J.~van Santen, {\it {IceCube-Gen2: the next-generation neutrino
  observatory for the South Pole}},  {\em PoS} {\bf ICRC2017} (2018) 991.

\bibitem{Berghaus:2009jb}
IceCube: P.~Berghaus, {\it {Muons in IceCube}},  {\em Nucl. Phys. Proc. Suppl.}
  {\bf 196} (2009) 261--266, [\href{http://arxiv.org/abs/0902.0021}{{\tt
  arXiv:0902.0021}}].

\bibitem{Okumura:2017wtz}
Super-Kamiokande: K.~Okumura, {\it {Measurements of the atmospheric neutrino
  flux by Super-Kamiokande: energy spectra, geomagnetic effects, and solar
  modulation}},  {\em J. Phys. Conf. Ser.} {\bf 888} (2017) 012116.

\bibitem{nufit4.1}
``{NuFIT} 4.1 (2019).'' \url{http://www.nu-fit.org/?q=node/211}.

\bibitem{Esteban:2018azc}
I.~Esteban, M.~C. Gonzalez-Garcia, A.~Hernandez-Cabezudo, M.~Maltoni, and
  T.~Schwetz, {\it {Global analysis of three-flavour neutrino oscillations:
  synergies and tensions in the determination of $\theta_23, \delta_CP$, and
  the mass ordering}},  {\em JHEP} {\bf 01} (2019) 106,
  [\href{http://arxiv.org/abs/1811.05487}{{\tt arXiv:1811.05487}}].

\bibitem{Toscano:2019qwd}
S.~Toscano, P.~Coppin, K.~D. de~Vries, N.~van Eijndhoven, and J.~A. Aguilar,
  {\it {Hybrid detection of high-energy cosmic neutrinos with the
  next-generation neutrino detectors at the South Pole}},  in {\em {HAWC
  Contributions to the 36th International Cosmic Ray Conference (ICRC2019)}}
  (2019) [\href{http://arxiv.org/abs/1908.09563}{{\tt arXiv:1908.09563}}].

\bibitem{Bradascio:2019eub}
IceCube: F.~Bradascio and T.~Glüsenkamp, {\it {An improved muon track
  reconstruction for IceCube}},  in {\em {HAWC Contributions to the 36th
  International Cosmic Ray Conference (ICRC2019)}} (2019)
  [\href{http://arxiv.org/abs/1908.07961}{{\tt arXiv:1908.07961}}].

\bibitem{ferrari2005fluka}
A.~Ferrari, P.~R. Sala, A.~Fasso, and J.~Ranft, {\it Fluka: A multi-particle
  transport code (program version 2005)},  tech. rep., 2005.

\bibitem{Battistoni:2007zzb}
G.~Battistoni, S.~Muraro, {\em et.~al.}, {\it {The FLUKA code: Description and
  benchmarking}},  {\em AIP Conf. Proc.} {\bf 896} (2007) 31--49.

\bibitem{Bedard:2018zml}
STRAW: J.~Bedard {\em et.~al.}, {\it {STRAW (STRings for Absorption length in
  Water): pathfinder for a neutrino telescope in the deep Pacific Ocean}},
  {\em JINST} {\bf 14} (2019) P02013,
  [\href{http://arxiv.org/abs/1810.13265}{{\tt arXiv:1810.13265}}].

\bibitem{nelles2019}
E.~Resconi, ``Towards a planetary neutrino monitoring system.'' XVIII
  International Workshop on Neutrino Telescopes, 2019.

\bibitem{Hiscock:1987hn}
W.~A. Hiscock, {\it {CAN BLACK HOLES NUCLEATE VACUUM PHASE TRANSITIONS?}},
  {\em Phys. Rev.} {\bf D35} (1987) 1161--1170.

\bibitem{Berezin:1987ea}
V.~A. Berezin, V.~A. Kuzmin, and I.~I. Tkachev, {\it {O(3) Invariant Tunneling
  in General Relativity}},  {\em Phys. Lett.} {\bf B207} (1988) 397--403.

\bibitem{Moss:1984zf}
I.~G. Moss, {\it {BLACK HOLE BUBBLES}},  {\em Phys. Rev.} {\bf D32} (1985)
  1333.

\bibitem{Cheung:2013sxa}
C.~Cheung and S.~Leichenauer, {\it {Limits on New Physics from Black Holes}},
  {\em Phys. Rev.} {\bf D89} (2014) 104035,
  [\href{http://arxiv.org/abs/1309.0530}{{\tt arXiv:1309.0530}}].

\bibitem{Gregory:2013hja}
R.~Gregory, I.~G. Moss, and B.~Withers, {\it {Black holes as bubble nucleation
  sites}},  {\em JHEP} {\bf 03} (2014) 081,
  [\href{http://arxiv.org/abs/1401.0017}{{\tt arXiv:1401.0017}}].

\bibitem{Burda:2015yfa}
P.~Burda, R.~Gregory, and I.~Moss, {\it {Vacuum metastability with black
  holes}},  {\em JHEP} {\bf 08} (2015) 114,
  [\href{http://arxiv.org/abs/1503.07331}{{\tt arXiv:1503.07331}}].

\bibitem{Burda:2016mou}
P.~Burda, R.~Gregory, and I.~Moss, {\it {The fate of the Higgs vacuum}},  {\em
  JHEP} {\bf 06} (2016) 025, [\href{http://arxiv.org/abs/1601.02152}{{\tt
  arXiv:1601.02152}}].

\bibitem{Cuspinera:2019jwt}
L.~Cuspinera, R.~Gregory, K.~M. Marshall, and I.~G. Moss, {\it {Higgs Vacuum
  Decay in a Braneworld}},  \href{http://arxiv.org/abs/1907.11046}{{\tt
  arXiv:1907.11046}}.

\bibitem{Cuspinera:2018woe}
L.~Cuspinera, R.~Gregory, K.~Marshall, and I.~G. Moss, {\it {Higgs Vacuum Decay
  from Particle Collisions?}},  {\em Phys. Rev.} {\bf D99} (2019) 024046,
  [\href{http://arxiv.org/abs/1803.02871}{{\tt arXiv:1803.02871}}].

\bibitem{Andreassen:2017rzq}
A.~Andreassen, W.~Frost, and M.~D. Schwartz, {\it {Scale Invariant Instantons
  and the Complete Lifetime of the Standard Model}},  {\em Phys. Rev.} {\bf
  D97} (2018) 056006, [\href{http://arxiv.org/abs/1707.08124}{{\tt
  arXiv:1707.08124}}].

\bibitem{Chigusa:2018uuj}
S.~Chigusa, T.~Moroi, and Y.~Shoji, {\it {Decay Rate of Electroweak Vacuum in
  the Standard Model and Beyond}},  {\em Phys. Rev.} {\bf D97} (2018) 116012,
  [\href{http://arxiv.org/abs/1803.03902}{{\tt arXiv:1803.03902}}].

\bibitem{koehne2013proposal}
J.-H. Koehne, K.~Frantzen, {\em et.~al.}, {\it Proposal: A tool for propagation
  of charged leptons},  {\em Computer Physics Communications} {\bf 184} (2013)
  2070--2090.

\bibitem{ArkaniHamed:1999dc}
N.~Arkani-Hamed and M.~Schmaltz, {\it {Hierarchies without symmetries from
  extra dimensions}},  {\em Phys. Rev.} {\bf D61} (2000) 033005,
  [\href{http://arxiv.org/abs/hep-ph/9903417}{{\tt hep-ph/9903417}}].

\bibitem{Aartsen:2019swn}
IceCube: M.~G. Aartsen {\em et.~al.}, {\it {Neutrino astronomy with the next
  generation IceCube Neutrino Observatory}},
  \href{http://arxiv.org/abs/1911.02561}{{\tt arXiv:1911.02561}}.

\bibitem{Mallamaci:2019mmo}
MAGIC: M.~Mallamaci, D.~Gora, and E.~Bernardini, {\it {MAGIC as a high-energy
  $\nu_\tau$ detector: performance study to follow-up IceCube transient
  events}},  in {\em {HAWC Contributions to the 36th International Cosmic Ray
  Conference (ICRC2019)}} (2019) [\href{http://arxiv.org/abs/1909.01314}{{\tt
  arXiv:1909.01314}}].

\bibitem{Vargas:2019kbb}
HAWC: H.~L. Vargas, {\it {Prospects of Earth-skimming neutrino detection with
  HAWC}},  in {\em {36th International Cosmic Ray Conference (ICRC 2019)
  Madison, Wisconsin, USA, July 24-August 1, 2019}} (2019)
  [\href{http://arxiv.org/abs/1908.07622}{{\tt arXiv:1908.07622}}].

\bibitem{Aab:2019auo}
Pierre Auger: A.~Aab {\em et.~al.}, {\it {Probing the origin of
  ultra-high-energy cosmic rays with neutrinos in the EeV energy range using
  the Pierre Auger Observatory}},  {\em JCAP} {\bf 1910} (2019) 022,
  [\href{http://arxiv.org/abs/1906.07422}{{\tt arXiv:1906.07422}}].

\bibitem{Otte:2019lbq}
A.~N. Otte, E.~Gazda, {\em et.~al.}, {\it {Development of a Cherenkov Telescope
  for the Detection of Ultrahigh Energy Neutrinos with EUSO-SPB2 and POEMMA}},
  in {\em {36th International Cosmic Ray Conference (ICRC 2019) Madison,
  Wisconsin, USA, July 24-August 1, 2019}}, {\em PoS} {\bf ICRC2019} (2019)
  [\href{http://arxiv.org/abs/1907.08728}{{\tt arXiv:1907.08728}}].

\bibitem{Tueros:2019phm}
GRAND: M.~Tueros, {\it {Grand, A Giant Radio Array For Neutrino Detection:
  Objectives, Design and Current Status}},  {\em EPJ Web Conf.} {\bf 216}
  (2019) 01006.

\bibitem{Dutta:2000jv}
S.~I. Dutta, M.~H. Reno, and I.~Sarcevic, {\it {Tau neutrinos underground:
  Signals of muon-neutrino ---\textgreater tau neutrino oscillations with
  extragalactic neutrinos}},  {\em Phys. Rev.} {\bf D62} (2000) 123001,
  [\href{http://arxiv.org/abs/hep-ph/0005310}{{\tt hep-ph/0005310}}].

\end{thebibliography}\endgroup

\appendix 
\section{Expected Topology as a function of Neutrino flavor}
\label{sec:xss}
In this section, we present the cross sections of Standard Model interactions classified by different event topologies.

\textbf{\textit{Tracks.}} The tracks can be produced in $\nu_\mu$ and $\nu_\tau$ charged current interactions. In the latter case, a tau produced in the interaction can either decay to muon or leave the detector. We generally require the tau energy to be higher than 5~PeV to clearly identify the tau track~\cite{Cowen:2007ny}. The number of track events from $\nu_\mu$ CC is given by Eq.~\eqref{eq:NXi} where $\Xi=$tr and the total $\nu_\mu-$nucleon interaction cross section is given by
\begin{equation}
    \sigma^{\nu_\mu N\rightarrow \mtr}(E_\nu)=\int_0^1 dy \dfrac{d\sigma_{\nu_\mu}^{CC}(E_\nu,y)}{dy}\Theta(E_\nu y-\Emin)\,,
\end{equation}
%\begin{equation}
%    N_{\nu_\mu}^{\mtr}=\dfrac{1}{2}TN_A\int_{\Emin}^{\Emax}dE_\nu g M_{\nu_\mu}^{CC}(E_\nu)\dfrac{d\phi_{\nu_\mu}}{dE_\nu}\int_0^1 dy \dfrac{d\sigma_{\nu_\mu}^{CC}(E_\nu,y)}{dy}\Theta(E_\nu y-\Emin)\,,
%    \label{eq:Nmutr}
%\end{equation}
where $d\sigma_{\nu_\mu}^{CC}/dy$ is the differential cross section in the interaction. The step function $\Theta$ is added to satisfy requirement of the minimum energy deposition (we require the cascade energy associated with the track to be above 20~PeV in our flavor composition analysis in Sec.~\ref{sec:flavorcomp}).

Taus from $\nu_\tau$ CC events can also leave a track. This can be realised either when the tau decays to a muon or when it escapes the detector without decaying. The number of track events from $\nu_\tau$ CC is also given by Eq.~\eqref{eq:NXi} but the total cross section is instead
\begin{equation}
    \sigma^{\nu_\tau N\rightarrow \mtr}(E_\nu)=\int_0^1 dy \dfrac{d\sigma_{\nu_\tau}^{CC}(E_\nu,y)}{dy}
        \times \left(f_\mu+(1-f_\mu)P_\mesc(E_\nu)\right)\Theta(E_\nu y-\Emin)\Theta(E_\nu (1-y)-\Emin^\tau)\,,
\end{equation}
% \begin{align}
%         N_{\nu_\tau}^{\mtr}=&\dfrac{1}{2}TN_A\int_{\Emin}^{\Emax}dE_\nu gM_{\nu_\tau}^{CC}(E_\nu)\dfrac{d\phi_{\nu_\tau}}{dE_\nu}\int_0^1 dy \dfrac{d\sigma_{\nu_\tau}^{CC}(E_\nu,y)}{dy}\nonumber\\
%         &\times \left(f_\mu+(1-f_\mu)P_\mesc(E_\nu)\right)\Theta(E_\nu y-\Emin)\Theta(E_\nu (1-y)-\Emin^\tau)\,,
%         \label{eq:Ntautr}
% \end{align}
where $\Emin^\tau=5\ $PeV is the minimum energy of the $\tau$, $f_\mu=0.174$ is the branching ratio of a tau decaying into muons, and $P_\mesc$ is the probability for the tau to leave the detector. In IceCube-Gen2, which we model as a 7.9 km$^3$ cylindrical detector with a height of 1.25 km and a radius of 1.42 km, the criterion for a partly contained tau track is roughly
\begin{equation}
    t>t_\mathrm{esc}=\dfrac{\langle L \rangle}{\gamma c}\,,
\end{equation}
where $t$ is the lifetime of a $\tau$ in the boosted frame and $\langle L \rangle=967$ m is the average distance the tau travels before leaving the detector. $\langle L \rangle$ is obtained from Eq.~(22) of Ref. \cite{palomares-Ruiz:2015mka} without the dust zone. The boost factor $\gamma=\bar{E}_\tau/m_\tau$ with the average tau energy is given by
\begin{equation}
    \bar{E}_\tau(E_\nu)=\dfrac{\int_0^1 dy \dfrac{d\sigma_{\nu_\tau}^{CC}(E_\nu,y)}{dy}\times E_\nu(1-y)\Theta(E_\nu y-\Emin)\Theta(E_\nu (1-y)-\Emin^\tau)}{\int_0^1 dy \dfrac{d\sigma_{\nu_\tau}^{CC}(E_\nu,y)}{dy}\times \Theta(E_\nu y-\Emin)\Theta(E_\nu (1-y)-\Emin^\tau)}\,,
\end{equation}
and $m_\tau=1.78\ $GeV is the mass of tau. The lifetime $t$ follows the distribution
\begin{equation}
    f(t|\bar{E}_\tau)=\dfrac{1}{\gamma \tau_0}\exp\left(-\dfrac{t}{\gamma \tau_0}\right)\,,
\end{equation}
where $\tau_0=2.91\times10^{-13}\ $s is the average lifetime of tau at rest. This yields
\begin{equation}
    P_\mesc(E_\nu)=\exp\left(-\dfrac{t_\mtr}{\gamma \tau_0}\right)\,.
\end{equation}

\textbf{\textit{Double Bang.}} A ``double bang'' signature is produced when the tau created from a $\nu_\tau$ charged current interaction decays electronically or hadronically inside the detector, which makes a visible second shower. We require the tau to travel farther than 100~m before it decays to clearly separate the second shower from the primary one~\cite{Cowen:2007ny}. Following~\cite{usner:2017aio} we construct the energy asymmetry $E_A=(E_1-E_2)/(E_1+E_2)$ where $E_1$ and $E_2$ are the energy of the first and second showers. We require $-0.98\le E_A \le 0.3$. Another criterion is that the minimum deposited energy must be at least 20~PeV, i.e. $E_1+E_2>20\ $PeV. In Eq.~\eqref{eq:NXi} the cross section of double bang events from $\nu_\tau$ CC interactions is
\begin{align}
    \sigma^{\nu_\tau N\rightarrow \mdb}(E_\nu)=&\int_0^1 dy \dfrac{d\sigma_{\nu_\tau}^{CC}(E_\nu,y)}{dy}
        \times \left(\int_0^1 dz \dfrac{dn(\tau\rightarrow\mathrm{had})}{dz}P_\mdb^\mathrm{had}\Theta(E_\nu(y+(1-y)(1-z)-\Emin)\right.\nonumber\\
        &+ \left.\int_0^1 dz \dfrac{dn(\tau\rightarrow\mathrm{e})}{dz}P_\mdb^\mathrm{e}\Theta(E_\nu(y+(1-y)z-\Emin)\right)H(E_A)\,,
    \label{eq:xsnutaudb}
\end{align}
% \begin{align}
%         N_{\nu_\tau}^{\mdb}=&\dfrac{1}{2}TN_A\int_{\Emin}^{\Emax}dE_\nu gM_{\nu_\tau}^{CC}(E_\nu)\dfrac{d\phi_{\nu_\tau}}{dE_\nu}\int_0^1 dy \dfrac{d\sigma_{\nu_\tau}^{CC}(E_\nu,y)}{dy}\nonumber\\
%         &\times \left(\int_0^1 dz \dfrac{dn(\tau\rightarrow\mathrm{had})}{dz}P_\mdb^\mathrm{had}\Theta(E_\nu(y+(1-y)(1-z)-\Emin)\right.\nonumber\\
%         &+ \left.\int_0^1 dz \dfrac{dn(\tau\rightarrow\mathrm{e})}{dz}P_\mdb^\mathrm{e}\Theta(E_\nu(y+(1-y)z-\Emin)\right)H(E_A)\,,
%         \label{eq:Ntaudb}
% \end{align}
where $dn/dz$ is the energy distribution of the daughter neutrino or electron in the tau decay~\cite{Dutta:2000jv} and
\begin{equation}
    H(E_A)=\begin{cases}
    1\,,&\mathrm{if }\ -0.98\le E_A \le 0.3\,,\\
    0\,,&\mathrm{otherwise}\,.
    \end{cases}
\end{equation}
The energy of the primary shower $E_1=E_\nu y$ while the energy of the second shower depends on the decay channel; $E_2=E_\nu (1-y)(1-z)$ for a hadronic decay and $E_2=E_\nu (1-y)z$ if the tau decays to an electron or positron instead. We neglect the energy loss when the tau propagates before decay. The probability for a tau to travel farther than 100~m and decay in the detector is
\begin{equation}
    P_\mdb^\mathrm{e, had}(E_\nu)=\exp\left(-\dfrac{t_\mdb}{\gamma \tau_0}\right)-\exp\left(-\dfrac{t_\mtr}{\gamma \tau_0}\right)\,,
\end{equation}
where $t_\mdb=100\ \mathrm{m}/\gamma c$ and the average energy of taus in hadronic and electronic decays reads
\begin{equation}
    \bar{E}_\tau^\mathrm{had}(E_\nu)=\dfrac{\int_0^1 dy \dfrac{d\sigma_{\nu_\tau}^{CC}(E_\nu,y)}{dy}\int_0^1 dz \dfrac{dn(\tau\rightarrow\mathrm{had})}{dz}E_\nu(1-y)\Theta(E_\nu(y+(1-y)(1-z)-\Emin)H(E_A)}{\int_0^1 dy \dfrac{d\sigma_{\nu_\tau}^{CC}(E_\nu,y)}{dy}\int_0^1 dz \dfrac{dn(\tau\rightarrow\mathrm{had})}{dz}\Theta(E_\nu(y+(1-y)(1-z)-\Emin)H(E_A)}\,, 
\end{equation}
and
\begin{equation}
    \bar{E}_\tau^\mathrm{e}(E_\nu)=\dfrac{\int_0^1 dy \dfrac{d\sigma_{\nu_\tau}^{CC}(E_\nu,y)}{dy}\int_0^1 dz \dfrac{dn(\tau\rightarrow\mathrm{e})}{dz}E_\nu(1-y)\Theta(E_\nu(y+(1-y)z-\Emin)H(E_A)}{\int_0^1 dy \dfrac{d\sigma_{\nu_\tau}^{CC}(E_\nu,y)}{dy}\int_0^1 dz \dfrac{dn(\tau\rightarrow\mathrm{e})}{dz}\Theta(E_\nu(y+(1-y)z-\Emin)H(E_A)}\,, 
    \label{eq:Ebartautoe}
\end{equation}
respectively.

\textbf{\textit{Showers.}} Showers are produced in neutral current interactions from all neutrino flavors where the rate is shown in Eq.~\eqref{eq:NNC} and the total cross section reads
\begin{equation}
    \sigma^\mathrm{NC}(E_\nu)=\int_0^1 dy \dfrac{d\sigma_{\nu_i}^{NC}(E_\nu,y)}{dy}\Theta(E_\nu y-\Emin)\,
\end{equation}
regardless of flavor.
% \begin{equation}
%     N_{\nu_\alpha}^{\msh}=\dfrac{1}{2}TN_A\int_{\Emin}^{\Emax}dE_\nu gM_{\nu_\alpha}^{NC}(E_\nu)\dfrac{d\phi_{\nu_\alpha}}{dE_\nu}\int_0^1 dy \dfrac{d\sigma_{\nu_\alpha}^{NC}(E_\nu,y)}{dy}\Theta(E_\nu y-\Emin)\,,  
%     \label{eq:Nncsh}
% \end{equation}
Showers are also produced in $\nu_e$ CC interactions where all the neutrino energy is deposited in the detector. The total cross section is
\begin{equation}
    \sigma^{\nu_e N\rightarrow \msh}(E_\nu)=\int_0^1 dy \dfrac{d\sigma_{\nu_e}^{CC}(E_\nu,y)}{dy}\,. 
\end{equation}
% \begin{equation}
%         N_{\nu_e}^{\msh}=\dfrac{1}{2}TN_A\int_{\Emin}^{\Emax}dE_\nu gM_{\nu_e}^{CC}(E_\nu)\dfrac{d\phi_{\nu_e}}{dE_\nu}\int_0^1 dy \dfrac{d\sigma_{\nu_e}^{CC}(E_\nu,y)}{dy}\,. 
%         \label{eq:Nesh}
% \end{equation}
The identification of shower events from $\nu_\tau$ CC interaction is more tricky. We consider two types of shower events. In the one case, the tau travels less than 100~m before it decays. In the other case, the tau travels farther than 100~m and decays in the detector, but the energy asymmetry condition for a double bang is not satisfied. The cross section of the former is given by
\begin{align}
    \sigma^{\nu_\tau N\rightarrow \msh}(E_\nu)=&\int_0^1 dy \dfrac{d\sigma_{\nu_\tau}^{CC}(E_\nu,y)}{dy}
        \times \left(\int_0^1 dz \dfrac{dn(\tau\rightarrow\mathrm{had})}{dz}P_\msh^\mathrm{had}\Theta(E_\nu(y+(1-y)(1-z)-\Emin)\right.\nonumber\\
        &+ \left.\int_0^1 dz \dfrac{dn(\tau\rightarrow\mathrm{e})}{dz}P_\msh^\mathrm{e}\Theta(E_\nu(y+(1-y)z-\Emin)\right)\,,
\end{align}
% \begin{align}
%         N_{\nu_\tau}^{\msh}=&\dfrac{1}{2}TN_A\int_{\Emin}^{\Emax}dE_\nu gM_{\nu_\tau}^{CC}(E_\nu)\dfrac{d\phi_{\nu_\tau}}{dE_\nu}\int_0^1 dy \dfrac{d\sigma_{\nu_\tau}^{CC}(E_\nu,y)}{dy}\nonumber\\
%         &\times \left(\int_0^1 dz \dfrac{dn(\tau\rightarrow\mathrm{had})}{dz}P_\msh^\mathrm{had}\Theta(E_\nu(y+(1-y)(1-z)-\Emin)\right.\nonumber\\
%         &+ \left.\int_0^1 dz \dfrac{dn(\tau\rightarrow\mathrm{e})}{dz}P_\msh^\mathrm{e}\Theta(E_\nu(y+(1-y)z-\Emin)\right)\,,
%         \label{eq:Ntaush}
% \end{align}
where 
\begin{equation}
    P_\msh^\mathrm{e, had}(E_\nu)=1-\exp\left(-\dfrac{t_\mdb}{\gamma \tau_0}\right)\,,
\end{equation}
and the average tau energy reads
\begin{equation}
    \bar{E}_\tau^\mathrm{had}(E_\nu)=\dfrac{\int_0^1 dy \dfrac{d\sigma_{\nu_\tau}^{CC}(E_\nu,y)}{dy}\int_0^1 dz \dfrac{dn(\tau\rightarrow\mathrm{had})}{dz}E_\nu(1-y)\Theta(E_\nu(y+(1-y)(1-z)-\Emin)}{\int_0^1 dy \dfrac{d\sigma_{\nu_\tau}^{CC}(E_\nu,y)}{dy}\int_0^1 dz \dfrac{dn(\tau\rightarrow\mathrm{had})}{dz}\Theta(E_\nu(y+(1-y)(1-z)-\Emin)}\,, 
\end{equation}
and
\begin{equation}
    \bar{E}_\tau^\mathrm{e}(E_\nu)=\dfrac{\int_0^1 dy \dfrac{d\sigma_{\nu_\tau}^{CC}(E_\nu,y)}{dy}\int_0^1 dz \dfrac{dn(\tau\rightarrow\mathrm{e})}{dz}E_\nu(1-y)\Theta(E_\nu(y+(1-y)z-\Emin)}{\int_0^1 dy \dfrac{d\sigma_{\nu_\tau}^{CC}(E_\nu,y)}{dy}\int_0^1 dz \dfrac{dn(\tau\rightarrow\mathrm{e})}{dz}\Theta(E_\nu(y+(1-y)z-\Emin)}\,. 
\end{equation}
For the latter case the cross section is given by \crefrange{eq:xsnutaudb}{eq:Ebartautoe} with $H(E_A)$ replaced by $\widetilde{H}(E_A)$ where $\widetilde{H}(E_A)=1-H(E_A)$.
% \onecolumngrid

\section{Probability distribution of $D$}
\label{sec:pdfd}
Here we show the probability distribution of the distance of the shower events from the Hadron Line $D$ defined in Eq.~\eqref{eq:D} by considering Standard Model NC, $\nu_e$ CC, and $\nu_\tau$ CC interactions, as well as black hole production. As the Planck scale $M_\star$ increases, the combined SM+BH distributions become indistinguishable from the SM.

\begin{figure}[!htb]
    \includegraphics[width=0.6\textwidth]{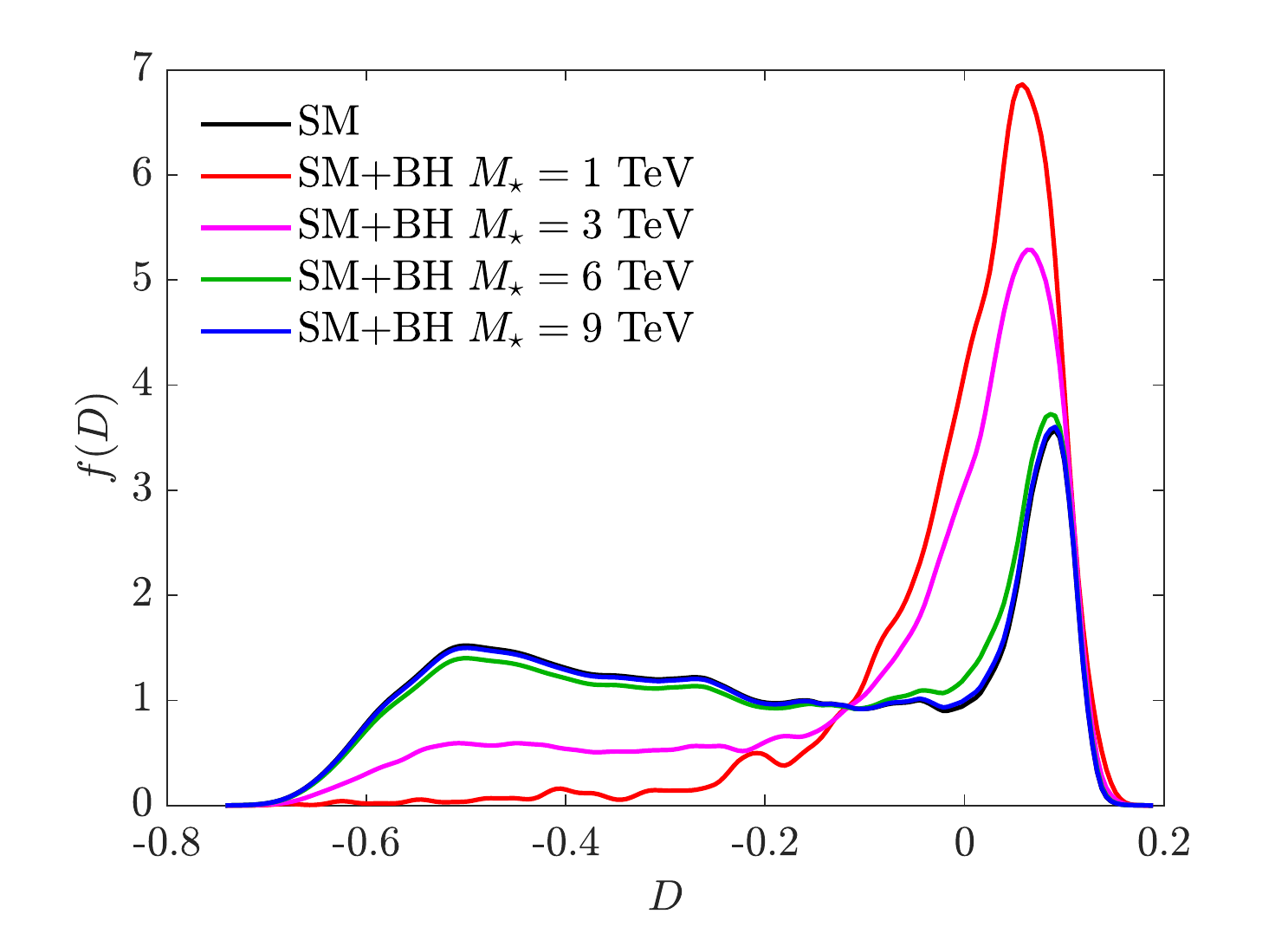} 
    \caption{Probability distribution of the distance from the hadron line $D$ assuming SM only processes (black) and SM with BH production in the cases where $M_\star=$ 1~TeV (red), 3~TeV (pink), 6~TeV (green) and 9~TeV (blue). $E^{-2}$ neutrino spectrum is applied with (1:1:1) flavor ratio and a minimum energy deposition of 20~PeV.}
    \label{fig:pdfd}
\end{figure}

\section{Inital energy loss and cross section}
\label{sec:iniloss}

\begin{figure}[!htb]
    \includegraphics[width=0.6\textwidth]{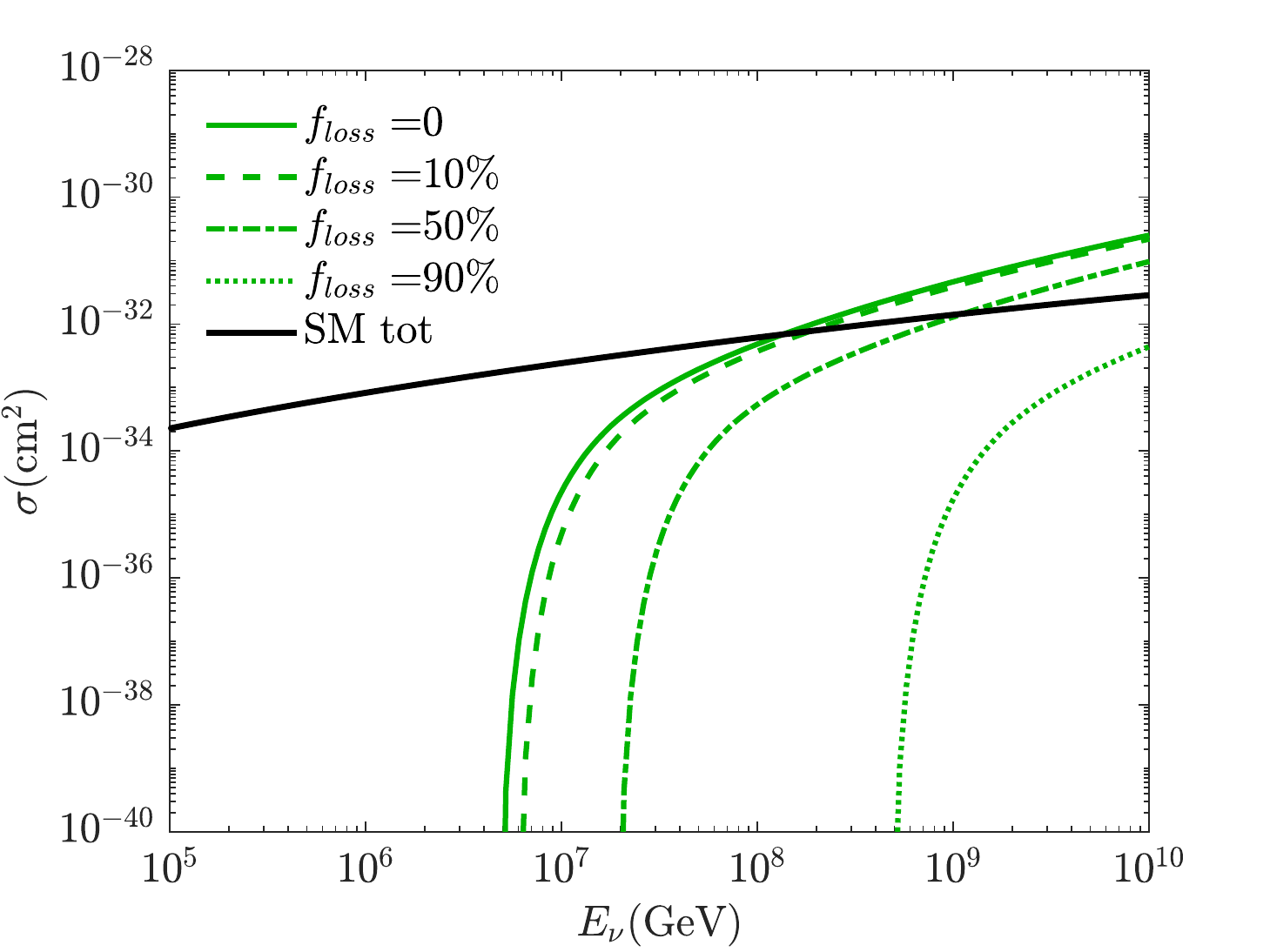} 
    \caption{The black hole production cross section from neutrino-nucleon interactions as a function of the incoming neutrino energy by including the initial energy loss. We assume $M_\star=3$~TeV and $n=6$ extra dimensions. The solid, dashed, dash-dotted and dotted green lines correspond to 0, 10\%, 50\%, 90\% energy loss respectively. The total Standard Model cross section is also shown by the black line.}
    \label{fig:xseloss}
\end{figure}

In case part of the initial particle energy is lost in BH production, Eq.~\eqref{eq:xsec} reads
\begin{equation}
    \sigma^{\nu N\rightarrow BH} = \int_{M_\star^2/(1-f_{loss})^2s}^1 du \pi b_{\mathrm{max}}^2 \sum_{i}f_i(u,Q),
    \label{eq:xseceloss}
\end{equation}
where $f_{loss}$ is the initial energy loss in BH formation. The BH production cross section including the initial energy loss is shown in Fig.~\ref{fig:xseloss}. We note that $f_{loss}>0$ increases the BH production threshold and a significant drop in BH production rate is observed when $f_{loss}$ is great than or equal to $50\%$.

\end{document}